\documentclass[a4paper,11pt]{article}
\pdfoutput=1

\usepackage{jcappub}

\usepackage[T1]{fontenc}

\usepackage[english]{babel}
\usepackage{amsfonts}
\usepackage{txfonts}
\usepackage{graphicx}
\usepackage{amssymb}
\usepackage{ae,aecompl}
\usepackage{fancyhdr}
\usepackage{multicol}
\usepackage{layout}
\usepackage{multirow}
\usepackage{times}
\usepackage{textcomp}
\usepackage{cprotect}

\usepackage[outdir=./figures/]{epstopdf}
\usepackage[export]{adjustbox}
    
\graphicspath{{./figures/}}

\newif\ifAMStwofonts
\AMStwofontstrue

\title{Comparison of different approaches to the quasi-static approximation in Horndeski models}

\author[a,1]{Francesco Pace,\note{Corresponding author.}}
\author[a]{Richard~A. Battye,}
\author[b,c]{Emilio Bellini,}
\author[b]{Lucas Lombriser,}
\author[d]{Filippo Vernizzi,}
\author[e]{and Boris Bolliet}

\affiliation[a]{Jodrell Bank Centre for Astrophysics, Department of Physics and Astronomy, University of Manchester, Manchester, M13 9PL, U.K.}
\affiliation[b]{D\'epartement de Physique Th\'eorique, Universit\'e de Gen\`eve, 24 quai Ernest Ansermet, 1211 Gen\`eve 4, Switzerland}
\affiliation[c]{Oxford Astrophysics, Department of Physics, Keble Road, Oxford, OX1 3RH, U.K.}
\affiliation[d]{Institut de physique theorique, Universit\'e Paris Saclay CEA, CNRS, 91191 Gif-sur-Yvette, France}
\affiliation[e]{Columbia Astrophysics Laboratory, Columbia University, 550 West 120th Street, New York, NY, 10027, USA}

\emailAdd{francesco.pace@manchester.ac.uk}
\emailAdd{richard.battye@manchester.ac.uk}
\emailAdd{emilio.bellini@unige.ch}
\emailAdd{lucas.lombriser@unige.ch}
\emailAdd{filippo.vernizzi@ipht.fr}
\emailAdd{boris.bolliet@gmail.com}

\abstract{
A quasi-static approximation (QSA) for modified gravity can be applied in a number of ways. We consider three different analytical formulations based on applying this approximation to: (1) the field equations; (2) the equations for the two metric potentials; (3) the use of the attractor solution derived within the Equation of State approach. We assess the veracity of these implementations on the effective gravitational constant ($\mu$) and the slip parameter ($\eta$), within the framework of Horndeski models. In particular, for a set of models we compare cosmological observables, i.e., the matter power spectrum and the CMB temperature and lensing angular power spectra, computed using the QSA, with exact numerical solutions. To do that, we use a newly developed branch of the \texttt{CLASS} code: \texttt{QSA\_class}. All three approaches agree exactly on very small scales. Typically, we find that, except for $f(R)$ models where all the three approaches lead to the same result, the quasi-static approximations differ from the numerical calculations on large scales ($k \lesssim 3 - 4 \times 10^{-3}\,h\,{\rm Mpc}^{-1}$). Cosmological observables are reproduced to within 1\% up to scales ${\rm K} = k/H_0$ of the order of a few and multipoles $\ell>5$ for the approaches based on the field equations and on the Equation of State, and we also do not find any appreciable difference if we use the scale-dependent expressions for $\mu$ and $\eta$ with respect to the value on small scales, showing that the formalism and the conclusions are reliable and robust, fixing the range of applicability of the formalism. We discuss why the expressions derived from the equations for the potentials have limited applicability. Our results are in agreement with previous analytical estimates and show that the QSA is a reliable tool and can be used for comparison with current and future observations to constrain models beyond $\Lambda$CDM.}


\arxivnumber{2011.05713}

\begin{document}

\label{firstpage}

\maketitle
\flushbottom

\section{Introduction}
The current accelerated expansion of the Universe has been established by many different probes \citep{Riess1998,Perlmutter1999,Riess2007,Planck2018_VI,Planck2018_VIII,Planck2018_X,DES_Y1_2018,Gruen2018}. Observations are all compatible with the presence of a cosmological constant $\Lambda$. However, its tiny value is disconcertingly smaller than what naturally expected in quantum field theory (see, for example \citep{Weinberg1989,Burgess2013,Padilla2015}). This has led to investigate many extensions of the standard cosmological model where dark energy and modified gravity are responsible for the cosmic acceleration \citep{Joyce2015,Joyce2016,Koyama2016,Sami2016,BeltranJimenez2018,Ishak2018,Euclid2018,Heisenberg2019}.

In models alternative to $\Lambda$CDM there is, in general, a different structure formation history which can be seen in modifications of the matter power spectrum $P(k)$ and angular temperature anisotropy power spectrum $C_{\ell}^{\rm TT}$. Since the field equations describe the evolution of two metric potentials, these modifications can be, in general, parameterised with two independent functions \citep{Zhang2007,Amendola2008a,Bertschinger2008,Bean2010,Pogosian2010}: the effective gravitational constant $\mu$ and the slip parameter $\eta$. For a generic model, these are functions of time (or of the scale factor $a$) and scale $k$. Often, though, simplified and phenomenological expressions that only depend on time are used to study deviations from the $\Lambda$CDM fiducial model.

These two functions are often thought of as being derived from the quasi-static approximation (QSA) \citep{Silvestri2013}. The basic idea is that time derivatives are subdominant with respect to spatial derivatives. This relies on the assumption that the relevant time scale for cosmological perturbations is the Hubble parameter (i.e., $\mathrm{d}/\mathrm{d}t\sim H$). This is the same assumption which leads to the Newtonian limit in general relativity, which on scales smaller than the horizon has proven to be an excellent approximation as demonstrated by $N$-body simulations. However, the application of the QSA formalism can be applied to different, but physically equivalent, equations of motion.

The QSA has been applied to the linearized field equations or gravitational potentials \citep{Silvestri2013,DeFelice2012,Bellini2014,Gleyzes2014}, and more recently, within the formalism of the Equation of State (EoS) approach \citep{Battye2007,Battye2012,Battye2013,Battye2013a,Battye2014,Battye2016a,Battye2017,Battye2018a,Battye2019} in \cite{Pace2019a} and time- and scale-dependent expressions for $\mu$ and $\eta$ were derived. These were verified to exactly recover expressions already presented in the literature \cite{Gubitosi2013,Bellini2014,Gleyzes2014,Lombriser2015a,Piazza2014} in the small-scale limit ${\rm K}=k/(aH)\rightarrow\infty$. Different methods appear to disagree in the limit ${\rm K}\rightarrow 0$, but this is expected because on these scales the QSA does not hold.

For the QSA to be valid, the scales considered have to be below the sound horizon ${\rm K}_{\rm sh}=c_{\rm s}k/(aH)=c_{\rm s}{\rm K}$, where $c_{\rm s}$ is the sound speed associated to the scalar field perturbations \cite{Gleyzes2014,Sawicki2015}. We require, in particular, that $k\gg aH/c_{\rm s}$. Hence, for models with a very small sound speed, the QSA might have very limited applicability or not be applicable at all.

In the literature, a few works concentrated on $f(R)$ models and studied when the exact results are recovered by applying the QSA. In \citep{deLaCruzDombriz2008}, the authors derived a fourth-order growth-factor equation in time\footnote{This is a consequence of the presence of two degrees of freedom.} and compared it to the QSA solution. The two, in general, differ unless $w\sim -1$ (as for the $\Lambda$CDM background) and if $\mathrm{d}f/\mathrm{d}R=f_R\ll1$ at $a=1$. This happens because when $f_R \ll 1$, the coefficients of the fourth- and third-time derivative become negligible and the equation reduces to a second-order one. Moreover, if $w\sim -1$, the coefficients of $\delta^{\prime}$ and $\delta$ have the right QSA limit, i.e., they reduce to the standard expression derived in other works.

In \citep{Noller2014,Sawicki2015}, the analysis is based on the evolution of the scalar field and it specifies the evolution of the background and of the perturbation part. For the latter, oscillations may arise. If oscillations are negligible on scales smaller than the horizon, the QSA works well, otherwise this is not the case. A fundamental assumption in this analysis is that $f_R\ll 1$ and the background is close to $\Lambda$CDM. Nevertheless, it has been shown by \cite{Silvestri2013} and \cite{Sawicki2015} that, in general, the QSA works in most viable models.

In the literature, there is an impressive body of work studying the properties and the consequences of the QSA, ranging from the determination of the expressions for selected models to its use to rule out dark energy and modified gravity models based on the properties of $\mu$, $\eta$ and $\Sigma$ (which we will define later) \cite{Pogosian2016}. We refer to \cite{Frusciante2020a} for a recent and exhaustive review of the phenomenology of the QSA.

The QSA is important for two related reasons: (i) the equations to be solved are much simpler than the full ones, being algebraic rather than differential, hence simplifying the numerical implementation in software; (ii) the formalism allows us to interpret results quickly and in a simple way in terms of quantities already known (e.g., a modified gravitational constant). Hence, on scales where it is reliable, it is a crucial tool to be applied to observations and to advance in our knowledge of the theory of gravity on cosmological scales.

In this work we will provide a detailed study of three different expressions (field equations, metric potentials and EoS approach) for the modified gravity functions $\mu$ and $\eta$ obtained by applying the QSA to the Horndeski theories \citep{Horndeski1974,Deffayet2011,Kobayashi2011}. Our goal is to understand how well they recover the correct behaviour as a function of scale ${\rm K}$, by comparing them to the exact numerical solution obtained from the code \texttt{EoS\_class} \citep{Pace2019a} which was tested against \texttt{hi\_class} \cite{Zumalacarregui2017,Bellini2020} and showed to agree at the sub-percent level. We do this by comparing to the exact forms of $\mu$ and $\eta$ extracted from the code and with observables, such as the matter power spectrum, $P(k)$, angular power spectrum of temperature anisotropies, $C_{\ell}^{\rm TT}$, and the CMB lensing power spectrum, $C_{\ell}^{\phi\phi}$.
This is done by implementing the equations of motion modified to include $\mu$ and $\eta$ as described in \cite{Zucca2019} in a newly designed branch of the Einstein-Boltzmann solver \texttt{CLASS} \citep{Lesgourgues2011a,Blas2011}, which we call \texttt{QSA\_class}.

The plan of the paper is as follows: in Section~\ref{sect:preliminaries} we provide a short introduction to the underlying mathematical framework which will serve as basis for the subsequent discussion, while in Section~\ref{sect:QSA} we present the different expressions for $\mu$ and $\eta$ using the QSA for different approaches. In Section~\ref{sect:comparison} we perform a detailed comparison of the expressions for the different approaches and specify them to selected classes of models. In Section~\ref{sect:numerical_comparison} we compare the analytical expressions derived in Section~\ref{sect:QSA} with the exact numerical expectations for the same classes of models studied in \cite{Pace2019a}, assess their regime of validity, and discuss when they break down and depart from the numerical solution. In Section~\ref{sect:observables} we show the spectra obtained from the different expressions for the modified gravity parameters and discuss their performance, comparing the approximated spectra with the exact ones. We finally conclude in Section~\ref{sect:conclusions}. In the Appendices ~\ref{sect:coefficientsEFE}, \ref{sect:coefficientsMP}, and \ref{sect:coefficientsEoS} we outline the application of the QSA to the field equations, to the metric potentials and to the EoS expressions, respectively. We also provide the coefficients required to derive the final expressions.

For this work, where necessary, we will use the same fiducial cosmological parameters used in \citep{Zumalacarregui2017,Pace2019a}: the CMB temperature $T_{\rm CMB}=2.725\,{\rm K}$, the Hubble parameter today $H_0=67.5\,{\rm km s}^{-1}{\rm Mpc}^{-1}$, flat spatial geometry $\Omega_{\rm k}=0$, baryon density parameter today $\omega_{\rm b}=\Omega_{\rm b}h^2=0.022$, cold dark matter density parameter today $\omega_{\rm CDM}=\Omega_{\rm CDM}h^2=0.12$, effective number of neutrino species $N_{\rm eff}=3.046$, dark sector density parameter today, as inferred by the closure relation ($\sum_i\Omega_i=1$), $\Omega_{\rm ds}=0.688$. We further assume that the normalisation of the amplitude of the initial density perturbations is $A_{\rm s}=2.215\times 10^{-9}$, the slope of the primordial power spectrum is $n_{\rm s}=0.962$ and the reionization redshift is $z_{\rm reio}=11.36$ under the assumption of instantaneous reionization. We will assume that the background equation of state for the dark sector is $w_{\rm ds}=-1$, as for a $\Lambda$CDM background. We also denote with $\Omega_{\rm m}$ the total matter density parameter. For model 1 (defined later), however, having $w_{\rm ds}=-1$ leads to conceptual problems, therefore, we will assume $w_{\rm ds}=-0.95$. Although the assumption $w_{\rm ds}=-1$ leads to a particular class of Lagrangians for Horndeski theories (see, for example \cite{Pirtskhalava2015}), it is possible for many others to have a background equation-of-state parameter arbitrary close to that of the $\Lambda$CDM model, justifying our assumption.

To facilitate the comparison of our results with other works in the literature, we provide an extensive dictionary between our notation and those of previous works in a Supplementary data document entitled: Supplementary Materials.

\section{Preliminaries}\label{sect:preliminaries}

\subsection{Basic notions}
For our calculations, we will closely follow the approaches and definitions of \cite{Gleyzes2014} (for the field equations and the two equations for the metric potentials) and \cite{Pace2019a} for the EoS approach, introduced before in \cite{Battye2016a} and used also in \cite{Battye2017,Battye2018a,Battye2019}. We define the perturbed Newtonian metric for scalar density perturbations \citep{Gleyzes2014}
\begin{equation}
 \mathrm{d}s^2 = -(1+2\Phi)\mathrm{d}t^2 + a(t)^2(1-2\Psi)\delta_{ij}\mathrm{d}x^i\mathrm{d}x^j\,,
\end{equation}
and the total matter (cold dark matter, baryons, photons, neutrinos) stress-energy tensor
\begin{align}
 T^{0}_{\phantom{0}0} & \equiv -\left(\rho_{\rm m}+\delta\rho_{\rm m}\right)\,,\\
 T^{0}_{\phantom{0}i} & \equiv \partial_i q_{\rm m} \equiv \left(\rho_{\rm m}+P_{\rm m}\right)\partial_i v_{\rm m} = -a^2T^{i}_{\phantom{i}0}\,,\\
 T^{i}_{\phantom{i}j} & \equiv \left(P_{\rm m}+\delta P_{\rm m}\right)\delta^i_j + 
     \left(\partial^i\partial_j-\frac{1}{3}\delta^i_j\partial^2\right)\sigma_{\rm m}\,,
\end{align}
where $\rho_{\rm m}$ and $P_{\rm m}$ denote the background matter density and pressure, and $\delta\rho_{\rm m}$ and $\delta P_{\rm m}$ the corresponding perturbed quantities; the velocity potential is denoted by $v_{\rm m}$ and $q_{\rm m}$ is the rescaled velocity. The matter anisotropic stress is $\sigma_{\rm m}$. At late times, matter pressure and anisotropic stress are, in general, negligible, but we will keep them for completeness, as they are important for the study of neutrinos, as shown in \cite{Zucca2019}.

Using the gauge-invariant notation of \cite{Battye2016a,Pace2019a}, the matter variables introduced in the stress-energy tensor become
\begin{equation*}
 q_{\rm m} \rightarrow -\tfrac{\rho_{\rm m}\Theta_{\rm m}}{3H}\,, \quad
 \delta_{\rm m} \rightarrow \Delta_{\rm m}-\Theta_{\rm m}\,, \quad 
 \delta P_{\rm m}/\rho_{\rm m} \rightarrow w_{\rm m}\Gamma_{\rm m}+c_{\rm a,m}^2(\Delta_{\rm m}-\Theta_{\rm m})\,, \quad 
 \sigma_{\rm m} \rightarrow -a^2P_{\rm m}\Pi_{\rm m}/k^2\,,
\end{equation*}
where $c_{\rm a,m}^2\equiv \mathrm{d}P_{\rm m}/\mathrm{d}\rho_{\rm m}$ is the matter adiabatic sound speed and $\delta_{\rm m}$ the matter density constrast $\delta\rho_{\rm m}/\rho_{\rm m}$. The (total) matter entropy perturbations and anisotropic stress are, respectively, $w_{\rm m}\Gamma_{\rm m}$ and $w_{\rm m}\Pi_{\rm m}$. 
We also identify, using the gauge-invariant notation,
\begin{equation*}
 \Phi \equiv Y\,, \quad 
 \Psi \equiv Z\,, \quad 
 W \equiv \frac{1}{2}\left(Z+Y\right)\,, \quad 
 X \equiv Z^{\prime}+Y = \frac{1}{2}\left(\Omega_{\rm m}\Theta_{\rm m}+\Omega_{\rm ds}\Theta_{\rm ds}\right)\,,
\end{equation*}
where $W$ is the Weyl potential which describes light deflection (i.e., gravitational lensing) and the variable $X$ is used later in the EoS approach. The prime represents the derivative with respect to $\ln{a}$ and $\Theta_{\rm ds}$ is the rescaled velocity of the dark sector. 
With respect to the notation used in the \texttt{CLASS} code, $\phi \rightarrow Z$ and $\psi \rightarrow Y$. For other variables used in \texttt{CLASS} the reader might be interested to, we refer to \cite{Pace2019a} for details.

\subsection{Parameterization of modified gravity models}
For each of the three potentials defined above, $Z$, $Y$ and $W$, we can associate an ``effective gravitational constant'' $G_{\rm eff}=G\mu_{x}$, with $G$ the Newton gravitational constant and $x\in\{Z, Y, W\}$, which is now a function of time and scale. In particular, if we write a ``Poisson-like'' equation, then we can define
\begin{align}\label{eqn:definitions}
 \mu_Z \equiv &\,-\frac{2}{3}\frac{{\rm K}^2Z}{\Omega_{\rm m}\Delta_{\rm m}}\,, & 
 \mu_Y \equiv &\, \mu = -\frac{2}{3}\frac{{\rm K}^2Y}{\Omega_{\rm m}\Delta_{\rm m}}\,, & 
 \mu_W \equiv &\, \Sigma = -\frac{2}{3}\frac{{\rm K}^2W}{\Omega_{\rm m}\Delta_{\rm m}} = 
 \frac{1}{2}\left(\mu_Z+\mu\right)\,, \\
 \eta \equiv &\, \frac{Z}{Y} = \frac{\mu_Z}{\mu}\,, & 
 \gamma \equiv &\, \frac{Y-Z}{Z} = \frac{1}{\eta} - 1\,, & 
 g \equiv & \frac{Z-Y}{Z+Y} = \frac{\mu_Z-\mu}{\mu_Z+\mu} = \frac{\eta-1}{\eta+1}\,.
\end{align}
For simplicity and for a better intuition of the physics involved, we will perform our calculations in the Newtonian gauge.

The field equations describe the evolution of two degrees of freedom, the Bardeen potentials $\Psi$, the space-space perturbation, and $\Phi$, the time-time perturbation, following the notation of \cite{Gleyzes2014}. We can, therefore, describe a generic cosmological model with two independent parameters among those in Eq.~(\ref{eqn:definitions}) and all the others can be derived from them. The function $G\mu$ is sometimes called $G_{\rm eff}$ (or $G_{\rm matter}$) in the literature \citep{Bean2010,Gleyzes2014} and it represents the effects of modifications of gravity on non-relativistic particles, as matter perturbations are sensitive to the gradient of the gravitational potential associated to this function.

The effect of $\mu_Z$ is not directly observable, and therefore it is customary to consider another function, the gravitational slip $\eta$ (also often called $\gamma$ in the literature \citep{Gleyzes2014,Zucca2019}), which parameterises the different evolution of  $\Phi$ and $\Psi$ relative to each other. In other words, when $\eta=1$, the gravitational potentials are the same and evolve exactly in the same way (as, for example, in general relativity and minimally coupled models) at late times, when no matter anisotropic stress is present, but they evolve differently when $\eta \neq 1$.

Another function which can be defined is $\Sigma$, associated to the Weyl potential, which probes the effect of weak gravitational lensing. In the literature, it is also sometimes called $G_{\rm light}=G\mu_W$ \citep{Pogosian2016}. In general relativity, for a $\Lambda$CDM model, these functions are constant and all equal to unity.

One can also define other functions related to the slip $\eta$ \citep{Hu2007a}: $\gamma$ (denoted with $\varpi$ in \citep{Caldwell2007}) and $g$. The function $g$ is of great importance for the studies of gravity in the Solar System, as it is the quantity constrained by the Cassini mission \cite{Will2014}.

What is the best choice of the pair of functions used is somehow arbitrary, so long as they are independent. Commonly studied pairs are: $(\mu_Z,\mu)$, $(\mu,\eta)$ and $(\mu,\Sigma)$. The first set is in general not used because only the effects of $\mu$ are more easily observed, but the other two are widely employed in cosmological studies. Here we focus on the last pair of parameters, $(\mu,\Sigma)$. One can measure $\mu$ by studying matter evolution and $\Sigma$ by light propagation. A nice property of the pair $(\mu,\Sigma)$ is observed by writing $\mu$ in terms of $\Sigma$ and $\eta$ \citep{Perenon2017}:
\begin{equation*}
 \mu - 1 = 2\frac{\Sigma-1}{1+\eta} - \frac{\eta-1}{1+\eta}\,, 
\end{equation*}
which shows a 45\textdegree correlation when $\eta\simeq 1$. This can help distinguishing specific classes of models, in particular late and early dark energy scenarios, as opposed to early modified gravity models, as discussed in depth in \cite{Perenon2017}. Focusing on the second pair of observables, $(\mu,\eta)$, would not change the conclusions of the paper.

Effects of these functions ($\mu$, $\eta$ and $\Sigma$) have been studied in the widely used code \texttt{MGCAMB} \citep{Zhao2009a,Hojjati2011} which has been recently extended to include effects of massive neutrinos \citep{Zucca2019}. This code, which represents a patch to the Einstein-Boltzmann code \texttt{CAMB} \citep{Lewis2000}, offers the user the possibility of choosing between two different sets of modified gravity parameters, $(\mu,\eta)$ and $(\mu,\Sigma)$,\footnote{The actual implementation does not change as $\eta=2\Sigma/\mu-1$.} and it also implements the time- and scale-dependent functions $R$ and $Q$ introduced in \cite{Bean2010}.\footnote{In the notation used in this work, $Q\equiv\mu$ and $R\equiv\eta$.} 
The \texttt{MGCAMB} code allows not only phenomenological parametrisations, but also expressions for specific models. We refer the reader to \cite{Zucca2019} for a thorough discussion of the several parametrisations implemented.

\subsection{Horndeski models}\label{sect:models}
We limit our analysis to the Horndeski models for modified gravity. Perturbation dynamics is described by four functions: the \textit{kineticity} $\alpha_{\rm K}$, the \textit{braiding} $\alpha_{\rm B}$, the \textit{rate of running of the Planck mass} $\alpha_{\rm M}$, and the \textit{tensor speed excess} $\alpha_{\rm T}$ \citep{Bellini2014}. 
Each of them has a precise physical meaning: $\alpha_{\rm K}$ only affects scalar perturbations and describes perfect fluid (no energy flow and anisotropic stress) dark energy models; $\alpha_{\rm B}$ also only affects scalar perturbations and describes the mixing of the kinetic terms of the scalar field and of the metric giving rise to a fifth-force; $\alpha_{\rm M}$ contributes to both scalar and tensor perturbations and to the anisotropic stress ($\eta\neq 1$); $\alpha_{\rm T}$ parameterises deviations of the speed of gravitational waves $c_{\rm T}$ from that of light: $c_{\rm T}^2=1+\alpha_{\rm T}$. Defining the effective Planck mass $M^2$, the relation between $M^2$ and $\alpha_{\rm M}$ is $\alpha_{\rm M} \equiv \tfrac{\mathrm{d}\ln{M^2}}{\mathrm{d}\ln{a}}$. For their definition in terms of the Horndeski functions, see references \cite{Bellini2014,Gleyzes2014,Pace2019a}.

Gravitational waves measurements \cite{LigoVirgo2017,LigoVirgoIntegral2017,LigoVirgo2017a} suggest that within the Horndeski class of models one should set $\alpha_{\rm T}\equiv 0$, but for completeness we will consider models with $\alpha_{\rm T}\neq 0$ to study the performance of the QSA.

We consider the same benchmark models discussed in \cite{Pace2019a}, to which we refer the reader for more details. Here, it suffices to remember which of the $\alpha$ are different from zero and the generic class of models they represent.

\begin{enumerate}
 \item \textit{$k$-essence-like models}: 
       $\alpha_{\rm K}\neq 0$, $\alpha_{\rm B}=\alpha_{\rm M}=\alpha_{\rm T}=0$ 
       \citep{ArmendarizPicon2000}.
 
 \item \textit{$f(R)$-like models}: 
       $\alpha_{\rm K}=\alpha_{\rm T}=0$, $\alpha_{\rm M}\neq 0$, $\alpha_{\rm B}\neq 0$. These models reduce to $f(R)$ cosmologies \citep{Silvestri2009,Sotiriou2010,DeFelice2010} when $\alpha_{\rm M}=2\alpha_{\rm B}$ and have been studied and compared in our previous works \cite{Battye2018a,Bellini2018}.
 
 \item \textit{KGB-like models}: 
       $\alpha_{\rm K}\neq 0$, $\alpha_{\rm B}\neq 0$, $\alpha_{\rm M}=\alpha_{\rm T}=0$ 
       \citep{Deffayet2010,Pujolas2011,Kobayashi2010,Kimura2011,Peirone2019}.
 
 \item $\alpha_{\rm K}\neq 0$, $\alpha_{\rm M}\neq 0$, $\alpha_{\rm B}=\alpha_{\rm T}=0$. 
       These are a particular subclass of the next more general class of models and satisfy a differential relation between the Horndeski functions $G_3$ and $G_4$ such that $XG_{3,X}+G_{4,\phi}=0$. We refer to \cite{Pace2019a} for more details. Note that here $X=\nabla_{\mu}\phi\nabla^{\mu}\phi$ is the kinetic term of the scalar field $\phi$.
 
 \item \textit{$c_{\rm T}=1$ models}: 
       $\alpha_{\rm K}\neq 0$, $\alpha_{\rm B}\neq 0$, $\alpha_{\rm M}\neq 0$, $\alpha_{\rm T}=0$. These represent the most generic Horndeski model compatible with GW constraints \citep{Creminelli2017,Ezquiaga2017,Baker2017}. 
       Within this class of models, there often exists a relation between $\alpha_{\rm B}$ and $\alpha_{\rm M}$, such as in the no slip gravity model proposed by \cite{Linder2018} and analysed in detail by \cite{Brush2019,Brando2019}, where $\alpha_{\rm B}=\alpha_{\rm M}$. Models with $\alpha_{\rm M} = 2\alpha_{\rm B}$ are conformally related to general relativity (i.e., they possess an Einstein frame where the gravitational kinetic term is described by the Einstein-Hilbert action; see, e.g., \cite{Crisostomi2019}). In this case, for $\alpha_{\rm K}=0$ one reduces to $f(R)$ models. However, differences with respect to $f(R)$ appear only on large scales \citep{Pace2019a}.
 
 \item \textit{Generic Horndeski models}: $\alpha_{\rm K}\neq 0$, $\alpha_{\rm B}\neq 0$, $\alpha_{\rm M}\neq 0$, $\alpha_{\rm T}\neq 0$. This is the most general Horndeski model.
\end{enumerate}

\section{Derivation of the QSA expressions for \texorpdfstring{$\mu$}{mu} and \texorpdfstring{$\eta$}{eta}}
\label{sect:QSA}

In this section, we discuss the three approaches to the QSA that we will study and present the functions $\mu$ and $\eta$. After a brief discussion of the steps required to correctly apply the approximation, we present the form of $\mu$ and $\eta$. For a more detailed derivation of the equations, we refer to the corresponding appendices \ref{sect:coefficientsEFE}, \ref{sect:coefficientsMP} and \ref{sect:coefficientsEoS}.

As we have already explained, the basic idea of the QSA is that the time derivatives are assumed to be negligible with respect to spatial derivatives, as we usually deal with scales smaller than the (sound) horizon ${\rm K}_{\rm sh}=c_{\rm s}k/(aH)\gg 1$, where $k$ is the wavelength mode, $c_{\rm s}$ the sound speed of the scalar field perturbations, $a$ the scale factor and $H$ the Hubble function. In this way, dynamical equations are turned into constraint equations and can be written as generalised ``Poisson equations'', where Newton's constant is replaced, in the Fourier space, by a function of space and time due to the presence of additional fluids or new degrees of freedom.

The QSA applied to different sets of equations can lead to different expressions in the large scale limit (${\rm K}\rightarrow 0$) as the approximations break down on these scales. On small scales (${\rm K}\rightarrow \infty$), however, it is easy to show that all the expressions have the same limit which we label $\mu_{\infty}$ and $\eta_{\infty}$ (and also $\mu_{Z,\infty}$ and $\mu_{W,\infty}$). This is because there is a precise hierarchy: ${\rm K}^2\Phi\sim{\rm K}^2\Psi\sim\Delta_{\rm m}$ and $\delta\phi/M_{\rm pl}\sim\Phi\sim\Psi$, where $\delta\phi$ is the perturbed scalar field. Hence, an important aspect of the analysis performed in this work is to understand where the different expressions coincide and results are robust and where they differ.

To derive the expressions for the modified gravity parameters under the QSA, in each case we apply the following procedure:
\begin{itemize}
 \item neglect time derivatives;
 \item consider only relevant terms on sub-horizon scales, i.e., assume ${\rm K}\gg 1$;
 \item for the equation of motion of the perturbed scalar field, take also into account (i.e., keep the corresponding coefficient of) the  mass term associated to the scalar degree of freedom;
 \item if dark sector variables are present, rewrite them in terms of matter variables;
 \item write the expressions as ``Poisson-like equations'';
 \item use Eqs.~(\ref{eqn:definitions}) to infer $\mu$, $\mu_Z$ and $\Sigma$;
 \item infer $\eta$ from the relation between the potentials.
\end{itemize}

We need to assume a functional form for $\mu$, $\mu_Z$, $\eta$ and $\Sigma$ and we will use a notation similar to that of \cite{Lombriser2015a}; in particular we shall use
\begin{align*}
 \mu_Z^{\rm QSA} & = \frac{\mu_{Z,+0}+\mu_{Z,+2}{\rm K}^2+\mu_{Z,+4}{\rm K}^4}
                          {\mu_{-0}+\mu_{-2}{\rm K}^2+\mu_{-4}{\rm K}^4}
                     \frac{1}{\bar{M}^2}\,, \\
 \mu^{\rm QSA} & = \frac{\mu_{+0}+\mu_{+2}{\rm K}^2+\mu_{+4}{\rm K}^4}
                        {\mu_{-0}+\mu_{-2}{\rm K}^2+\mu_{-4}{\rm K}^4}
                   \frac{1}{\bar{M}^2}\,, \\
 \eta^{\rm QSA} & = \frac{\mu_{Z,+0}+\mu_{Z,+2}{\rm K}^2+\mu_{Z,+4}{\rm K}^4}
                         {\mu_{+0}+\mu_{+2}{\rm K}^2+\mu_{+4}{\rm K}^4}\,,
\end{align*}
where $\bar{M}^2=M^2/M_{\rm pl}^2$, $M_{\rm pl}^{-2}=8\pi G$ and $M^2$ is the effective Planck mass squared. In the previous three expressions, the different coefficients used represent time-dependent functions, written in terms of the $\alpha_{\rm X}$, which can be derived by applying one of the three QSA approaches described in this work. A positive (negative) subscript is used for the numerator (denominator) and its numerical value refers to the corresponding power of ${\rm K}$, while the index $Z$ is used for $\mu_{Z}$ (and $\eta$ as a derived quantity).

The superscript $\mathrm{QSA}$ will be replaced by the acronym referring to the particular approximation scheme used: $\mathrm{EFE}$ for the effective field equations, $\mathrm{MP}$ for the two equations for the metric potentials and $\mathrm{EoS}$ when using the attractor solution based on the EoS approach. In the semi-dynamical approach of \cite{Lombriser2015a}, the above expressions contain terms proportional to ${\rm K}^6$, as time derivatives of the potentials are taken into account. The expressions for the semi-dynamical approach have the same limit on small scales $\bar{M}^2\mu_{\infty}=\mu_{+6}/\mu_{-6}$ as the one found in this and in previous works. In fact $\mu_{+4}=\mu_{+6}$ and $\mu_{-4}=\mu_{-6}$ and our expressions with lower powers in ${\rm K}$ can be derived from those in \cite{Lombriser2015a} as there at least $\mu_{+0}$ and $\mu_{-0}$ are zero. However, it is not guaranteed that $\mu_{+0}$ ($\mu_{-0}$) in this work coincides with $\mu_{+2}$ ($\mu_{-2}$) in \cite{Lombriser2015a}.

\subsection{Effective Field Equations (EFE) approach}\label{sect:EFE}
Our starting point are the four field equations augmented by the equation of motion for the perturbed scalar field, which, following the notation of \cite{Gleyzes2014}, is denoted by $\pi$. In a more common notation where the perturbations of the scalar field $\phi$ are denoted by $\delta\phi$, one has $\pi = \delta\phi/\dot{\phi}$. We anticipate that choosing $\pi$ or $\delta\phi$ does not affect the functional form of $\mu$ and $\eta$, but it changes the mass term for the scalar field, i.e., $M_{\pi}^2\neq M_{\delta\phi}^2$. We will discuss this point later on in the section.

Here, we briefly outline the procedure followed to derive the expressions for $\mu$ and $\eta$ and we refer the reader to appendix~\ref{sect:coefficientsEFE} for a detailed derivation. We start from Eqs.~(\ref{eqn:EFE}) and neglect all terms containing time derivatives. We further assume a sub-horizon limit where ${\rm K}\gg 1$ and we, therefore, only consider terms with explicit dependence on ${\rm K}^2$. Among the terms without a dependence on scale, we consider the term $C^{\pi}_{\pi}$ in Eq.~(\ref{eqn:EoM-pi}) as it represents a mass term associated with the perturbed scalar field which can be comparable to ${\rm K}^2$, as is the case of $f(R)$ models \cite{DeFelice2010}.

This procedure leads to Eqs.~(\ref{eqn:QSA_EFE}) and we can solve the system for the two potentials and the perturbed scalar field $\pi$. Using the gauge invariant notation and the definitions in Eq.~(\ref{eqn:definitions}), it is easy to find the expressions for $\mu$ and $\eta$, whose coefficients are
\begin{subequations}\label{eqn:muetaEFE}
 \begin{align}
  \mu_{+0} = &\, (1+\alpha_{\rm T})\,\mu_{\rm p}\,, & 
  \mu_{-0} = &\, \mu_{\rm p}\,, & 
  \mu_{Z,+0} = &\, \mu_{\rm p}\,, \\
  \mu_{+2} = &\, \alpha c_{\rm s}^2\bar{M}^2\mu_{\infty}\,, & 
  \mu_{-2} = &\, \alpha c_{\rm s}^2\,, & 
  \mu_{Z,+2} = &\, \alpha c_{\rm s}^2\bar{M}^2\mu_{Z,\infty}\,,
 \end{align}
\end{subequations}
where
\begin{equation}\label{eqn:mup}
 \mu_{\rm p} = 6\left\{\left(\dot{H}+\frac{\rho_{\rm m}+P_{\rm m}}{2M^2}\right)\dot{H} + \dot{H}\alpha_{\rm B}\left[H^2(3+\alpha_{\rm M})+\dot{H}\right]+H\frac{\partial\left(\dot{H}\alpha_{\rm B}\right)}{\partial t}\right\}/H^4\,,
\end{equation}
and $\mu_{+4}=\mu_{-4}=\mu_{Z,+4}=0$. We also defined $\alpha=\alpha_{\rm K}+6\alpha_{\rm B}^2$ and $c_{\rm s}^2$ represents the sound speed of the perturbations. Its explicit expression is given in Appendix~\ref{sect:coefficientsEoS}, Eq.~(\ref{eqn:cs2}).

For completeness, we also consider the relationship between the perturbed scalar field $\pi$ and the gauge-invariant matter density perturbation $\Delta_{\rm m}$
\begin{equation}\label{eqn:QSApi}
 H\pi = - \frac{\mu_{\pi}}{\mu_{-0}+\mu_{-2}{\rm K}^2}\frac{1}{\bar{M}^2}\Omega_{\rm m}\Delta_{\rm m}\,,
\end{equation}
where $\mu_{\pi}=3[\alpha_{\rm B}(1+\alpha_{\rm T})+\alpha_{\rm T}-\alpha_{\rm M}]$.

It is interesting to consider the limits of $\mu$, $\eta$ and $\pi$ on small and large scales. On small scales (${\rm K}\rightarrow\infty$), $\mu=\mu_{\infty}$ and $\eta=\eta_{\infty}$ in agreement with \cite{Pace2019a} and references therein, while $\pi=\pi_{\infty}\propto\Delta_{\rm m}/{\rm K}^2$, i.e., on small scales the perturbed 
scalar field is of the same order of magnitude of the potentials and the velocity perturbations. This can be understood by looking at Eq.~(\ref{eqn:EFE-0i}). 
On large scales (${\rm K}\rightarrow 0$), we find $\mu=\mu_0=1+\alpha_{\rm T}$ and $\eta=\eta_0=1/(1+\alpha_{\rm T})$, that is, models where gravitational waves do not propagate luminally will be clearly distinguishable from general relativity. This is also one of the conclusions reached in \cite{Pogosian2016}, to which we refer for an in-depth discussion. However, in \cite{Lombriser2015,Lombriser2016a,Lombriser2017} the authors discuss a class of models where (in the notation of this work) $\mu=\eta=1$ with all the $\alpha$ functions different from zero. This is achieved by fixing the sound speed $c_{\rm s}^2=1$ and deducing a relation between $\alpha_{\rm T}$ and $\alpha_{\rm M}$ (see their Eq.~(2.22)). Assuming an exact $\Lambda$CDM background, these models are completely degenerate with $\Lambda$CDM, both at the level of the background and linear perturbations. In other words, linear perturbations are indistinguishable from the $\Lambda$CDM, even at large ${\rm K}$. To break the degeneracy in measurement of the large scale structure, $\alpha_{\rm T}$ has to be inferred, with the help of gravitational waves. Our result, based on the use of a QSA, is clearly at odds with these works. This can be interpreted as an artefact of making a QSA, as on the horizon scale the assumptions behind it break down.

The resulting expressions for $\mu$ and $\eta$ are quadratic in ${\rm K}$, consistent with previous works. In particular, we find that the values of the coefficients of the expressions for $\mu$ and $\eta$ are in agreement with those provided by 
\cite{Gubitosi2013,Bloomfield2013a,Bloomfield2013b,Gleyzes2013,Piazza2014,Perenon2015,Pogosian2016,Perenon2017,Amendola2019}. We also note that the coefficients for $\mu$ and $\eta$ discussed in \cite{Gleyzes2013} reduce to Eqs.~(\ref{eqn:muetaEFE}) in the Horndeski limit.

Similar expressions were first derived by \cite{DeFelice2011} and later confirmed by \cite{Arjona2019b}. The main difference with respect to our approach is that we have chosen a different variable for the scalar field perturbation; we used $\pi$, whereas \cite{DeFelice2011} used $\delta\phi$. These choices lead to the same result on small scales (we verified that their expressions are in agreement with the coefficients $\mu_{\pm 2}$ and $\mu_{Z,+2}$) and on the same limit on large scales $\mu_0$ and $\eta_0$, but the  coefficients $\mu_{\pm 0}$ and $\mu_{Z,+0}$ differ from ours, as these terms are proportional to the mass associated to the scalar degree of freedom, which is different when using $\pi$ or $\delta\phi$. One can show that, in general,
\begin{equation}
 M_{\delta\phi}^2 = \left\{
  \left[2\left(\frac{\ddot{\phi}}{\dot{\phi}}\right)^2 - \frac{\dddot{\phi}}{\dot{\phi}}\right]
  C_{\ddot{\pi}}^{\pi} - \frac{\ddot{\phi}}{\dot{\phi}}C_{\dot{\pi}}^{\pi} + 
  C_{\pi}^{\pi}\right\}\frac{M^2}{\dot{\phi}^2}\,,
\end{equation}
where, as before $\phi$ represents the value of the scalar field at the background level and $C_{\pi}^{\pi}$ is the mass of the scalar field using the variable $\pi$.

To understand why this is the case, it suffices to consider again the relation between $\pi$ and $\delta\phi$, $\delta\phi=\dot{\phi}\,\pi$. When considering the equation of motion of the scalar field, there are terms involving its derivatives and since they contain terms proportional to $\pi$, it is clear that $M_{\delta\phi}^2\neq C_{\pi}^{\pi}$. We will give more details on the relation between the set of coefficients in the field equations in the Supplementary data. As the limits on both large and small scales coincide, we will only consider the expression for $\pi$, as in this case there is no dependence on the background evolution of the scalar field.

\subsection{Metric potentials approach}\label{sect:MP}
The field equations can be combined into two independent equations, one describing the evolution of the potential $\Psi$ ($Z$ in the gauge-invariant notation) in terms of the matter variables and a constraint equation relating $\Phi$ ($Y$ in the gauge-invariant notation) to $\Psi$.

For compactness, here we only describe, as before, the necessary steps to derive the expressions for the QSA. In Appendix~\ref{sect:coefficientsMP} we will provide a more detailed derivation, together with the relevant coefficients required to derive our expressions.

We start from Eqs.~(\ref{eqn:MP}) and we apply a QSA by neglecting the time derivatives of the potential. We also neglect matter anisotropic stress $\sigma_{\rm m}$, pressure perturbations $\delta P_{\rm m}$ (as negligible for the dark matter component) and combine $\delta\rho_{\rm m}$ and $q_{\rm m}$ into the gauge-invariant density perturbation $\rho_{\rm m}\Delta_{\rm m}=\delta\rho_{\rm m}-3Hq_{\rm m}$.

These approximations lead to Eqs.~(\ref{eqn:QSA_MP}) and solving for $Z$ and $Y$, it is straightforward to derive the expressions for $\mu_Z$ and $\mu$:
\begin{equation}
 \mu_Z = \frac{{\rm K}^2C_{\delta\rho_{\rm m}}}{C_{\Psi}}\frac{1}{\bar{M}^2}\,, \quad 
 \mu = \left[\frac{\alpha_{\rm B}(1+\alpha_{\rm T})+\alpha_{\rm T}-\alpha_{\rm M}}{\alpha_{\rm B}}
             \frac{{\rm K}^2C_{\delta\rho_{\rm m}}}{C_{\Psi}} - 
             \frac{\alpha_{\rm T}-\alpha_{\rm M}}{\alpha_{\rm B}}\right]\frac{1}{\bar{M}^2}\,.
\end{equation}
The expression for $\mu_Z$ is obtained directly from Eq.~(\ref{eqn:MP_Psi_QSA}) while the expression for $\mu$ is derived by inserting Eq.~(\ref{eqn:MP_Psi_QSA}) into (\ref{eqn:MP_Phi_QSA}). The ratio of these two functions gives the slip parameter $\eta$:
\begin{equation}
 \eta = \left[\frac{\alpha_{\rm B}(1+\alpha_{\rm T})+\alpha_{\rm T}-\alpha_{\rm M}}{\alpha_{\rm B}} - 
             \frac{\alpha_{\rm T}-\alpha_{\rm M}}{\alpha_{\rm B}}
             \frac{C_{\Psi}}{{\rm K}^2C_{\delta\rho_{\rm m}}}\right]^{-1}\,.
\end{equation}

Inserting the relevant coefficients leads to quartic expressions in ${\rm K}$ and the relevant coefficients for $\mu$ and $\eta$ are
\begin{subequations}\label{eqn:MPlist}
 \begin{align}
  \mu_{+0} = &\, \beta_1\beta_4(\alpha_{\rm M}-\alpha_{\rm T})/\alpha_{\rm B}\,, \quad 
  \mu_{+2} =  \beta_1[(1+\alpha_{\rm T})\beta_6+(\alpha_{\rm T}-\alpha_{\rm M})
                      (\beta_6-\beta_5)/\alpha_{\rm B}]\,,\\
  \mu_{-0} = &\, \beta_1\beta_4\,, \quad 
  \mu_{-2} = \beta_1\beta_5\,, \quad 
  \mu_{Z,+0} = 0 \,, \quad 
  \mu_{Z,+2} = \beta_1\beta_6 \,,\\
  \mu_{-4} = &\, \alpha_{\rm B}^2c_{\rm s}^2\,, \quad 
  \mu_{+4} = \alpha_{\rm B}^2c_{\rm s}^2\bar{M}^2\mu_{\infty}\,, \quad 
  \mu_{Z,+4} = \alpha_{\rm B}^2c_{\rm s}^2\bar{M}^2\mu_{Z,\infty}\,,
 \end{align}
\end{subequations}
and the $\beta_i$ are given in Appendix~\ref{sect:coefficientsMP}.

These expressions cannot be used when $\alpha_{\rm B}=0$ (such as in $k$-essence models), as $\mu_{+0}$ diverges. To avoid this, we can first set $\alpha_{\rm B}=0$ in Eqs.~(\ref{eqn:MP}) and then apply the QSA assuming negligible contribution from $\delta P_{\rm m}$ and $\sigma_{\rm m}$. This leads to a straightforward derivation of the expression for $\mu_Z$, but $\mu$ does not have the correct small-scale limit, as it is easy to verify. This shows a failure of this QSA, as discussed in \cite{Bellini2014,Sawicki2015}. A simple way to see this is to set $\alpha_{\rm B}=0$ in (\ref{eqn:MP_Phi}); in the resulting equation there is no explicit ${\rm K}^2$ term and the remaining leading terms are of the same order of those neglected before.

Therefore, in this case, we assume that $\eta=\eta_{\infty}$, i.e., the small scale limit is always valid also on large scales, and $\mu=\mu_Z/\eta_{\infty}$:
\begin{equation}
 \mu_Z = \frac{c_{\rm s}^2\mu_{Z,\infty}{\rm K}^2}{\beta_4+c_{\rm s}^2{\rm K}^2}\,, \quad 
 \mu = \frac{c_{\rm s}^2\mu_{\infty}{\rm K}^2}{\beta_4+c_{\rm s}^2{\rm K}^2}\,.
\end{equation}
In contrast to the general case of Eq.~(\ref{eqn:MPlist}), these expressions are only quadratic (and not quartic) in ${\rm K}$.

To see why the QSA is broken for models where the braiding is not present, it is useful to note that when $\alpha_{\rm B}\neq 0$ there exists a new scale, called \textit{braiding scale} ${\rm K}_{\rm B}$ defined as ${\rm K}_{\rm B}^2=\beta_1/\alpha_{\rm B}^2$ \citep{Bellini2014}. This scale appears in both the dynamical and constraint equations and it is easy to see that the gauge-invariant potential $Z$ can cluster on small scales only if $\alpha_{\rm B}\neq 0$ ($\mu_{Z}\neq 1$). When there is no braiding ($\alpha_{\rm B}=0$), there is no scale dependence as well as no dependence on $\delta\rho_{\rm m}$ in the constraint equation and it is, therefore, not correct to neglect the time derivative of the potential (especially if $c_{\rm s}^2\ll 1$), as this is the only scale dependence which can appear in the system.

When limiting ourselves to the case of quintessence and $k$-essence, the expressions above lead to $\mu_Z=\mu\neq 1$ and $\eta=1$ (as $\gamma_9=0$). We will comment more on the differences with the expressions from the field equations (Section~\ref{sect:EFE}) and from the EoS (Section~\ref{sect:EoS}) approach later.

To the best of our knowledge, the expressions in Eq.~(\ref{eqn:MPlist}) have not been derived before, even if in \cite{Bellini2014,Gleyzes2014} the scale-independent small-scale limit has been derived.

\subsection{Equation of State approach}\label{sect:EoS}
In this section, we briefly discuss the derivation of the modified gravity parameters $\mu$ and $\eta$ obtained in \cite{Pace2019a} within the EoS approach. As in the previous sections, we will just sketch the general procedure, and leave to appendix~\ref{sect:coefficientsEoS} a more detailed derivation of the equations.

In this formalism, modifications to gravity are identified with an effective fluid described by a non-trivial stress-energy tensor $U_{\mu\nu}$. 
Its background evolution is completely described by the knowledge of the equation of state 
$w_{\rm ds}=P_{\rm ds}/\rho_{\rm ds}$, where $\rho_{\rm ds}$ and $P_{\rm ds}$ are the background density and pressure, respectively, while at the linear perturbation level two new gauge-invariant equations of state are introduced, the entropy perturbations $w_{\rm ds}\Gamma_{\rm ds}$ and the anisotropic stress $w_{\rm ds}\Pi_{\rm ds}$.

The evolution of the perturbations can be derived by considering the linearly perturbed stress-energy tensor $\delta U_{\mu\nu}$ and from the condition $\nabla^{\mu}\delta U_{\mu\nu}=0$, one derives the continuity and Euler equations (\ref{eqn:EoM}). It is useful to combine Eqs.~(\ref{eqn:EoM}) into a single second order equation for $\Delta_{\rm ds}$ which describes the evolution of perturbations and it is analogous to the standard growth factor equation for matter perturbations. We only consider scales where ${\rm K}\gg 1$, which correspond to $k\gtrsim 10^{-3}~{\rm Mpc}^{-1}$ at $z=0$. In this regime, \cite{Pace2019a} showed that $\Theta\ll\Delta$, for both matter and dark sector perturbations, and it is, therefore, safe to neglect any velocity contribution.

Applying a QSA to Eq.~(\ref{eqn:gf}) implies neglecting the time derivatives of $\Delta_{\rm ds}$. From a physical point of view, we are imposing that the time variation on cosmological time scales is small. We are then left with a relation between dark sector and matter density perturbations which manifests in the form of an attractor solution
\begin{equation}
 \Omega_{\rm ds}\Delta_{\rm ds} = -\frac{C_{\zeta\Delta_{\rm m}}}{c_{\rm a,ds}^2+C_{\zeta\Delta_{\rm ds}}}\Omega_{\rm m}\Delta_{\rm m}\,.
\end{equation}
In Fig.~6 of \cite{Pace2019a}, it has been shown that this analytical relation is in excellent agreement with the full numerical solution over a wide range of scales and times for all the models investigated, as long as ${\rm K}$ is of the order of a few. When ${\rm K}\simeq 1$, the attractor solution becomes progressively less accurate.

To relate the modified gravity parameters to the attractor solution, we use Einstein field equations \citep{Battye2016a}
\begin{equation*}
 -\frac{2}{3}{\rm K}^2Z = \Omega_{\rm m}\Delta_{\rm m} + \Omega_{\rm ds}\Delta_{\rm ds}\,, \quad
 \frac{1}{3}{\rm K}^2(Y-Z) = \Omega_{\rm m}w_{\rm m}\Pi_{\rm m} + \Omega_{\rm ds}w_{\rm ds}\Pi_{\rm ds}\,,
\end{equation*}
so that
\begin{equation}
 \mu_Z = 1 + \frac{\Omega_{\rm ds}\Delta_{\rm ds}}{\Omega_{\rm m}\Delta_{\rm m}} 
       = 1 - \frac{C_{\zeta\Delta_{\rm m}}}{c_{\rm a,ds}^2+C_{\zeta\Delta_{\rm ds}}}\,, \qquad
 \mu = \mu_Z - 2\frac{\Omega_{\rm ds}w_{\rm ds}\Pi_{\rm ds}}{\Omega_{\rm m}\Delta_{\rm m}}\,,
\end{equation}
and $\eta = \mu_Z/\mu$. As discussed before, we can safely neglect the matter anisotropic stress $w_{\rm m}\Pi_{\rm m}$.

The coefficients for the modified gravity parameters are
\begin{subequations}
 \begin{align}
  \mu_{+0} = &\, \gamma_1(\gamma_2-\gamma_7)(1+\alpha_{\rm T})\,, & 
  \mu_{-0} = &\, \gamma_1(\gamma_2-\alpha_{\rm T}/3)\,, & 
  \mu_{Z,+0} = &\, \gamma_1(\gamma_2-\gamma_7)\,,\\
  \mu_{+2} = &\, \alpha_{\rm B}^2c_{\rm s}^2\bar{M}^2\mu_{\infty}\,, & 
  \mu_{-2} = &\, \alpha_{\rm B}^2c_{\rm s}^2\,, & 
  \mu_{Z,+2} = &\, \alpha_{\rm B}^2c_{\rm s}^2\bar{M}^2\mu_{Z,\infty}\,,
 \end{align}
\end{subequations}
where the functions $\gamma_i$ are given in Appendix~\ref{sect:coefficientsEoS} and $\mu_{+4}=\mu_{-4}=\mu_{Z,+4}=0$.

\section{Comparison of the different expressions}\label{sect:comparison}
\subsection{Small- and large-scale limit of the expressions}
In this section we provide a comprehensive comparison between the different expressions for $\mu$ and $\eta$. As discussed already, all three approaches lead to the same result for ${\rm K}\rightarrow\infty$, as the MP and the EoS expressions directly result from combining the field equations. We do not discuss this limit here, as it was addressed previously in \cite{Pace2019a}, but we only report the expressions for $\mu_{\infty}$, $\mu_{Z,\infty}$, and $\eta_{\infty}$ for completeness \citep{Gubitosi2013,Bellini2014,Perenon2015,Perenon2017,Pace2019a}:
\begin{align}
 \mu_{\infty} = &\, \frac{\alpha c_{\rm s}^2(1+\alpha_{\rm T})+2[\alpha_{\rm B}(1+\alpha_{\rm T})+
                          \alpha_{\rm T}-\alpha_{\rm M}]^2}{\alpha c_{\rm s}^2\bar{M}^2}\,, \label{eqn:muY_infty}\\
 \mu_{Z,\infty} = &\, \frac{\alpha c_{\rm s}^2+2\alpha_{\rm B}[\alpha_{\rm B}(1+\alpha_{\rm T})+
                            \alpha_{\rm T}-\alpha_{\rm M}]}{\alpha c_{\rm s}^2\bar{M}^2}\,, \label{eqn:muZ_infty}\\
 \eta_{\infty} = &\, \frac{\alpha c_{\rm s}^2+2\alpha_{\rm B}[\alpha_{\rm B}(1+\alpha_{\rm T})+
                           \alpha_{\rm T}-\alpha_{\rm M}]}
                          {\alpha c_{\rm s}^2(1+\alpha_{\rm T})+2[\alpha_{\rm B}(1+\alpha_{\rm T})+
                           \alpha_{\rm T}-\alpha_{\rm M}]^2}\,. \label{eqn:eta_infty}
\end{align}

We, instead, focus on the limit for ${\rm K}\rightarrow 0$, which represents the regime of interest for this work, and consider the coefficients for the different quantities previously defined. A summary of our results is presented in Table~\ref{tab:MG0}, where we present expressions for $\mu_0$, $\mu_{Z,0}$, $\eta_0$, $\Sigma_0$, $\gamma_0$ and $g_0$. We remind the reader that the functions $g$ and $\gamma$ are derived from the slip parameter $\eta$, and $\Sigma$ from the knowledge of $\mu$ and $\mu_{Z}$ (or alternatively $\mu$ and $\eta$). Below we investigate $\mu$ and $\eta$ in more detail, as they are the quantities also implemented in our code.

\begin{table}[!ht]
 \centering
 \begin{tabular}{|ccccc|}
  \hline
  & EFE & MP & MP ($\alpha_{\rm B}=0$) & EoS \\
  \hline
  $\bar{M}^2\mu_{Z,0}$ & $1$ 
                       & $0$ 
                       & $0$ 
                       & $\frac{\gamma_2-\gamma_7}{\gamma_2-\alpha_{\rm T}/3}$\\

  $\bar{M}^2\mu_{0}$ & $1+\alpha_{\rm T}$ 
                     & $\frac{\alpha_{\rm M}-\alpha_{\rm T}}{\alpha_{\rm B}}$ 
                     & 0 
                     & $\bar{M}^2\mu_{Z,0}(1+\alpha_{\rm T})$ \\

  $\eta_{0}$ & $\frac{1}{1+\alpha_{\rm T}}$
             & $0$ 
             & $\eta_{\infty}$ 
             & $\frac{1}{1+\alpha_{\rm T}}$ \\

  $\bar{M}^2\Sigma_{0}$ & $1+\alpha_{\rm T}/2$ 
                        & $\frac{\alpha_{\rm M}-\alpha_{\rm T}}{2\alpha_{\rm B}}$ 
                        & $0$ 
                        & $\frac{1}{2}\left(\mu_{Z,0}+\mu_{Y,0}\right)$ \\

  $\gamma_{0}$ & $\alpha_{\rm T}$ 
               & $\infty$ 
               & $\gamma_{\infty}$ 
               & $\alpha_{\rm T}$ \\

  $g_{0}$ & $-\frac{\alpha_{\rm T}}{2+\alpha_{\rm T}}$ 
          & $-1$ 
          & $g_{\infty}$ 
          & $-\frac{\alpha_{\rm T}}{2+\alpha_{\rm T}}$ \\
  \hline
 \end{tabular}
 \caption{Limit on large scales (${\rm K}\rightarrow 0$) of the phenomenological MG functions.}
 \label{tab:MG0}
\end{table}

Different approaches give a different limit when ${\rm K}\rightarrow 0$. For the EFE and EoS approaches, $\eta_0=1/(1+\alpha_{\rm T})$, but as we will see in the next section, the exact numerical solution for $\eta$ differs from this value on large scales, as, not unexpectedly, the approximations we have made break down.

In the MP approach, for a generic model, $\mu_{0}=0$ and this is a consequence of the particular structure of the dynamical equation for $Z$: as density perturbations and potentials differ by a factor ${\rm K}^2$, this term will dominate over higher powers on large scales, leading to $\eta_{0}=\mu_{0}=0$. For the reasons discussed in Section~\ref{sect:MP}, models with $\alpha_{\rm B}=0$ need special care as the QSA breaks, therefore, in this case, we assume $\eta_{0}=\eta_{\infty}$, even if we anticipate that this will not be necessarily the case.

Although the expressions in Table~\ref{tab:MG0} are relatively simple and we can appreciate similarities especially for the EFE and EoS approaches (they are both quadratic in ${\rm K}$), it is instructive to investigate them in detail for particular Horndeski sub-classes, where one or more $\alpha$s are set to zero. We will not give the full expressions (they can be easily derived from the coefficients given in each section), but we will discuss generic features for selected models, to appreciate differences and similarities of the different approaches.

In the following, we will discuss in more detail models 1-5. We do not present model 6 as its expressions have already been presented in Table~\ref{tab:MG0} and no further simplifications are possible.

\subsection{Model 1 - \textit{\texorpdfstring{$k$}{k}-essence-like}}
For model 1 ($k$-essence-like models), only $\alpha_{\rm K}\neq 0$ and the expressions simplify considerably, making it a good test case for a theoretical analysis and comparison. In this case, it is easy to show that $\mu=\eta=1$ \textit{identically} for the EFE and EoS expressions, but for the MP equations, we find that $\eta=1$ \textit{identically} and $\mu=0$ on large scales and $\mu=1$ on small scales
\begin{equation*}
 \mu^{\rm MP} = \frac{c_{\rm s}^2{\rm K}^2}{\beta_4+c_{\rm s}^2{\rm K}^2}\,,
\end{equation*}
where $\beta_4 = 3\left(1+c_{\rm a,ds}^2\right) + 2\dot{H}/H^2$. Note that without a link to a specific Lagrangian, in these models $c_{\rm s}^2=0$ for a $\Lambda$CDM background where $w_{\rm ds}=-1$. According to our previous discussion, the QSA is valid only on scales smaller than the sound horizon of the scalar field, which means that in this case the QSA is hardly applicable. For this model only, therefore, we will consider a wCDM background where $w_{\rm ds}=-0.95$ in Sections~\ref{sect:numerical_comparison} and \ref{sect:observables}.

\subsection{Model 2 - \textit{\texorpdfstring{$f(R)$}{f(R)}-like models}}
Model 2 is characterised by having $\alpha_{\rm K}=\alpha_{\rm T}=0$. We shall first discuss the general expressions and then specialise them to $f(R)$ models where $\alpha_{\rm M}=2\alpha_{\rm B}\ll 1$, with $\alpha_{\rm B}=\tfrac{\dot{f}_{R}}{2H(1+f_R)}$.

The expressions for the modified gravity parameters are
\begin{align*}
 \mu^{\rm EFE} = &\, \frac{\mu_{+0}^{\rm EFE}+\alpha_{\rm B}^2c_{\rm s}^2\bar{M}^2\mu_{\infty}{\rm K}^2}{\mu_{+0}^{\rm EFE}+\alpha_{\rm B}^2c_{\rm s}^2{\rm K}^2}\frac{1}{\bar{M}^2}\,, & 
 \eta^{\rm EFE} = &\, \frac{\mu_{+0}^{\rm EFE}+\alpha_{\rm B}^2c_{\rm s}^2\bar{M}^2\mu_{Z,\infty}{\rm K}^2}{\mu_{+0}^{\rm EFE}+\alpha_{\rm B}^2c_{\rm s}^2\bar{M}^2\mu_{\infty}{\rm K}^2}\,,\\
 \mu^{\rm EoS} = &\, \frac{\gamma_1(\gamma_2-\gamma_7)+\alpha_{\rm B}^2c_{\rm s}^2\bar{M}^2\mu_{\infty}{\rm K}^2}{\gamma_1\gamma_2+\alpha_{\rm B}^2c_{\rm s}^2{\rm K}^2}\frac{1}{\bar{M}^2}\,, & 
 \eta^{\rm EoS} = &\, \frac{\gamma_1(\gamma_2-\gamma_7)+\alpha_{\rm B}^2c_{\rm s}^2\bar{M}^2\mu_{Z,\infty}{\rm K}^2}{\gamma_1(\gamma_2-\gamma_7)+\alpha_{\rm B}^2c_{\rm s}^2\bar{M}^2\mu_{\infty}{\rm K}^2}\,,
\end{align*}
where $\mu_{+0}^{\rm EFE}=\mu_{\rm p}$ has been defined in Eq.~(\ref{eqn:mup}). For the MP expressions, the coefficients do not simplify enough to obtain a concise expressions, therefore, we do not report them here, but we refer the reader to the expressions in Appendix~\ref{sect:coefficientsMP}.

For $f(R)$ models, the QSA has been studied and expressions are given by \cite{DeFelice2010,DeFelice2011}. Due to the importance of these models in the scientific literature and their relative simplicity which allows to obtain several exact results, we will consider their expressions in detail, as it was shown that the QSA works well for these models \citep{Hu2007}. Taking into account that $\bar{M}^2=1+f_R\approx 1$, $c_{\rm s}^2=1$, and $1/\alpha_{\rm B}\gg 1$, we find the following expressions for all the three approaches investigated in this work
\begin{equation*}
 \mu = \frac{\mu_{+0}+\frac{4}{3}{\rm K}^2}{\mu_{+0}+{\rm K}^2}\frac{1}{1+f_R}\,, \qquad
 \eta = \frac{\mu_{+0}+\frac{2}{3}{\rm K}^2}{\mu_{+0}+\frac{4}{3}{\rm K}^2}\,,
\end{equation*}
where
\begin{equation*}
 \mu_{+0} = \frac{4+\Gamma}{\alpha_{\rm B}}\frac{\dot{H}}{H^2} = 
 \frac{1}{3}\frac{1+f_{R}}{H^2f_{RR}}\,, \qquad
 \Gamma=\frac{\ddot{H}}{H\dot{H}}\,.
\end{equation*}
To go from the first to the second equality in the definition of $\mu_{+0}$, we used $\dot{f}_{R}=\dot{R}f_{RR}$ and the definition of the Ricci scalar $R=6(2H^2+\dot{H})$.

This result is a consequence of the high mass of the scalaron ($\propto 1/\alpha_{\rm B}$), which dominates on large scales (see also \cite{Battye2018a}). This is not the case in general for model 2, though.
Note that while this result is exact for the EFE approach, for the MP and EoS we only considered the leading terms.

Since both \cite{DeFelice2010} and \cite{DeFelice2011} provided similar expressions, it is useful to make a comparison with our results and verify whether they match or not. \cite{DeFelice2010}, starting from the perturbed equations written in terms of the degree of freedom $\delta R$ found
\begin{equation}
 \mu = \frac{M_{\delta\phi}^2/H^2 + \frac{4}{3}{\rm K}^2}{M_{\delta\phi}^2/H^2 + {\rm K}^2}\frac{1}{1+f_R}\,,
\end{equation}
where the mass squared of the scalaron is $M_{\delta\phi}^2=\tfrac{R}{3}\left(\tfrac{1}{m}-1\right)$ \citep{Silvestri2009}, and $m=\tfrac{Rf_{RR}}{1+f_R}$ so that $M_{\delta\phi}^2\approx\tfrac{1}{3}\tfrac{1+f_R}{f_{RR}}$, in perfect agreement 
with our results.

In \cite{DeFelice2011}, the authors write the perturbed field equations for a generic Horndeski model in terms of the perturbed scalar field $\delta\phi$. Applying the QSA, they find
\begin{equation}
 \mu = \frac{\frac{2}{3}\frac{\phi}{M_{\rm pl}}M_{\delta\phi}^2/H^2 + \frac{4}{3}{\rm K}^2}
            {\frac{2}{3}\frac{\phi}{M_{\rm pl}}M_{\delta\phi}^2/H^2 + {\rm K}^2}
       \frac{M_{\rm pl}}{\phi}\,,
\end{equation}
where $\phi=(1+f_R)M_{\rm pl}$ and the mass squared of the scalar degree of freedom is 
$M_{\delta\phi}^2=-G_{2\phi\phi}=(2f_{RR})^{-1}$, with $G_2=-\tfrac{1}{2}M_{\rm pl}^2[Rf_R-f(R)]$. Also in this case, the expressions provide the same result. In the limit of $f_{RR}\rightarrow 0$, that is infinite scalaron mass, $f(R)$ cosmologies reduce to the standard $\Lambda$CDM dynamics with $\mu=\eta=1$ at all scales.

\subsection{Model 3 - \textit{KGB-like models}}
In model 3, KGB-like models, $\alpha_{\rm K}$ and $\alpha_{\rm B}$ are both non zero and the braiding scale defined in section~\ref{sect:MP} plays a role. For this model, $\mu$ will take on different forms in the three approaches, but $\eta\equiv 1$ for all. More in detail, the expressions for $\mu$ are
\begin{align*}
 \mu^{\rm EFE} = &\, \frac{\mu_{+0}^{\rm EFE}+\alpha c_{\rm s}^2\mu_{\infty}{\rm K}^2}
                          {\mu_{+0}^{\rm EFE}+\alpha c_{\rm s}^2{\rm K}^2}\,, \\
 \mu^{\rm EoS} = &\, \frac{\gamma_1(\gamma_2-\gamma_7)+\alpha_{\rm B}^2c_{\rm s}^2\mu_{\infty}{\rm K}^2}
                          {\gamma_1(\gamma_2-\gamma_7)+\alpha_{\rm B}^2c_{\rm s}^2{\rm K}^2}\,, \\
 \mu^{\rm MP} = &\, \frac{\beta_1\beta_6{\rm K}^2+\alpha_{\rm B}^2c_{\rm s}^2\mu_{\infty}{\rm K}^4}
                         {\beta_1\beta_4+\beta_1\beta_5{\rm K}^2+\alpha_{\rm B}^2c_{\rm s}^2{\rm K}^4}\,,
\end{align*}
where $\mu_{+0}^{\rm EFE} = \mu_{\rm p}$ has been defined in Eq.~(\ref{eqn:mup}) and the coefficients $\beta_i$ and $\gamma_i$ can be inferred from the general expressions in Appendices~\ref{sect:coefficientsMP} and \ref{sect:coefficientsEoS}, respectively. We note that the expressions for the  EFE and EoS approaches are both quadratic (top and middle) in ${\rm K}$ and have the same large scale limit ($\mu_0=1$), while in the MP approaches the expressions are quartic in ${\rm K}$, and tend to zero on large scales.

\subsection{Model 4}
Model 4 is a particular subclass of model 5, where the functions $G_3$ and $G_4$ satisfy the differential relation $XG_{3,X}+G_{4,\phi}=0$. Linear dynamics is described by two functions, $\alpha_{\rm K}$ and $\alpha_{\rm M}$, while $\alpha_{\rm B}=\alpha_{\rm T}=0$. Since $G_4$ is a function of the scalar field $\phi$, to achieve $\alpha_{\rm T}=0$ we are free to write it as $G_4(\phi)=\tfrac{1}{2}M_{\rm pl}^2f(\phi/M_{\rm pl})$, where $f(\phi/M_{\rm pl})$ is a dimensionless function of the scalar field $\phi$. The solution of the differential equation is therefore $G_3 = -\tfrac{1}{2}M_{\rm pl}\left[f^{\prime}\left(\phi/M_{\rm pl}\right)\ln{\left(X/m^4\right)}+g\left(\phi/M_{\rm pl}\right)\right]$, where the prime represents the derivative with respect to $\phi/M_{\rm pl}$, $m$ is an arbitrary mass scale and $g(\phi/M_{\rm pl})$ a dimensionless function of $\phi$.

Since $\alpha_{\rm B}=0$, the expressions for $\mu$ and $\eta$ simplify significantly, not only at the level of the coefficients (as for the EFE approach), but also regarding the functional form. For example, the MP expressions are now quadratic and not quartic. A comment is necessary for the EoS approach: formally, the expressions would be scale-independent as the coefficients of the ${\rm K}^2$ terms are proportional to $\alpha_{\rm B}^2$. This would mean that the small-scale limit will not be reached, in contrast to its derivation which sees first taking the limit of ${\rm K}\rightarrow\infty$ and later on specifying the values of the $\alpha$ functions. Therefore, in our numerical implementation, we set the coefficients of the ${\rm K}^2$ terms to a small, but finite and different from zero, value.

The coefficients of interest then read
\begin{align*}
 \mu^{\rm EFE} = &\, \frac{\mu_{+0}^{\rm EFE}+\alpha c_{\rm s}^2\bar{M}^2\mu_{\infty}{\rm K}^2}
                          {\mu_{+0}^{\rm EFE}+\alpha c_{\rm s}^2{\rm K}^2}\frac{1}{\bar{M}^2}\,, & 
 \eta^{\rm EFE} = &\, \frac{\mu_{+0}^{\rm EFE}+\alpha c_{\rm s}^2{\rm K}^2}
                           {\mu_{+0}^{\rm EFE}+\alpha c_{\rm s}^2\bar{M}^2\mu_{\infty}{\rm K}^2}\,,\\
 \mu^{\rm EoS} = &\, \frac{\gamma_1\gamma_2+\alpha_{\rm B}^2c_{\rm s}^2\bar{M}^2\mu_{\infty}{\rm K}^2}
                          {\gamma_1\gamma_2+\alpha_{\rm B}^2c_{\rm s}^2{\rm K}^2}\frac{1}{\bar{M}^2}\,, & 
 \eta^{\rm EoS} = &\, \frac{\gamma_1\gamma_2+\alpha_{\rm B}^2c_{\rm s}^2{\rm K}^2}
                           {\gamma_1\gamma_2+\alpha_{\rm B}^2c_{\rm s}^2\bar{M}^2\mu_{\infty}{\rm K}^2}\,,\\
 \mu^{\rm MP} = &\, \frac{c_{\rm s}^2\bar{M}^2\mu_{\infty}{\rm K}^2}{\beta_4+c_{\rm s}^2{\rm K}^2}
                    \frac{1}{\bar{M}^2}\,, &
 \eta^{\rm MP} = &\, \eta_{\infty}\,,
\end{align*}
where $\mu_{+,0}^{\rm EFE}=6\left(\dot{H}+\frac{\rho_{\rm m}+P_{\rm m}}{2M^2}\right)\dot{H}/H^4$ and $\eta=\eta_{\infty}$ due to the breakdown of the QSA, as previously discussed.

\subsection{Model 5 - \textit{\texorpdfstring{$c_{\rm T}=1$}{cT1} models}}
Finally, in model 5 ($c_{\rm T}=1$ models), only $\alpha_{\rm T}=0$ and this represents the most general model allowed by current observations of gravitational waves. Expressions and numerical values of the effective gravitational constant and the slip differ, in general, for all the three approaches. It is nevertheless useful to consider a particular model, the no slip gravity model \citep{Linder2018} where $\alpha_{\rm B}=\alpha_{\rm M}$, 
while $\alpha_{\rm K}$ is independent from them. The model is given his name as $\eta_{\infty}=1$, and the relevant expressions
\begin{align*}
 \mu^{\rm EFE} = \, \mu^{\rm EoS} = \mu^{\rm MP} = \frac{1}{\bar{M}^2}\,, \quad 
 \eta^{\rm EFE} = \, \eta^{\rm EoS} = 1\,, \quad
 \eta^{\rm MP} = \, \frac{\beta_1\beta_6{\rm K}^2+\alpha_{\rm B}^2c_{\rm s}^2{\rm K}^4}
                         {\beta_1\beta_4+\beta_1\beta_5{\rm K}^2+\alpha_{\rm B}^2c_{\rm s}^2{\rm K}^4}\,.
\end{align*}
This model has the interesting property that $\mu=\mu_{\infty}$ for all the approaches and $\eta^{\rm EFE}=\eta^{\rm EoS}=\eta_{\infty}$ at all scales and times. For $\bar{M}^2\approx 1$, the phenomenology of the no slip model is similar to that of $k$-essence. Note that these results do not apply to the slip parameter derived from the metric potentials ($\eta^{\rm MP}_0=0$ and $\eta^{\rm MP}_{\infty}=1$), but it is interesting to see that for selected models, the different approaches can lead to the same effective gravitational constant.

\section{Comparison with the exact numerical results}\label{sect:numerical_comparison}
In this section we compare the analytical predictions for $\mu$ and $\eta$ from Sections~\ref{sect:EFE}, \ref{sect:MP} and \ref{sect:EoS} with the exact numerical results obtained with our numerical code \texttt{EoS\_class} \cite{Pace2019a}. 
We present results for $\mu$ and $\eta$ as a function of the scale $k$ only, assuming the widely used phenomenological parameterization for the $\alpha$ functions $\alpha_{\rm X}=\alpha_{\rm X,0}\Omega_{\rm ds}(a)$, where ${\rm X}\in\{{\rm K}, {\rm B}, {\rm M}, {\rm T}\}$ and $\Omega_{\rm ds}(a)$ represents the evolution of the dark sector component. In our previous work \cite{Pace2019a}, we considered a range of values for $\alpha_{\rm X,0}$, but here we specialise to the following values: $\alpha_{\rm K,0}=1$, $\alpha_{\rm B,0}=0.625$, $\alpha_{\rm M,0}=1$ and $\alpha_{\rm T,0}=1$. For model 6, we set  $\alpha_{\rm M,0}=0.47$ as in \cite{Noller2019}. This is because $\alpha_{\rm M}=\alpha_{\rm T}$ leads to additional cancellations in the coefficients which make the model not sufficiently general. We remind the reader that for model 2 $\alpha_{\rm K,0}=0$. Note that $\alpha_{\rm K}$ is usually unconstrained by observations and important only on scales larger than the sound horizon. For constraints on $\alpha_{\rm K}$, see \cite{Gleyzes2016,Alonso2017}. We also limit ourselves to study $\mu$ and $\eta$ at $z=0$, as this is the epoch with largest differences, since at earlier times the $\alpha_X$ are closer to the $\Lambda$CDM values within the framework we have used. For all the models we assume $w_{\rm ds}=-1$, except for model 1, where we set $w_{\rm ds}=-0.95$.

From Fig.~7 of \cite{Pace2019a}, we can infer the value of the sound speed for the models considered here at $z=0$. This ranges from $c_{\rm s}^2\simeq 0.5$ for model 3 to $c_{\rm s}^2\simeq 2.5$ for model 4, with $c_{\rm s}^2\simeq 2$ for all the other models.\footnote{We observe that the choice of these parameters sometimes leads to a superluminal speed of propagation \cite{Bonvin2006}. This does not necessarily imply closed timelike curves \cite{Babichev2008} but can nevertheless be problematic in theories with a Lorentz-invariant UV completion \cite{Adams2006}.} In \cite{Pace2019a}, $c_{\rm s}^2=0$ for model 1, while here, having assumed $w_{\rm ds}=-0.95$, we have $c_{\rm s}^2=0.15$. These numbers will be useful in the next section when discussing the regime of validity of the QSA in recovering the observables.

\begin{figure}[!t]
 \centering
 \includegraphics[width=7.65cm,height=5.9cm]{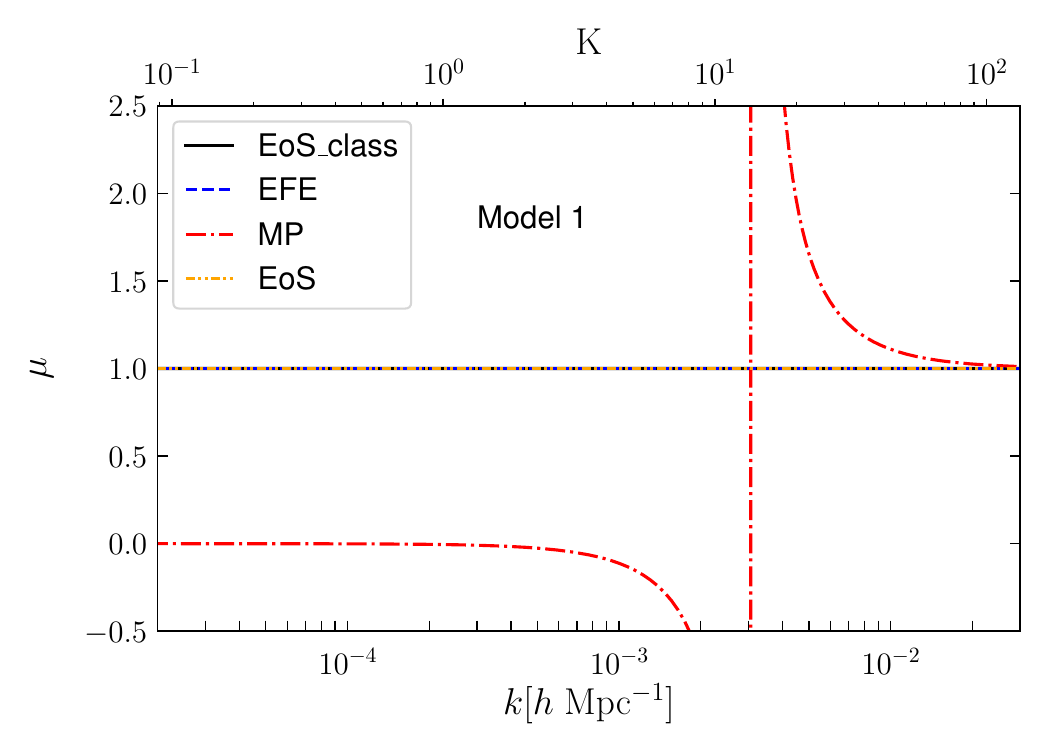}
 \includegraphics[width=7.65cm,height=5.9cm]{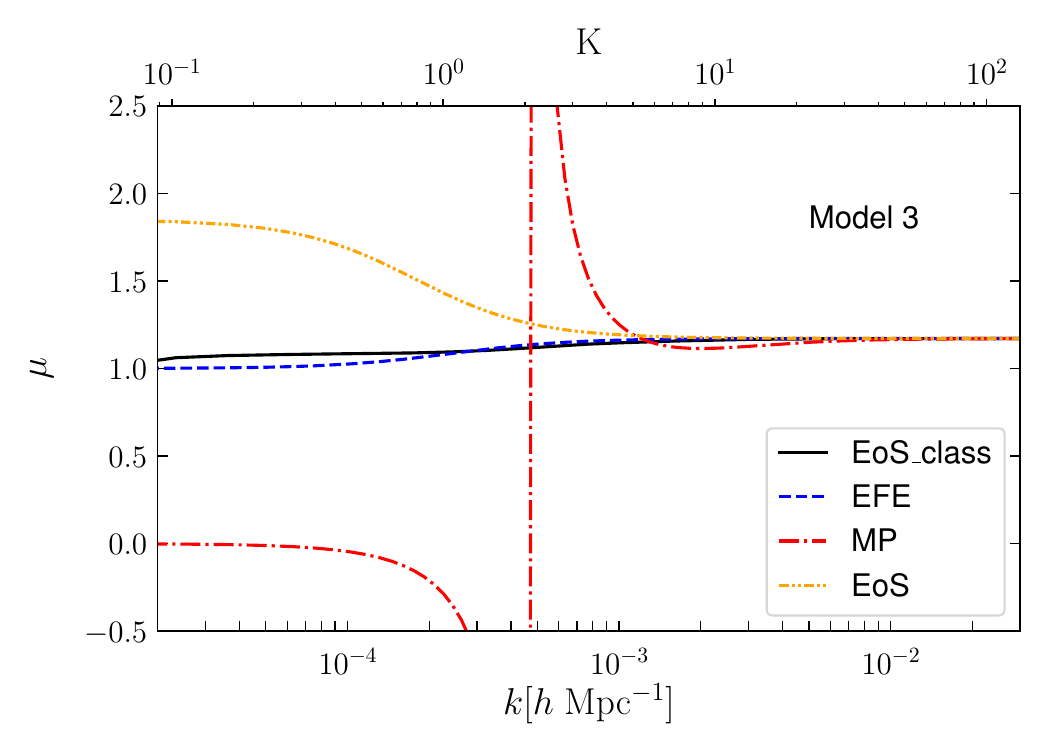}
 \includegraphics[width=7.65cm,height=5.9cm]{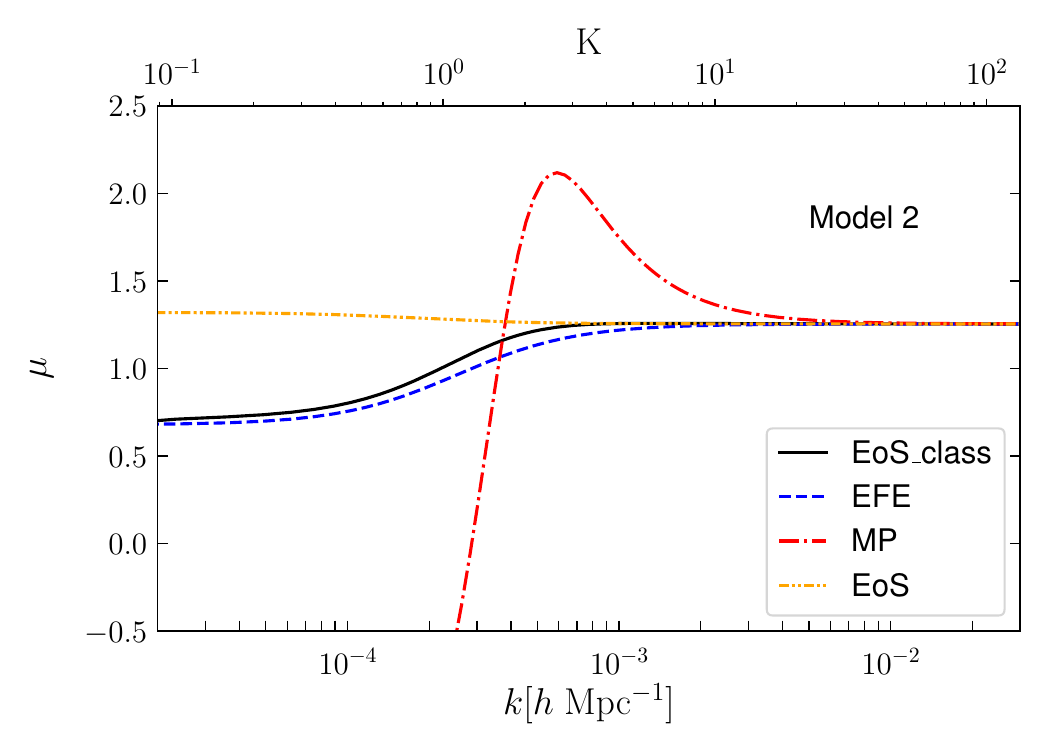}
 \includegraphics[width=7.65cm,height=5.9cm]{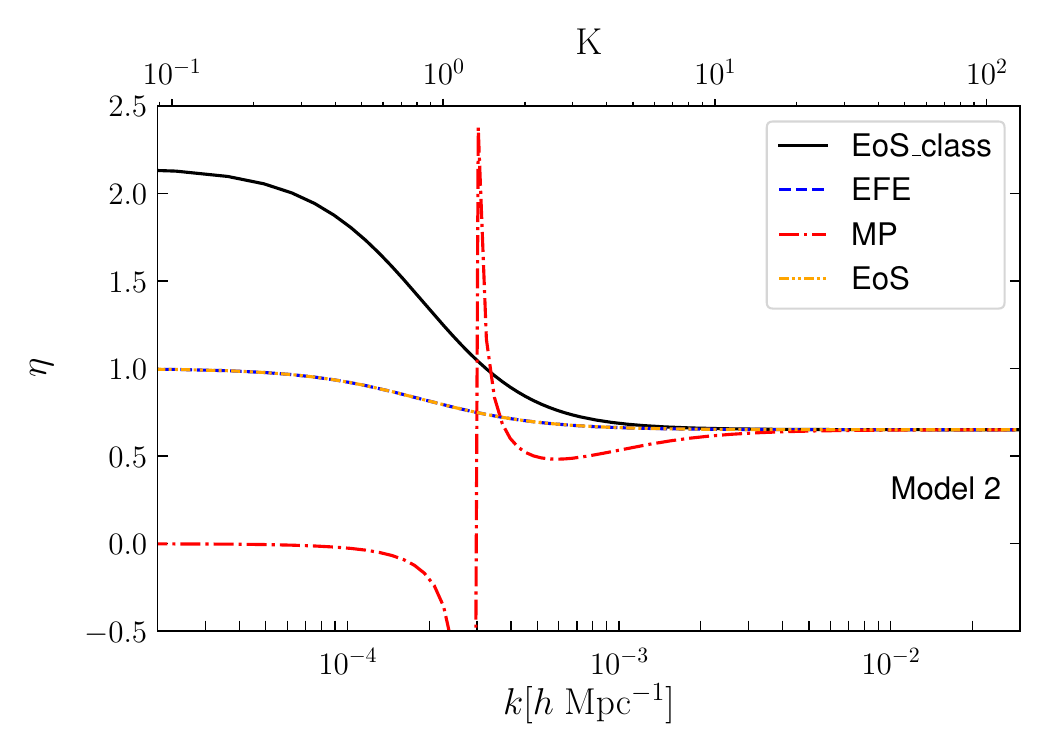}
 \includegraphics[width=7.65cm,height=5.9cm]{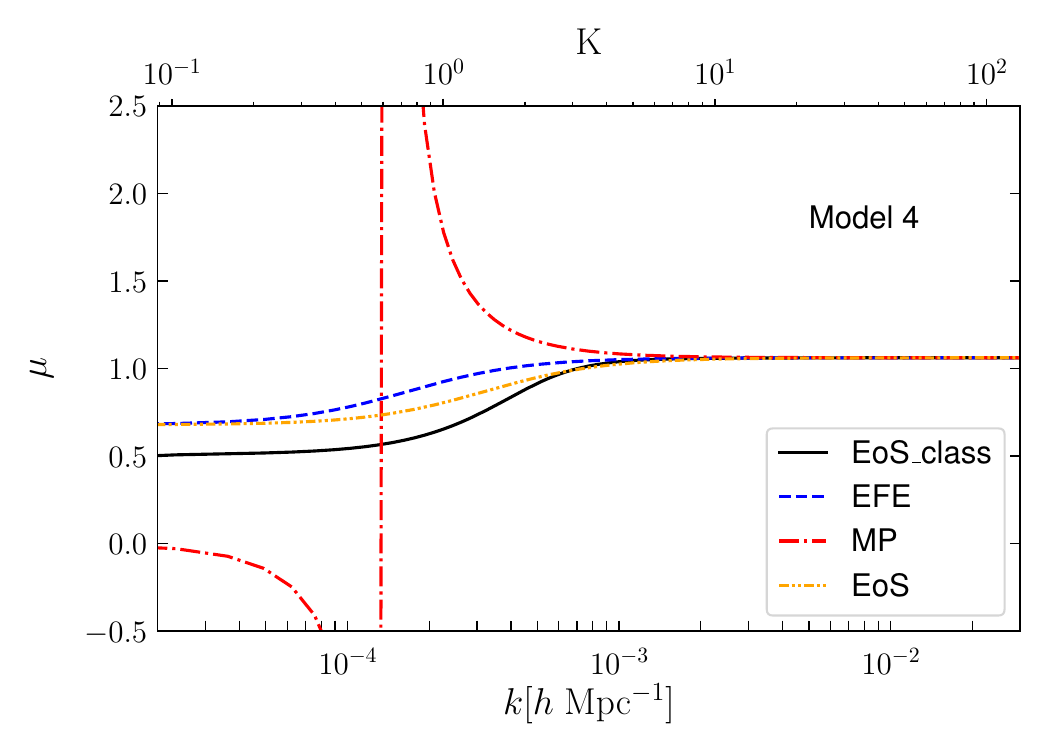}
 \includegraphics[width=7.65cm,height=5.9cm]{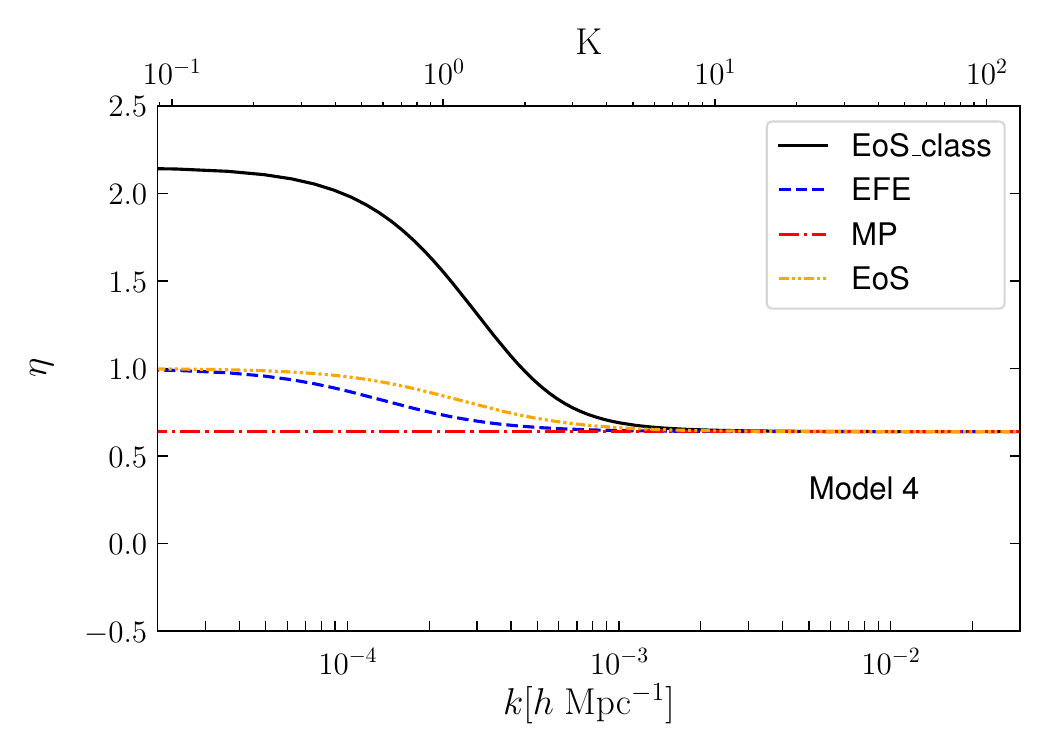}
 \caption[justified]{Scale-dependence of the effective gravitational constant, $\mu$, and the slip, $\eta$, at $z=0$ for different models. The black solid line represents the exact numerical solution of the code \texttt{EoS\_class}, the blue dashed line the prediction for EFE, the red dot-dashed line the MP prediction and the orange dashed-dot-dotted line the EoS solution. 
 \textit{Top left panel:} Model 1 ($k$-essence-like) for $\alpha_{\rm K,0}=1$ and 
 $\alpha_{\rm B}=\alpha_{\rm M}=\alpha_{\rm T}=0$. 
 \textit{Top right panel:} Model 3 (KGB-like) for $\alpha_{\rm K,0}=1$, $\alpha_{\rm B,0}=0.625$ and $\alpha_{\rm M}=\alpha_{\rm T}=0$. 
 \textit{Middle panels:} Model 2 ($f(R)$-like) for $\alpha_{\rm K,0}=\alpha_{\rm T}=0$, $\alpha_{\rm B,0}=0.625$ and $\alpha_{\rm M,0}=1$. Left (right) panel shows $\mu$ ($\eta$).
 \textit{Bottom panels:} Model 4 for $\alpha_{\rm K,0}=1$, $\alpha_{\rm M,0}=1$ and $\alpha_{\rm B}=\alpha_{\rm T}=0$. Left (right) panel shows $\mu$ ($\eta$).}
 \label{fig:MGparam1}
\end{figure}

\begin{figure}[!t]
 \centering
 \includegraphics[width=7.65cm,height=5.85cm]{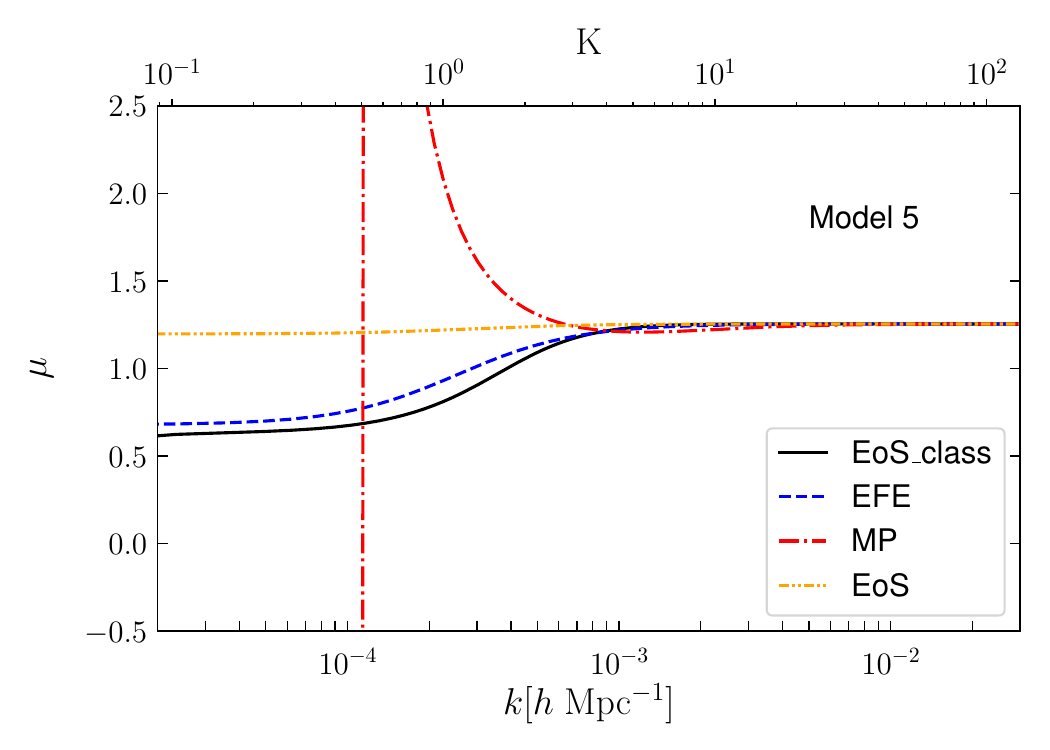}
 \includegraphics[width=7.65cm,height=5.85cm]{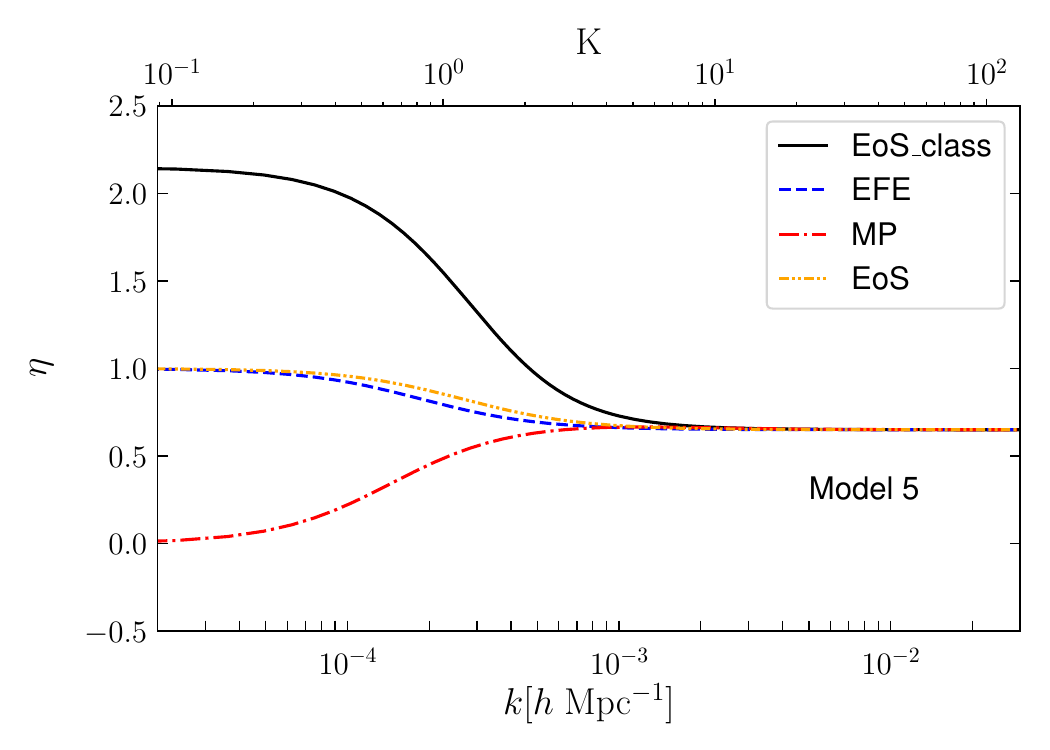}
 \includegraphics[width=7.65cm,height=5.85cm]{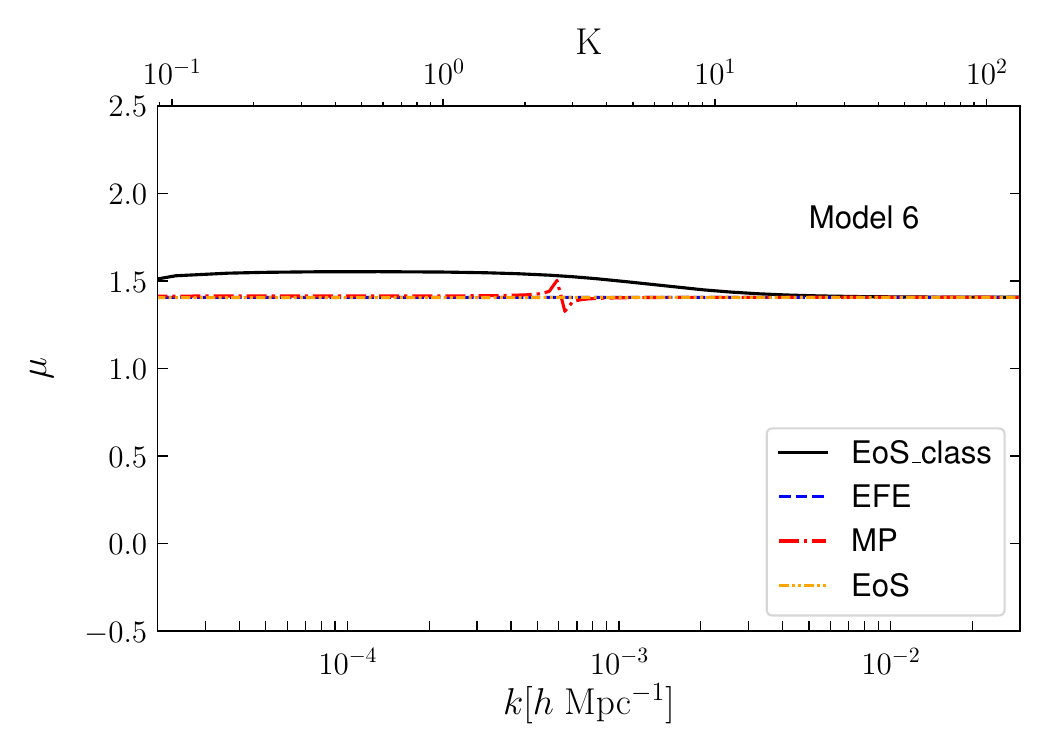}
 \includegraphics[width=7.65cm,height=5.85cm]{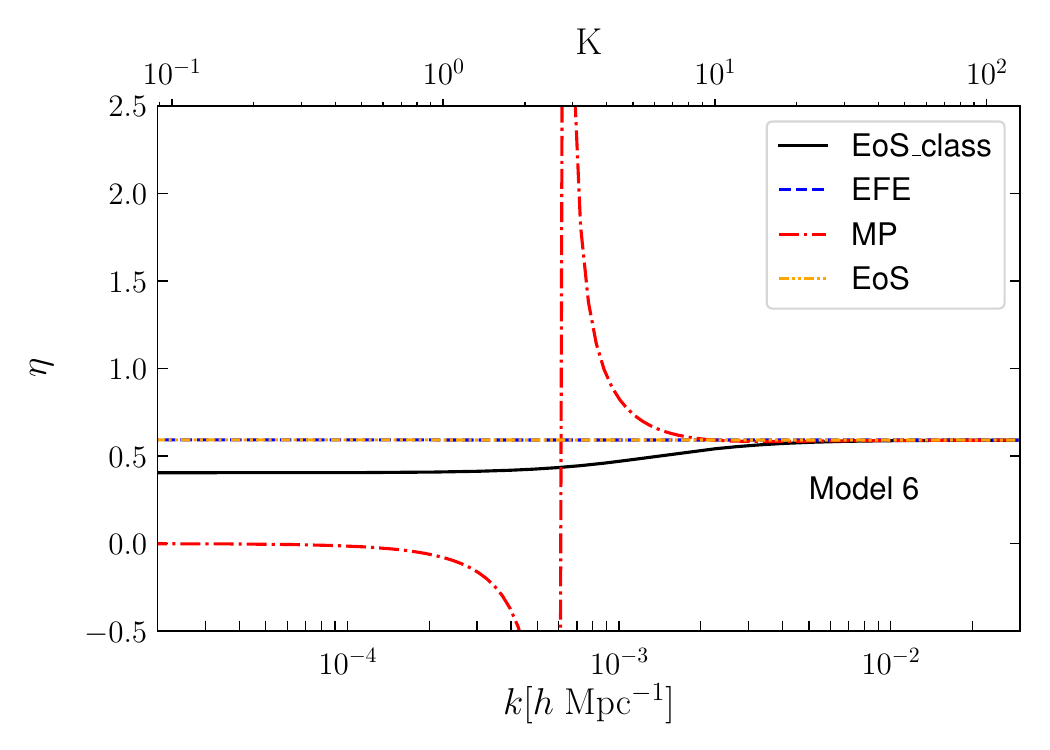}
 \caption[justified]{Scale-dependence of the effective gravitational constant, $\mu$, and the slip, $\eta$, at $z=0$ for different models. Line styles and colours are as in Fig.~\ref{fig:MGparam1}. 
 \textit{Top panels:} Model 5 for $\alpha_{\rm K,0}=1$, $\alpha_{\rm B,0}=0.625$,  $\alpha_{\rm M,0}=1$ and  $\alpha_{\rm T}=0$. 
 \textit{Bottom panels:} Model 6 for $\alpha_{\rm K,0}=\alpha_{\rm T,0}=1$, $\alpha_{\rm B,0}=0.625$ and $\alpha_{\rm M,0}=0.47$. Left (right) panel shows $\mu$ ($\eta$).}
 \label{fig:MGparam2}
\end{figure}

Our results are shown in Figures~\ref{fig:MGparam1} (for models 1--4) and \ref{fig:MGparam2} (for models 5 and 6). For models 1 and 3 we do not show the slip $\eta$ as all the approaches predict $\eta=1$ exactly, in agreement with the numerical result. Note that this is expected, as the anisotropic stress for the dark sector is null in these models.

For model 1 (top left panel), the EFE and EoS approach show an excellent agreement over all scales, while the MP expression reproduces the exact numerical values only on small scales and goes to zero on large scales, as already discussed. The divergence appearing for $\mu^{\rm MP}$ is due to the denominator $\beta_4+c_{\rm s}^2{\rm K}^2\to 0$, as $\beta_4 \le 0$. We verified that a very similar behaviour is also present at higher redshifts. Smaller values of the sound speed result in deviations between the analytical and the numerical solution on smaller scales. For the expected value of quintessence models, $c_{\rm s}^2=1$, the analytical solution for the MP approach starts deviating from the numerical solution at $k\approx 10^{-3}h~$Mpc$^{-1}$ (${\rm K}\approx 4$).

In the middle panels we show results for model 2 with $\alpha_{\rm K}=\alpha_{\rm T}=0$, $\alpha_{\rm B,0}=0.625$ and $\alpha_{\rm M,0}=1$. For a generic $f(R)$-like model, the analytical expressions exactly recover the small-scale limit (${\rm K}\gg 1$), but we notice substantial differences on large scales (${\rm K}\lesssim\mathcal{O}(1)$) which diminish at earlier times when deviations from general relativity are less important. These deviations are a clear sign of the breaking of the approximations made (sub-horizon scales). We notice a general better agreement between \texttt{EoS\_class} and the EFE expressions, while EoS overestimates this quantity. At the same time though, the EoS expression agrees longer with the numerical expectation with respect to EFE, but this is likely a coincidence due to the parameter choice. The predictions of the MP approach show a departure from the exact solution at $k\approx 7\times 10^{-3}h~$Mpc$^{-1}$ (${\rm K}\simeq 20$) and on very large scales it becomes negative. 
The analytical predictions for $\eta$ are identical for EFE and EoS, in agreement with expressions in Table~\ref{tab:MG0} and $\eta_0=1$, a factor of two lower than the numerical solution, showing that the analytical expressions under-predict the true result. Differences start to arise for ${\rm K}$ of order of a few, as one approaches the horizon scale. Once again the MP expressions show the strongest differences with respect to the numerical solution, both at large and intermediate scales.

As explained in detail in Section~\ref{sect:comparison}, for $f(R)$ models all the three approaches lead to the same expression for $\mu$ and $\eta$ as on large scales the mass of the scalaron is the dominating term. When comparing the analytical predictions with the numerical ones for a model with $B_0=10^{-5}$, we find an excellent agreement for both $\mu$ and $\eta$, justifying the widely adopted QSA for this class of models. Thanks to the above considerations, it is easy to understand that this agreement holds also at high redshifts. Note though, that at high redshifts, the dominant component is the mass of the scalaron, therefore, the regime of applicability of $\mu_{\infty}$ and $\eta_{\infty}$ is pushed to smaller scales. This ensures that the correct general relativistic limit is reached. At $z=0$, both $\mu$ and $\eta$ do not depend on scale as the transition to $\mu_{\infty}$ and $\eta_{\infty}$ takes place on very small scales, i.e., ${\rm K}\simeq 200$. For this reason, being $\mu=\eta=1$ over the scales of the plot, we do not show the results for $f(R)$ models.

For model 3 (KGB-like) with $\alpha_{\rm B,0}=0.625$, the EFE prediction is in very good agreement with the numerical one, while EoS overestimates $\mu$ and starts deviating from the exact solution at $k\approx 10^{-3}h~$Mpc$^{-1}$. 
Once again, the MP expression diverges as the numerator goes to zero. On very large scales, $\mu^{\rm MP}_0\to 0$. To explain our results, it is useful to refer to Table~\ref{tab:MG0}. There, we see that $\mu_{0}^{\rm EFE}=1$, $\mu_{0}^{\rm MP}=0$ and $\mu_{0}^{\rm EoS}=1-\tfrac{\gamma_7}{\gamma_2}$, where $\gamma_7=-2\alpha_{\rm B}^2/\alpha$. It is the coefficient $\gamma_7$ responsible for the differences between EFE and EoS. We also verified that for higher values of $\alpha_{\rm B,0}$, differences between EoS and \texttt{EoS\_class} are more pronounced.

In the bottom panels of Figure~\ref{fig:MGparam1} we present the results for model 4 with $\alpha_{\rm M,0}=1$. The behaviour of the MP result is very similar to the models already discussed, including the divergence which seems to be quite a generic feature for this approach for the parameters adopted. Note that for this model, $\alpha_{\rm B}=0$ and as discussed in the previous section, the QSA is no longer satisfied. We derived $\mu_Z$, and assuming $\eta=\eta_{\infty}$, we inferred $\mu=\mu_Z/\eta$. In contrast to previous cases, EoS and EFE exhibit a similar qualitative behaviour, with differences appearing at the transition scale between small and large scales.

To understand why this is the case, we refer again to Table~\ref{tab:MG0}. It is straightforward to see that $\mu_0^{\rm EFE}=1/\bar{M}^2$, and since $\bar{M}^2>1$, $\mu_0^{\rm EFE}<1$ on very large scales. Regarding the EoS expression, we now have $\gamma_7=0$ and this forces $\mu_0^{\rm EoS}=\mu_0^{\rm EFE}$. 
Interestingly enough, for stronger deviations from general relativity, $\alpha_{\rm M,0}=4$, the numerical solution approaches 0, being therefore in agreement with the MP prediction, while the EFE and EoS expectations differ from each other in the range $10^{-4}h~{\rm Mpc}^{-1}\lesssim k \lesssim 2\times 10^{-3}h~{\rm Mpc}^{-1}$. Similar conclusions can be deduced analysing the slip parameter. All the models underpredict the numerical expectation and depart from it at $k \lesssim 2\times 10^{-3}h~{\rm Mpc}^{-1}$. We remind the reader that we cannot make a proper comparison for $\eta^{\rm MP}$ due to the breakdown of the QSA in this class of models.

In the top panels of Figure~\ref{fig:MGparam2} we consider model 5 ($c_{\rm T}=1$ models), which is the most generic Horndeski model allowed by gravitational wave observations. We assume $\alpha_{\rm B,0}=0.625$ and $\alpha_{\rm M,0}=1$. The behaviour of the analytical predictions is very similar to what we found for models 3 and 4. We see, once again, a divergence for $\mu^{\rm MP}$ and very good agreement between the numerical solution and the EFE prediction. The EoS prediction, instead, is about a factor of two higher than the numerical solution on large scales, once again due to the term $\gamma_7/\gamma_2$. This term is also responsible for the major deviations we observe between the numerical and the EoS solution when we increase the parameters of the model, while the EFE prediction stays always very close to the numerical one and the MP is no longer diverging. However, as for the others, this departs from the exact result for $k \lesssim \times 10^{-2}h~{\rm Mpc}^{-1}$. It is also easy to see that $\mu_0^{\rm MP}\propto\alpha_{\rm M}/\alpha_{\rm B}<0$, explaining its behaviour on large scales.\footnote{We remind the reader that our definition of $\alpha_{\rm B}$ differs by a factor $-2$ with that adopted by \cite{Bellini2014,Zumalacarregui2017}. In our code, therefore, the braiding is negative.} At higher redshifts, all the analytical predictions, except for MP, agree with the results of our code \texttt{EoS\_class}. When considering the slip parameter $\eta$, right panel, we clearly see that none of the theoretical models reproduces the numerical behaviour on large scales and deviates from it for $k \lesssim 5\times 10^{-3}h~{\rm Mpc}^{-1}$.

Finally, in the bottom panels of Figure~\ref{fig:MGparam2}, we present the behaviour for model 6, the most general Horndeski model, assuming $\alpha_{\rm K,0}=\alpha_{\rm T,0}=1$, $\alpha_{\rm B,0}=0.625$ and $\alpha_{\rm M,0}=0.47$.\footnote{We change the value of $\alpha_{\rm M,0}$ with respect to our previous work, as when $\alpha_{\rm M}=\alpha_{\rm T}$ some of coefficients go to zero, leading to simplifications which reduce the generality of the model.} 
All the approaches show an identical behaviour for the effective gravitational constant $\mu$, but underpredict the numerical value for ${\rm K}\lesssim 30$. The EFE and EoS approaches predict the same behaviour for the slip $\eta$ and differ from that of the MP. All three approaches deviate from the numerical solution for $k\lesssim 5\times 10^{-3}h\,{\rm Mpc}^{-1}$.

This discussion shows that, in general, the analytical predictions do not reproduce the numerical behaviour of $\mu$ and $\eta$ on large scales, even if for some particular models the agreement is better than for others. This normally happens for models where modifications to gravity are small, a condition which can be realised, for example, at early times or when $\alpha_{\rm X,0}$ is small. Deviations between the analytical predictions and the numerical results arise when ${\rm K}$ is of the order unity, as expected, as the sub-horizon condition is violated and it is no longer correct to neglect time derivatives and scale-independent terms. For $f(R)$ models, however, we find an excellent agreement at all scales, due to the large mass for the scalaron. When this is not the case, as for a generic model 2, the agreement on large scales is lost. Therefore, in the next section we will investigate in detail how strongly the differences between the analytical predictions and the numerical expectation of $\mu$ and $\eta$ affect the spectra derived by solving the equations of the QSA, rather than the full ones.

We would also like to comment further on the divergence seen for the expressions derived within the MP approach and the consequences it carries. We will discuss in detail models 1 and 4, as the expressions are considerably simpler than other cases. For these models, the only relevant function is $\beta_4$, which, for a $k$-essence-like model reads $\beta_4=2\dot{H}/H^2+3(1+c_{\rm a,ds}^2)$. Assuming a $\Lambda$CDM background, $c_{\rm a,ds}^2=w_{\rm ds}=-1$, this coefficient reduces to $\beta_4=2\dot{H}/H^2\propto-(\rho_{\rm m}+P_{\rm m})<0$ at all times and smaller at earlier times, when matter dominates. This implies that at earlier times the divergence is shifted towards smaller scales, which are those of interest, being in the regime of validity of the quasi-static approximation. The scale where the divergence takes place is therefore ${\rm K}^2\propto1/c_{\rm s}^2$. To shift the divergence to scales not affecting the whole analysis (${\rm K}\to 0$), the sound speed of perturbations must be very large, $c_{\rm s}^2\gg 1$. 
This is the case, for example, of the cuscuton model \citep{Afshordi2007a,Afshordi2007b}, where the sound speed of perturbations is infinite. This model can be realised as the incompressible limit of a $k$-essence theory. Since it can be shown that perturbations do not introduce any additional dynamical degree of freedom but simply satisfy a constraint equation, the theory is causal in that no microscopic information is carried. Similar conclusions can be reached for model 4, where additional terms involving $\alpha_{\rm M}$ will be present. This implies that for the set of parameters and models considered, the formulation of the metric potentials is not viable.

To see this from another point of view, we can refer to the equations for the potentials. When the denominator goes to zero, it means that the coefficient $C_{\Psi}$ in Eq.~(\ref{eqn:MP}) is zero and the only terms surviving are those with $\ddot{\Psi}$ and $\dot{\Psi}$, illustrating a break down of the QSA in this case. These divergences are also present in the EoS approach, but only for models which are very different from the $\Lambda$CDM, which are of much less interest as they are already ruled out.

One might wonder what happens, in general, to $\mu_{-0}$ for the EFE approach. While this is difficult to establish in general due to the interplay between $\alpha_{\rm M}$, $\alpha_{\rm B}$ and the $\ddot{H}$ term, we can consider, for simplicity, model 4 where $\alpha_{\rm B}=0$. It turns out that $\mu_{-0}=\dot{H}^2(M^2-M_{\rm pl}^2)/M^2$ which is always positive, explaining why the expression for $\mu$ does not show any divergence.

For the EoS approach, it is not that simple to establish, in general, whether $\mu_{-0}$ is positive or negative. We can make some progress though, considering again model 4 with $\alpha_{\rm K}\neq 0$ and $\alpha_{\rm M}\neq0$, so that $\mu_{-0}=\gamma_1\gamma_2=(\rho_{\rm m}+P_{\rm m}+2M^2\dot{H})^2/(4H^4M^4)>0$. There exist, therefore, models where the approach is viable and does not lead to divergences/negative values for $\mu$ and/or $\eta$. We stress that this is not necessarily true in general, due to the complexity of the coefficients.

After this discussion, one can ask whether including terms with the derivatives to derive the expressions for $\mu$ (and $\eta$) as done in the semi-dynamical approach \citep{Lombriser2015a} can avoid the divergences. It turns out that the answer depends on the pivot scale: if the pivot scale is nonzero then $\mu\to0$, however, choosing the pivot scale at ${\rm K}=0$, $\mu$ can become undefined (of the form $0/0$). This is, though, not very accurate as one should also consider the fact that $f_{\zeta}$ scales as ${\rm K}^2$ (see \cite{Hu2007a}) and a proper determination of the limit on large scales require to take into account that correction.

We conclude this section by commenting more in detail upon the differences between the expressions for EFE and EoS. We only consider these two as they are quadratic in the scale and results easier to interpret. We saw that, for example, the slip has the same limit on small and large scales, while this is not necessarily the case on intermediate scales. 
This implies that the transition scale ${\rm K}_{\ast}$ between small and large scales is different for the two approaches. Since the expressions for EFE and EoS are quadratic in ${\rm K}$, it is easy to infer it. 
Rewriting the expression for $\mu$ (and similarly for $\mu_Z$) as
\begin{equation}
 \mu = \frac{\mu_0+\mu_{\infty}({\rm K}/{\rm K}_{\ast})^2}{1+({\rm K}/{\rm K}_{\ast})^2}\,,
\end{equation}
where
\begin{equation*}
 \mu_0 \equiv \frac{\mu_{+0}}{\mu_{-0}}\frac{1}{\bar{M}^2}\,, \quad 
 {\rm K}_{\ast}^2 \equiv \frac{\mu_{-0}}{\mu_{-2}}\,,
\end{equation*}
we find that the transition scale reads
\begin{equation}
 \begin{split}
  {\rm K}_{\ast}^{\rm EFE} \equiv &\, \left[
    \frac{6\left\{\left(\dot{H}+\frac{\rho_{\rm m}+P_{\rm m}}{2M^2}\right)\dot{H} + \dot{H}\alpha_{\rm B} 
                  \left[H^2(3+\alpha_{\rm M})+\dot{H}\right]+H(\dot{H}\alpha_{\rm B})^{\hbox{$\cdot$}}\right\}
                  /H^4}{\alpha c_{\rm s}^2}\right]^{1/2} \,, \\
  {\rm K}_{\ast}^{\rm EoS} \equiv &\, \left[
  \frac{\gamma_1(\gamma_2-\alpha_{\rm T}/3)}{\alpha_{\rm B}^2c_{\rm s}^2}\right]^{1/2}\,,
 \end{split}
\end{equation}
for EFE and EoS, respectively. These expressions, in general, differ from each other, but reduce to the same value for $f(R)$ models, as $\alpha_{\rm B}\ll 1$.

\section{Observable spectra}\label{sect:observables}
In the previous section, we compared our analytical results with the exact numerical calculation obtained with our code \texttt{EoS\_class}. We saw that, in general, the exact numerical behaviour for $\mu$ and $\eta$, obtained by solving the full equations, is not reproduced correctly on large scales by any of the three different QSA recipes. This is a direct consequence of the QSA which turns differential equations into constraint equations, by neglecting terms in the dynamical equations.

In this section, we explore the observational consequences of the differences between exact numerical and approximated analytical expression discussed in the previous section and understand how well the spectra derived by applying the QSA reproduce the spectra obtained by solving the full equations. In particular, we consider $C_{\ell}^{\rm TT}$, $C_{\ell}^{\phi\phi}$, and $P(k)$. We implemented the analytical expressions for $\mu$ and $\eta$ derived in Section~\ref{sect:QSA} in a suitably modified version of the code \texttt{CLASS} which we call \texttt{QSA\_class}. For a better quantification of the impact of the differences on large scales, we also approximate the effective gravitational constant and the slip parameter with their value on small scales. We do not report results for $f(R)$ models as we saw that there is an excellent agreement between the numerical and the analytical prediction and for MP expressions, due to the problems arising with the divergences and explained in detail in Section~\ref{sect:numerical_comparison}. We also do not show results for $k$-essence-like models, as $\mu=\eta=1$ identically and the approximated dynamics reproduce the exact ones.

\begin{figure*}[!t]
 \centering
 \includegraphics[scale=0.2865]{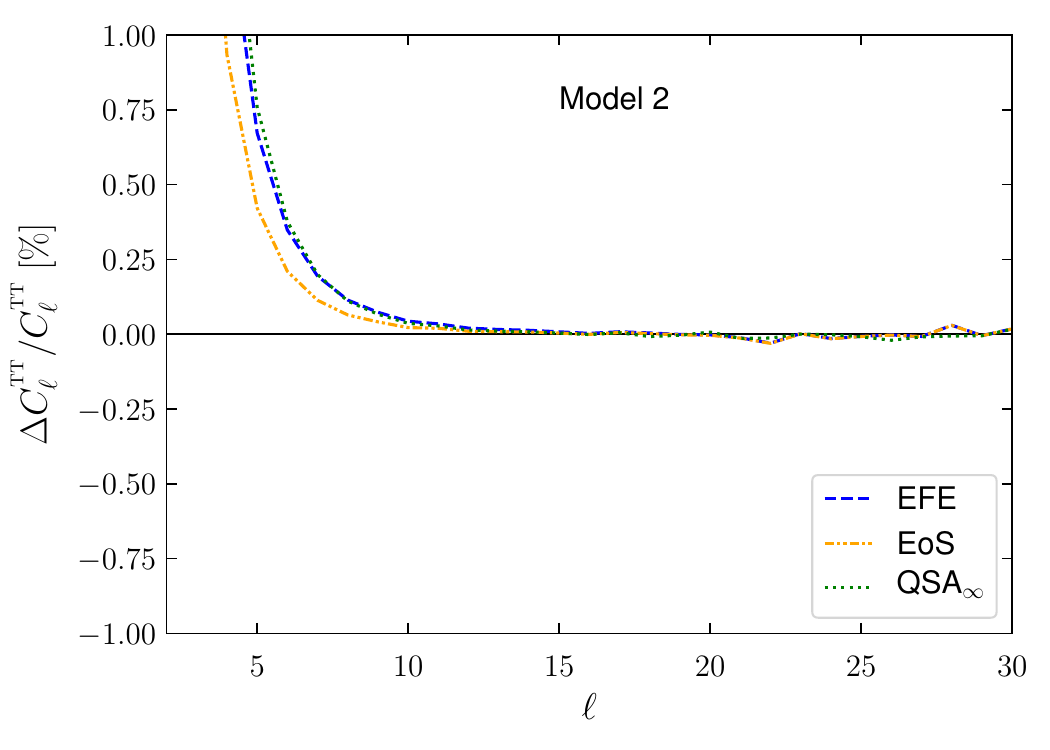}
 \includegraphics[scale=0.2865]{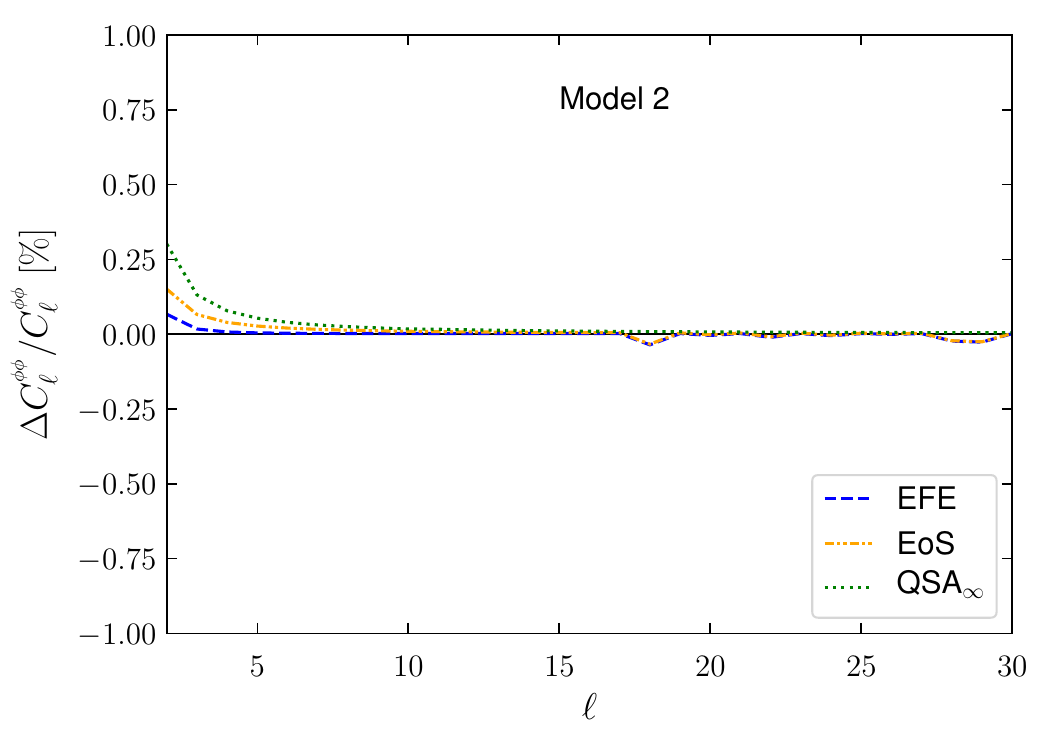}
 \includegraphics[scale=0.2865]{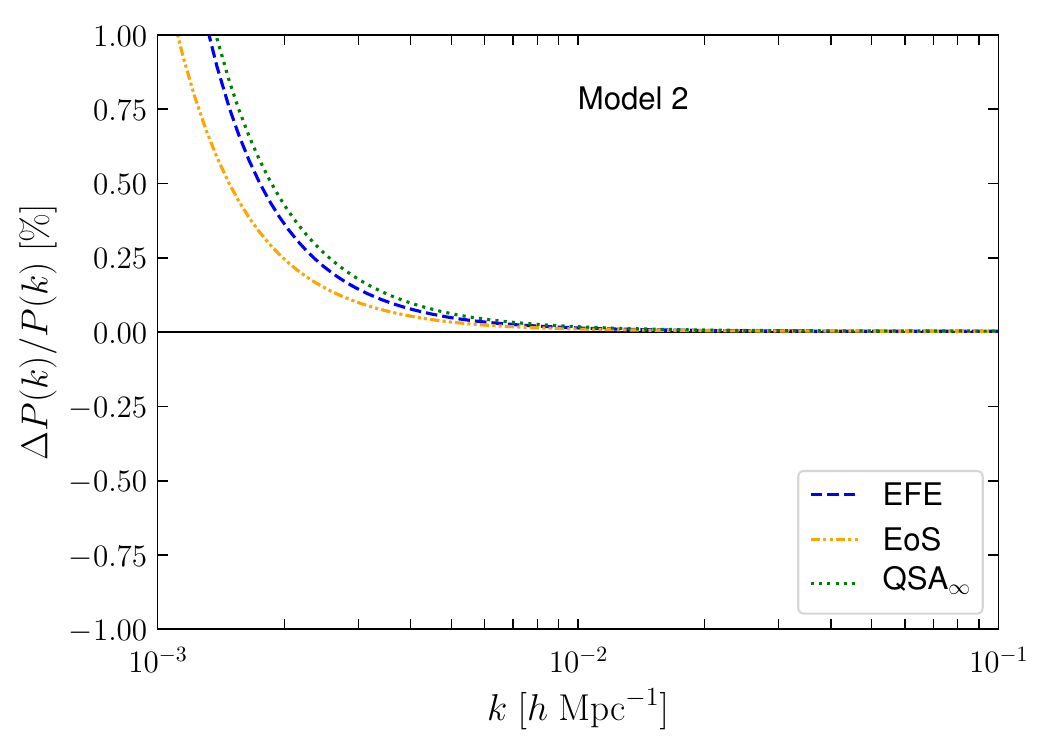}
 \includegraphics[scale=0.2865]{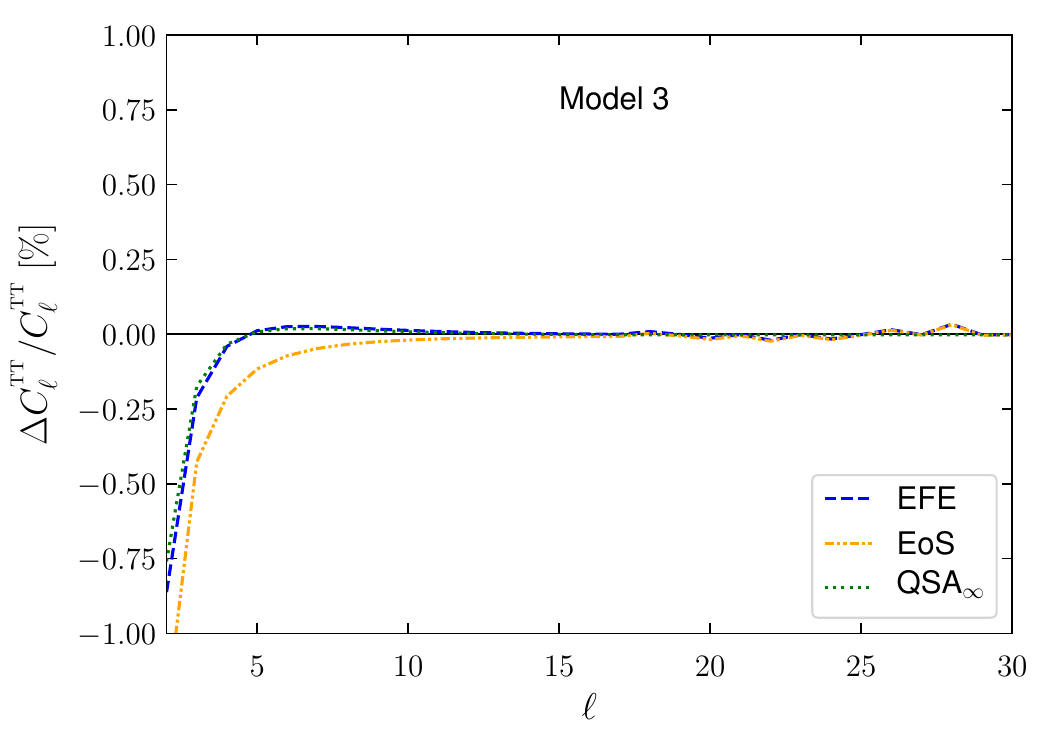}
 \includegraphics[scale=0.2865]{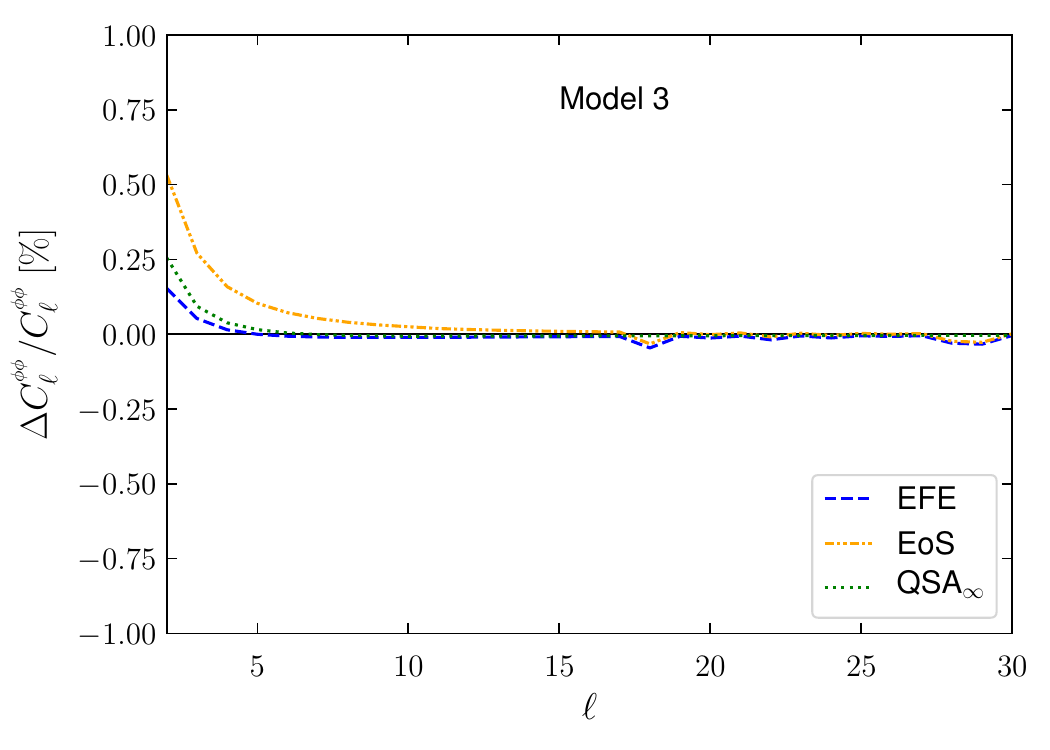}
 \includegraphics[scale=0.2865]{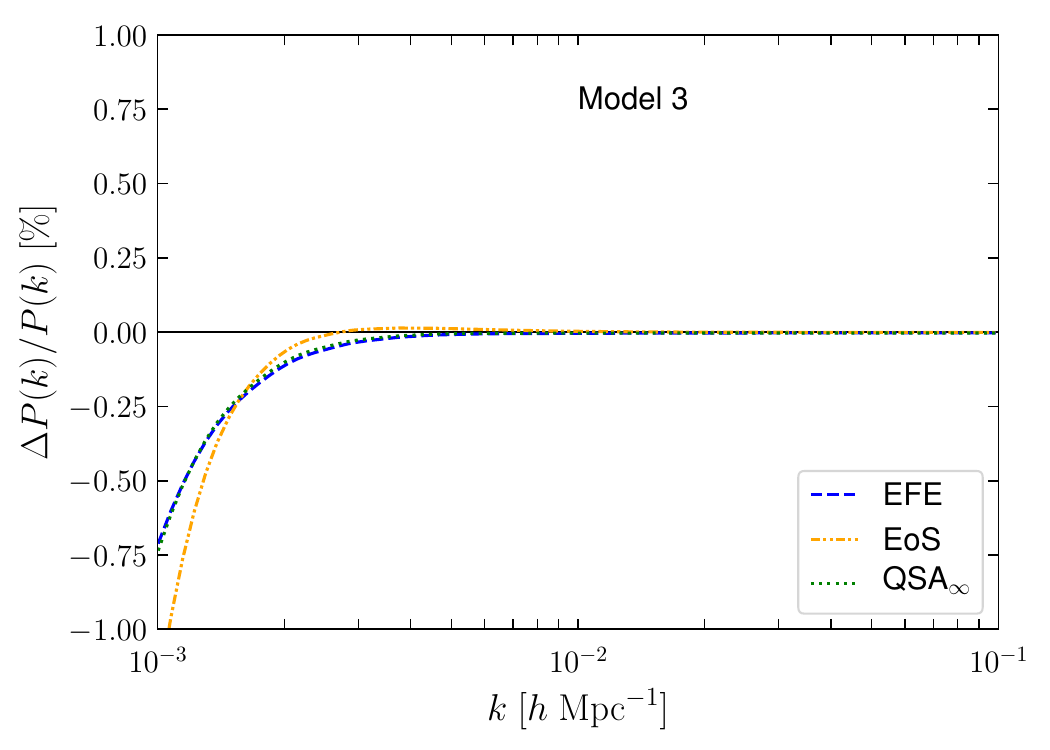}
 \includegraphics[scale=0.2865]{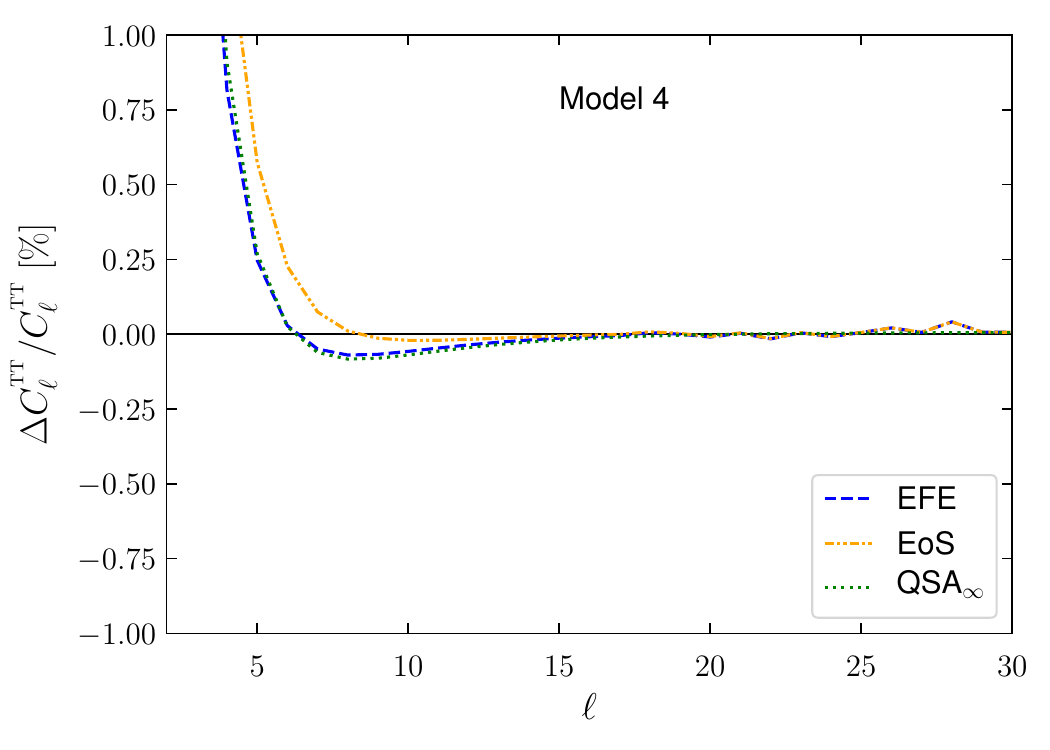}
 \includegraphics[scale=0.2865]{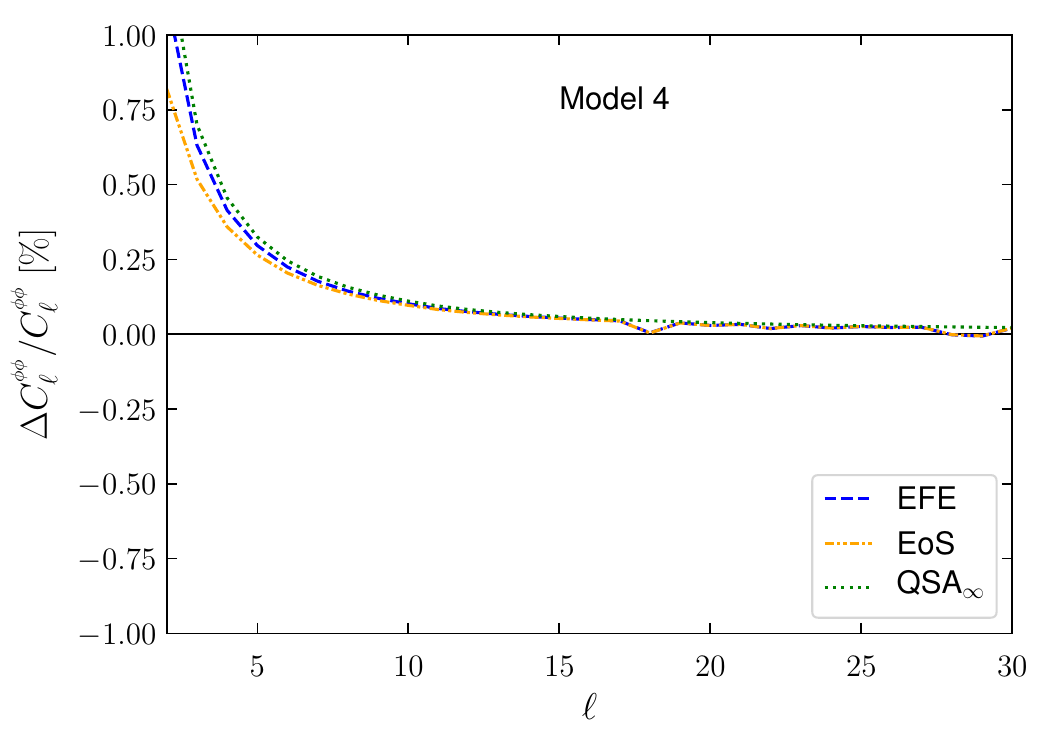}
 \includegraphics[scale=0.2865]{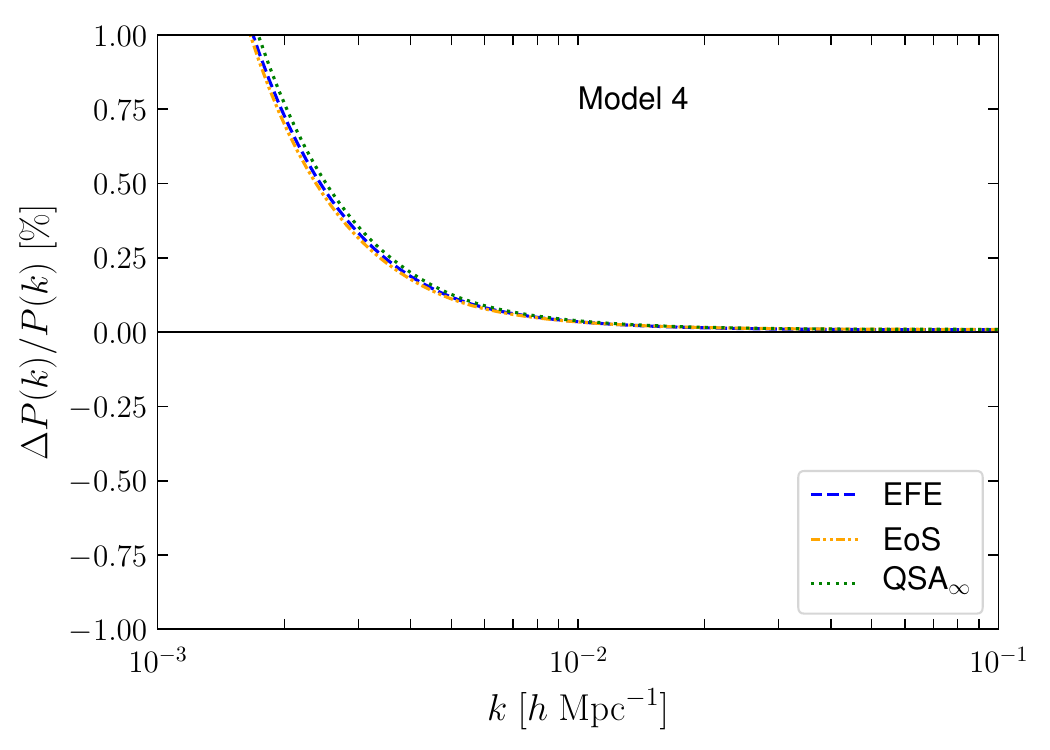}
 \includegraphics[scale=0.2865]{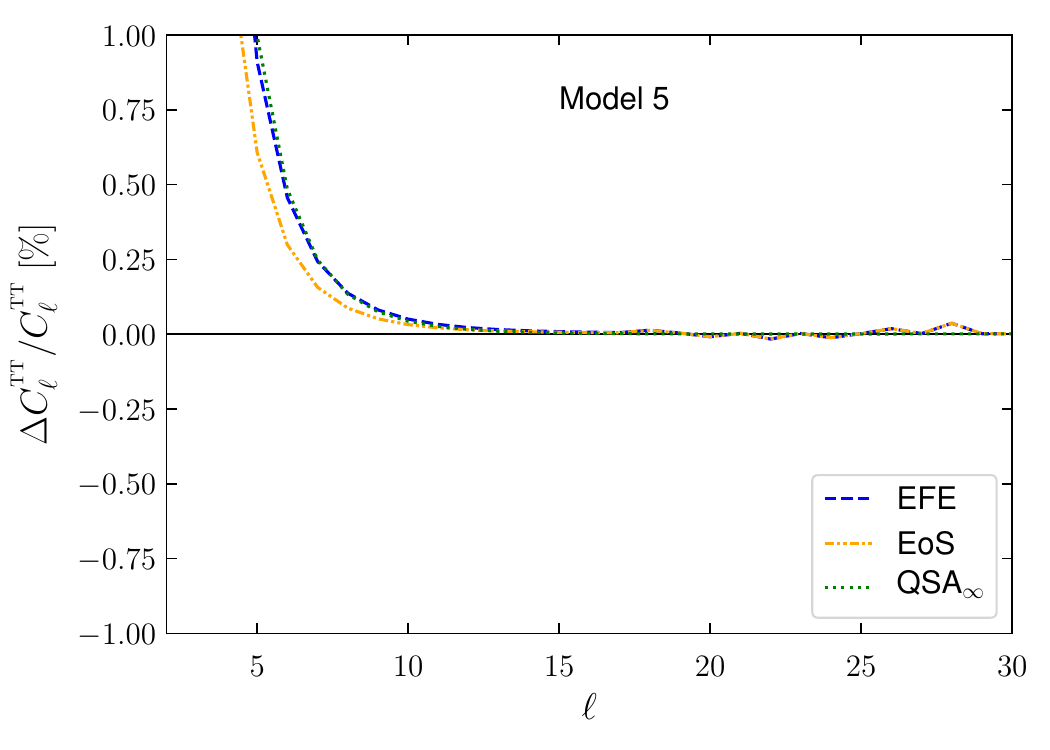}
 \includegraphics[scale=0.2865]{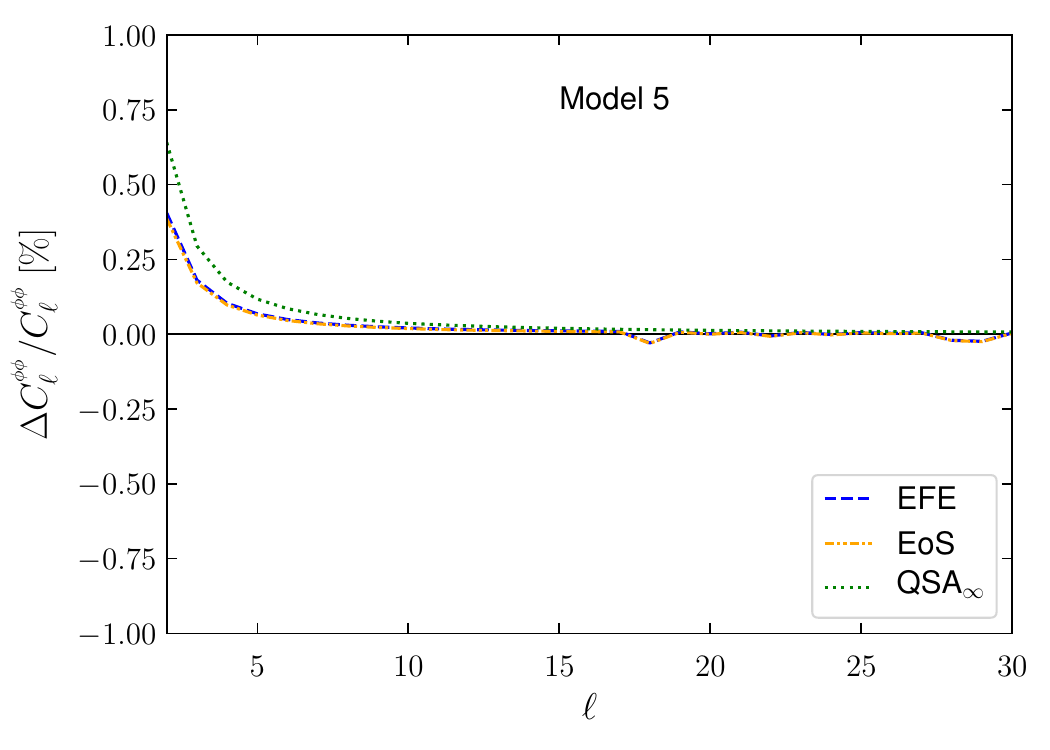}
 \includegraphics[scale=0.2865]{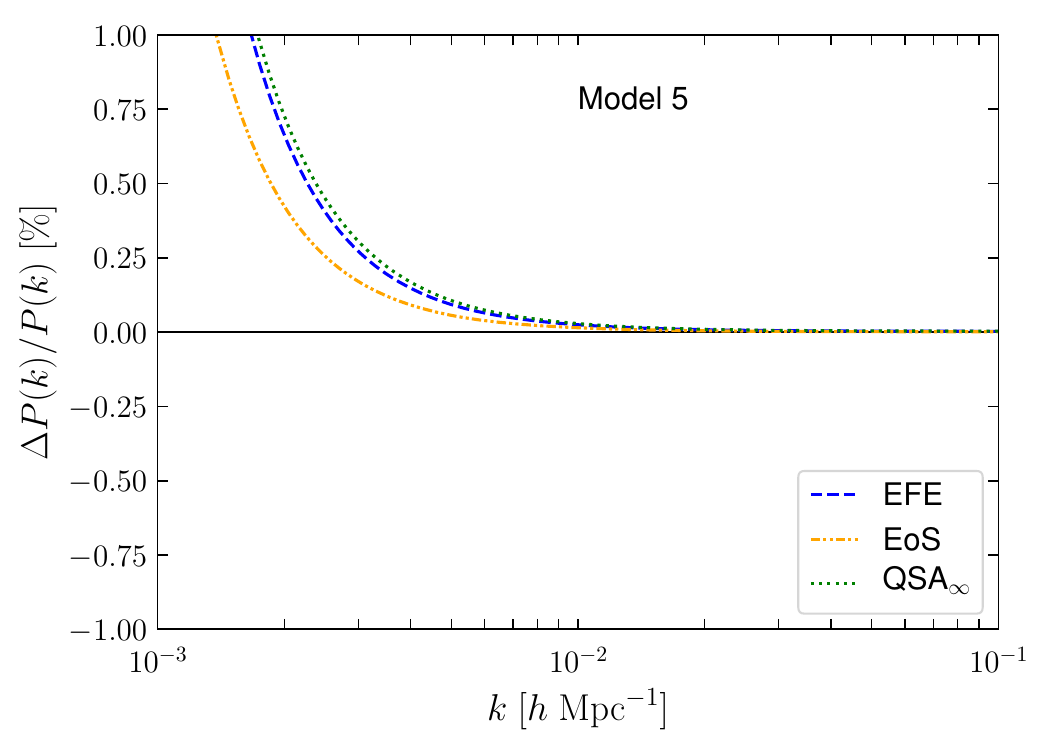}
 \includegraphics[scale=0.2865]{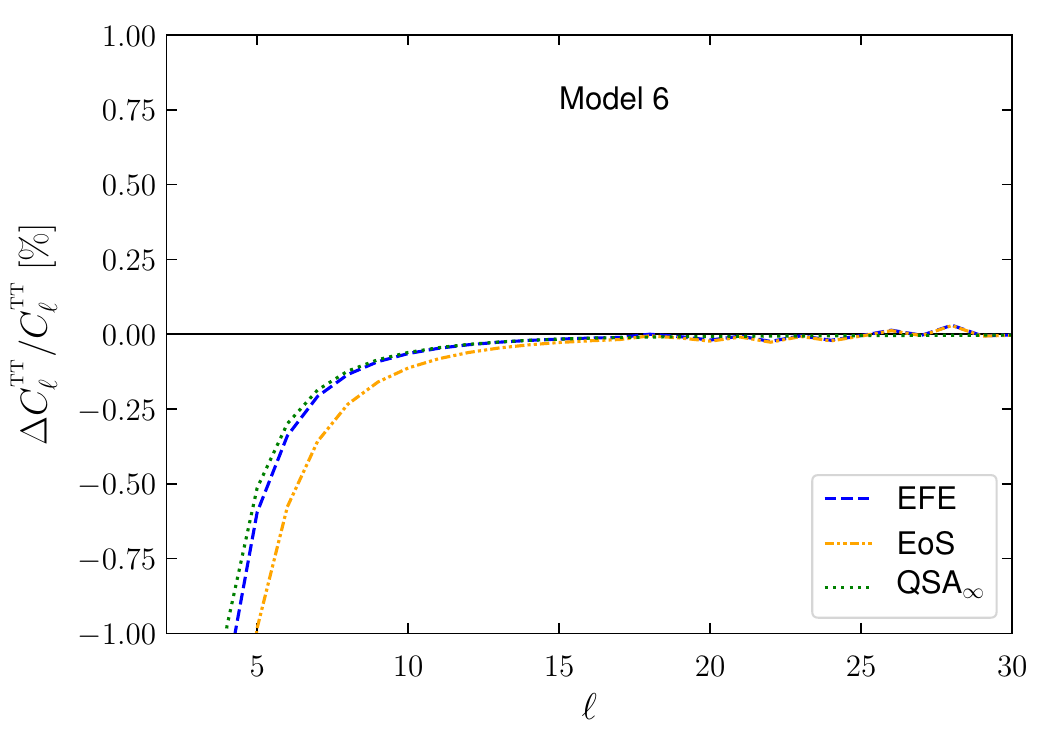}
 \includegraphics[scale=0.2865]{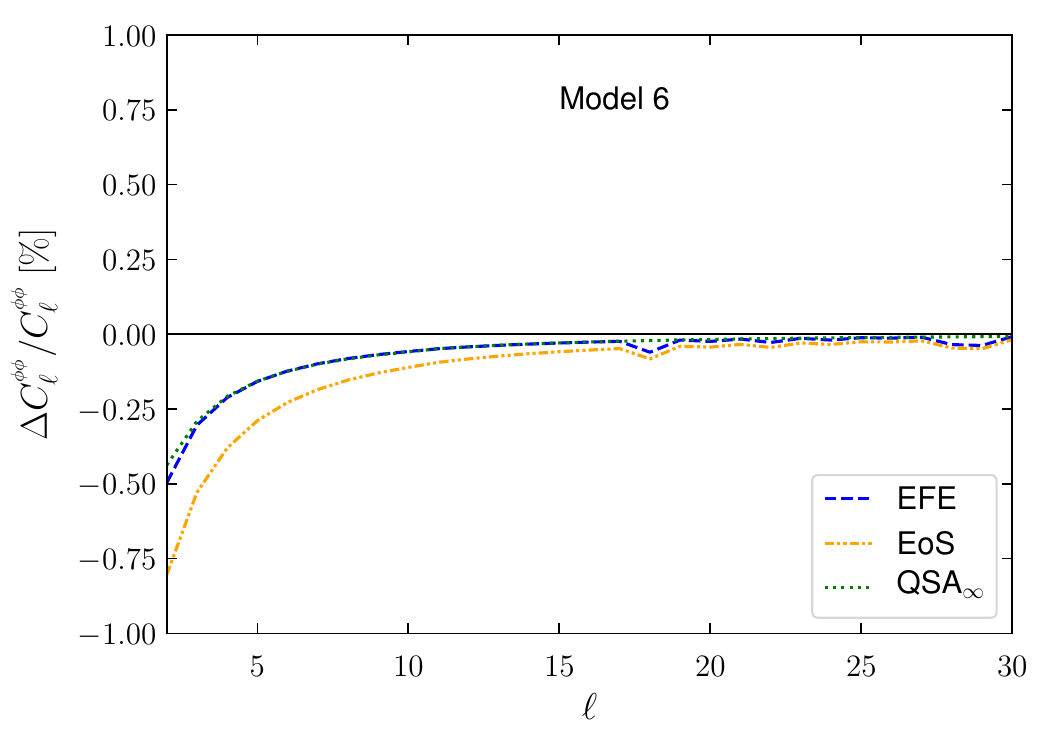}
 \includegraphics[scale=0.2865]{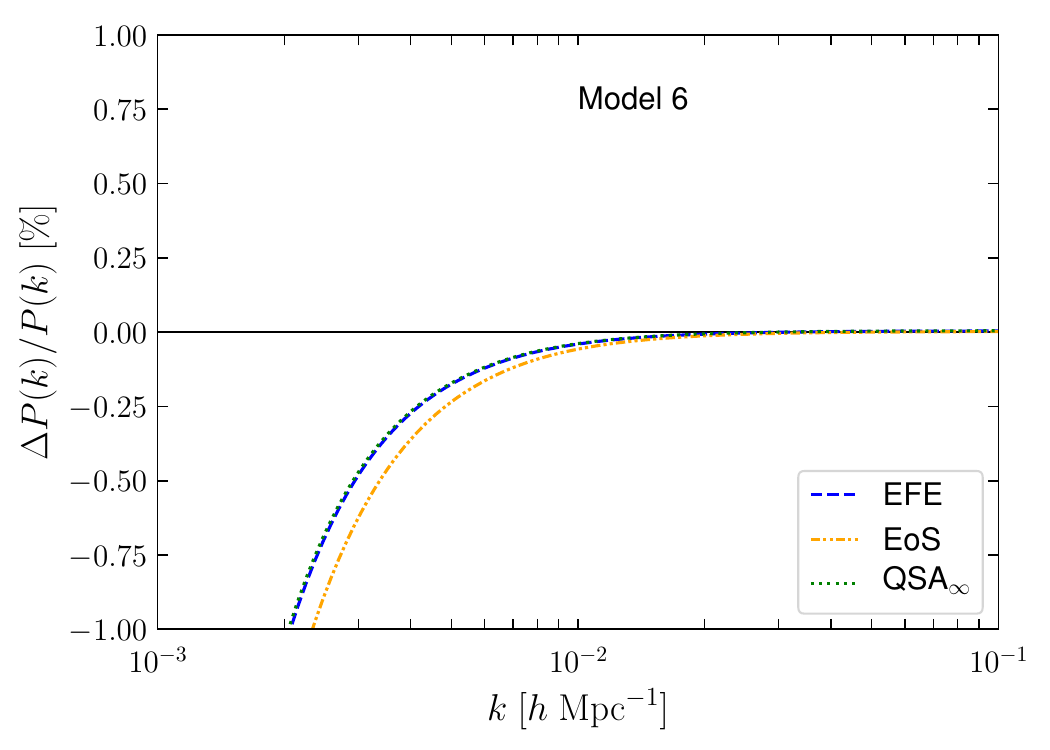}
 \caption[justified]{Relative difference between the approximated (obtained with \texttt{QSA\_class}) and the exact spectra (obtained with \texttt{EoS\_class}). Left panels show the angular temperature anisotropy power spectrum $C_{\ell}^{\rm TT}$, middle panels the angular power spectrum of the lensing potential $C_{\ell}^{\phi\phi}$ and the rightmost panels present the linear matter power spectrum $P(k)$. From top to bottom, we present models 2 to 6. Different QSA schemes are shown with different line styles and colours: blue dashed for EFE, orange dashed-dot-dotted for EoS and green dotted for the approximation $\mu=\mu_{\infty}$ and $\eta=\eta_{\infty}$ on all scales.}
 \label{fig:QSA_spectra}
\end{figure*}

Our results are presented in Figure~\ref{fig:QSA_spectra}, where we show the relative difference $\Delta C/C$, where $C=\{C_{\ell}^{\rm TT}, C_{\ell}^{\phi\phi}, P(k)\}$, $\Delta C=C^{\rm QSA}-C^{\rm EoS\_class}$ and $C=C^{\rm EoS\_class}$ in the denominator. 
For all the models considered, we found an excellent agreement, in general at the subpercent level, on most scales and multipoles. Differences appear only on very large scales, $\ell\lesssim 4-5$ and $k\lesssim2\times10^{-3}\,h\,{\rm Mpc}^{-1}$, where they might exceed the percent level, but are having a little impact, as shown in Figure~\ref{fig:QSA_spectra}. This is because such scales are cosmic-variance-dominated and beyond the range of scales observed by future surveys. For example, Euclid\footnote{\href{https://www.euclid-ec.org/}{https://www.euclid-ec.org/}} \citep{Euclid2018} will be able to probe scales up to $\ell=10$ and $k\approx 6\times 10^{-3}\,h\,{\rm Mpc}^{-1}$ for galaxies up to $z=1.8$ \cite{Euclid2019} using the cosmology of this work. We also verified that the spectra obtained with the expressions in the QSA do not modify the value found for the normalization of the matter power spectrum, as differences of about 0.1\% are more likely due to numerics. Larger deviations between the spectra obtained by solving the full dynamical equations and the approximated ones appear for $C_{\ell}^{\rm TT}$ and $P(k)$, while for $C_{\ell}^{\phi\phi}$ they can be smaller by a factor of a few. The model deviating more in the angular power spectrum of the lensing potential $C_{\ell}^{\phi\phi}$ is model 4, but it differs only by about 1\% at $\ell=2$ from the exact spectrum.

It is also noteworthy that all the recipes studied in this work are virtually indistinguishable from each other, regardless of the particular cosmological model or probe considered. This leads to the strong suggestion that the particular limit on large scales is irrelevant, as are the differences between the two analytical predictions (EFE and EoS) and the exact numerical expectation of the modified gravity parameters. This conclusion is supported by the spectra derived by approximating the modified gravity parameters $\mu$ and $\eta$ with their value on small scales, i.e., $\mu=\mu_{\infty}$ and $\eta=\eta_{\infty}$. As this limit is identical to all the approaches, we do not need to specify any of them in particular. We immediately see that there is no appreciable difference between $\mu=\mu_{\infty}$ and any of the more complete approaches, indicating, once more, that large scales have only weak influence on the observables. This leads to the conclusion that it is possible to use the simplest small-scale limit to recover, with very good accuracy, the full dynamics of the model. This has two advantages: from a theoretical point of view, it is easy to understand the phenomenology of a model knowing the behaviour of $\mu_{\infty}$ and $\eta_{\infty}$ as demonstrated in \cite{Pace2019a}; from a numerical point of view, instead, the implementation is much simpler and less error-prone.

In fact, the numerical implementation of $\mu_{\infty}$ and $\eta_{\infty}$ is much simpler than any of the other recipes. Also, the expressions for MP and EoS are, in general, more complex than for EFE. While we cannot say anything about the performance for MP, there is no appreciable difference in running time for EFE and EoS, but the simple use of the small scale limit can result in a general speedup of the code. In general, though, due to simpler dynamics, \texttt{QSA\_class} is faster than \texttt{EoS\_class}. However, an exact determination of how faster (or slower) a recipe is with respect to the others goes beyond the purpose of this work, but we can safely suggest the use of the small scale limit for a quick determination of the QSA on observables.

An important question one might ask is: why does the QSA appear to work so well for generic subclasses of the Horndeski models? An answer to this question is detailed in \cite{Sawicki2015}, to which we refer to for an in-depth analysis.\footnote{This reference also shows how to consistently extend the QSA to larger scales expanding order by order the solutions to the equations of motion.} In the following, we will briefly outline the procedure, translating their notation into that used in this work. Note that although in \cite{Sawicki2015} only KGB-like models were considered, their conclusions have more general validity, as the relation between the two potentials is given by a constraint equation, rather than by a differential equation. In other words, the QSA is either a good approximation, or not, for \textit{both} potentials.

The starting point is Eq.~(\ref{eqn:MP_Psi}) in the limit ${\rm K}\gg 1$, such that the coefficient of the term $\Psi$ is written as $M_{\pi}^2/H^2+c_{\rm s}^2{\rm K}^2$, where $M_{\pi}^2$ is a mass term. Hence, by construction, the assumption fails for ${\rm K}\simeq 1$. The potential $\Psi$ can be written as $\Psi\equiv\Psi_{\rm QS}+\psi$, where $\psi$ satisfies a homogeneous equation and has an oscillating behaviour, while $\Psi_{\rm QS}$ is the particular solution of the non-homogeneous equation (\ref{eqn:MP_Psi}) and satisfies the QSA approximation. Under the (conservative) assumption that at the sound horizon crossing the amplitude of the oscillations is at most as large as the amplitude of the quasi-static solution, the oscillations of $\psi$ decay fast enough to leave $\Psi_{\rm QS}$. While solving the equations for the two components is difficult, it is though rather straightforward to show that the maximum range of validity of QSA is within the sound-horizon, rather than the cosmological horizon.

The main conclusion, based on the analysis in \cite{Sawicki2015}, is that the QSA is a good approximation up to corrections of the order $\mathcal{O}(1/{\rm K}^2)$ for modes inside the sound horizon. This allows us to infer up to which scales the QSA can be trusted. For models with $c_{\rm s}\sim\mathcal{O}\left(1\right)$ (which is indeed the case for the models studied here as discussed in the previous section), being $H\simeq 3\times 10^{-4}\,h\,{\rm Mpc}^{-1}$ today, the QSA is supposed to work well for $k\gtrsim 3\times10^{-3}\,{\rm Mpc}\,h^{-1}$, in perfect agreement with our numerical results. This scale translates to $\ell\simeq 8$, again in agreement with our findings, hence explaining why the approximate expressions for $\mu$ and $\eta$ reproduce well the observed spectra.

Finally, we can determine, using Eq.~(14) in \cite{Sawicki2015}, whether the QSA can be applied to a survey such as Euclid with an accuracy of 1\%. Whilst in \cite{Sawicki2015} a few surveys are considered, here we limit ourselves to Euclid, which will probe much larger scales. The result is that models with $c_{\rm s}\gtrsim 0.1$ can be safely analysed in the quasi-static limit. This condition is satisfied by all the models analysed in this work.

We can, therefore, conclude that, in general, the QSA is a good approximation for scales below the sound horizon and can be safely applied when the oscillations of the solution of the homogeneous equation for the potential $\Psi$ decay fast enough and their amplitude never exceeds that of the quasi-static solution.

When the conditions discussed above are not satisfied, then we should expect the QSA to be inaccurate. Some additional conditions can be derived for specific models. This has been done, for example, for $f(R)$ models \cite{deLaCruzDombriz2008,Noller2014}, where it was pointed out that in these cosmologies the QSA fails whenever $f_{R,0}\sim\mathcal{O}(1)$ and the background substantially differs from $w_{\rm ds}\simeq -1$.

\section{Conclusions}\label{sect:conclusions}
In this work we studied the effects of the QSA limit on observables such as the angular temperature anisotropy power spectrum $C_{\ell}^{\rm TT}$, the angular power spectrum of the lensing potential $C_{\ell}^{\phi\phi}$ and the linear matter power spectrum $P(k)$ within the framework of the Horndeski models.

The QSA transforms dynamical equations into constraint equations and it is based on the assumption that time derivatives are smaller than spatial derivatives. This means that the metric potentials are slowly varying in time and one can neglect their oscillations 
\citep{Silvestri2013,Noller2014,Sawicki2015}. The QSA introduces a hierarchy in the system: the metric potentials are of the order of the scalar field perturbations and the velocity perturbations while density perturbations are a factor ${\rm K}^2$ larger. For consistency, as discussed in \cite{Silvestri2013}, neglecting time derivatives requires one to consider only scales smaller than the scalar-field sound horizon ($c_{\rm s}^2{\rm K}^2\gg 1$).

We considered three different approaches: the first one is based on the field equations augmented with the equation of motion of the scalar field $\pi=\delta\phi/\dot{\phi}$; the second one is based on the equations describing the evolution of the metric potentials (a dynamical equation and a constraint one); the third is based on the Equation of State approach which interprets the modifications of gravity in terms of a fluid endowed with pressure perturbations and anisotropic stress.

The equations for the metric potentials are obtained by combining the field equations and expressing the perturbations of the scalar field and its derivatives in terms of the potentials and of the matter variables, while the expressions for the Equation of State are obtained by rewriting the metric potentials and the scalar degree of freedom in terms of the fluid variables, matter and dark sector.

For the field equations, the QSA is achieved by neglecting the time derivatives of the metric potentials and the perturbations of the scalar degree of freedom, but retaining the terms proportional to ${\rm K}^2$ and the mass of the scalar field (for models such as $f(R)$, the mass of the scalaron can be of the same order of magnitude of ${\rm K}$). 
For the approach based on the metric potentials, one neglects the time derivatives of the potentials and keeps only the terms proportional to ${\rm K}^2$ and matter density perturbations, while neglecting velocity perturbations as, on sub-horizon scales, they are negligible with respect to density perturbations. Finally, for the EoS approach, one derives a growth-factor-like equation for the dark sector density perturbations, and neglecting its time derivatives, one can establish a relation between dark sector and matter density perturbations.

The derivation of the modified gravity parameters $\mu$ and $\eta$ in the QSA in these different approaches is given in Section~\ref{sect:QSA} and details are provided, respectively for the three approaches, in Appendixes~\ref{sect:coefficientsEFE}, \ref{sect:coefficientsMP} and \ref{sect:coefficientsEoS}. We compared these approaches in Section~\ref{sect:comparison}. The expressions differ, with respect to each other, in the limit for ${\rm K}\rightarrow 0$, but agree in the small-scale limit ${\rm K}\rightarrow \infty$. A notable exception is represented by $f(R)$ models: all the approaches return the same result which perfectly agrees with the numerical solution.

In Section~\ref{sect:numerical_comparison}, we compared the analytical expressions with the exact numerical solutions and found that while they agree on small scales, they differ from the numerical solution when ${\rm K}\lesssim\mathcal{O}(1)$, as the sub-horizon approximation is violated and time derivatives and velocity perturbations cannot be neglected with respect to other terms. The transition takes place at a scale of $k\approx 10^{-3}h~{\rm Mpc}^{-1}$ (at $z=0$) which corresponds to ${\rm K}$ of order of a few. These differences, which appear only for low multipoles and small values of $k$, are expected exactly because of the break down of the QSA, but one can confidently rely on the small scale limit. Going from the perturbations to the observables, though, large scales have little influence and the small-scale limit captures the whole physics of the model, as shown in Section~\ref{sect:observables}. While on one hand this means that none of the above calculations might indeed be necessary to understand the physics of the model, it is actually a welcome fact that the small scale limit of the modified gravity parameters suffices in recovering the spectra, as the numerical implementation and the physical interpretation are simpler and less error-prone than for the full expressions. These differences appearing on large scales are not problematic from an observational point of view, as even future surveys such as Euclid will not be able to probe these scales.

In Section~\ref{sect:observables} we also discussed the reasons behind the very good performance of the QSA in recovering the exact spectra. As explained in \cite{Sawicki2015}, this is because the scales considered are well inside the sound horizon of the model and because oscillatory solutions for Eq.~(\ref{eqn:MP_Psi}) decay fast enough and have a smaller amplitude than the quasi-static solution. The main result is thus that the QSA is exact up to corrections of the order $1/{\rm K}^2$, which correspond to scales of a few times $10^{-3}\,h\,{\rm Mpc}^{-1}$, in perfect agreement with our numerical findings.

In Section~\ref{sect:numerical_comparison}, we also discussed the problems associated with the expressions derived from the equations of the metric potentials, as we explained a distinctive feature is that they diverge on scales of interest, ${\rm K}\simeq\mathcal{O}(1)$. We explained this for some simple models which allow us to perform the calculations, and linked the problem to the failure of the QSA approximation: since the coefficient of the potential $\Psi$ becomes zero, it is not justified to apply the QSA and neglect time derivatives, as these are the only terms surviving.

We finally comment more in detail upon the differences we see among the three different approaches. The final expressions differ from each other because the QSA is applied differently in the three approaches. However, if we make sure to apply the QSA in \textit{exactly} the same way for all the approaches, i.e., we force, for example, dropping exactly the same terms of the QSA for EFE on the other two approaches, then the expressions of the other approaches turn out to be identical to those of EFE. This can be done by establishing a hierarchy among the coefficients in EFE comparing terms in an equation and establishing which one is dominant and it is a direct consequence of the fact that the starting equations are the same (although written differently) and the same coefficients are now kept (or neglected).

Finally, in the Supplementary data, we provide an extensive comparison between results in this and in previous works, translating the relevant expressions in the notation used here.

\acknowledgments
\noindent RAB and FP acknowledge support from Science and Technology Facilities Council (STFC) grant ST/P000649/1. BB and FP acknowledge financial support from the European Research Council (ERC) Consolidator Grant 725456. EB acknowledges support from the European Research Council Grant No: 693024 and the Beecroft Trust. LL acknowledges the support by a Swiss National Science Foundation (SNSF) Professorship grant (No. 170547). FP thanks Savvas Nesseris and Rub\'en Arjona for useful discussions and for giving access to the material which allowed us to extend their expressions to the case where $\alpha_{\rm T}\neq 0$. The authors thank an anonymous referee for their valuable comments which helped improving the manuscript.

\appendix

\section{Derivation and coefficients from the effective field equations (EFE)}
\label{sect:coefficientsEFE}

The field equations and the equation of motion for the perturbed scalar field read
\begin{subequations}\label{eqn:EFE}
 \begin{align}
  & C^{00}_{\dot{\Psi}}\dot{\Psi} + C^{00}_{\Phi}\Phi + C^{00}_{\Psi}{\rm K}^2\Psi + C^{00}_{\dot{\pi}}\dot{\pi} + 
    \left(C^{00}_{\pi}+C^{00}_{\pi2}{\rm K}^2\right)\pi = 
    -\frac{\delta\rho_{\rm m}}{M^2}\,,\label{eqn:EFE-00}\\
  & C^{0i}_{\dot{\Psi}}\dot{\Psi} + C^{0i}_{\Phi}\Phi + C^{0i}_{\dot{\pi}}\dot{\pi} + C^{0i}_{\pi}\pi = 
    -\frac{q_{\rm m}}{M^2}\,, \label{eqn:EFE-0i}\\
  & C^{ij,{\rm trl}}_{\Phi}\Phi + C^{ij,{\rm trl}}_{\Psi}\Psi + C^{ij,{\rm trl}}_{\pi}\pi = 
    -\frac{\sigma_{\rm m}}{M^2}\,, \label{eqn:EFE-ij-trl}\\
  & C^{ij,{\rm tr}}_{\ddot{\Psi}}\ddot{\Psi} + C^{ij,{\rm tr}}_{\dot{\Psi}}\dot{\Psi} + 
    C^{ij,{\rm tr}}_{\dot{\Phi}}\dot{\Phi} + C^{ij,{\rm tr}}_{\Phi}\Phi + C^{ij,{\rm tr}}_{\ddot{\pi}}\ddot{\pi} + 
    C^{ij,{\rm tr}}_{\dot{\pi}}\dot{\pi} + C^{ij,{\rm tr}}_{\pi}\pi = 
         \frac{1}{M^2}\left(\delta P_{\rm m} - \frac{2}{3}H^2{\rm K}^2\sigma_{\rm m}\right)\,, \label{eqn:EFE-ij-tr}\\
  & C^{\pi}_{\ddot{\pi}}\ddot{\pi} + C^{\pi}_{\dot{\pi}}\dot{\pi} + 
    \left(C^{\pi}_{\pi}+C^{\pi}_{\pi2}{\rm K}^2\right)\pi + C^{\pi}_{\ddot{\Psi}}\ddot{\Psi} + 
    C^{\pi}_{\dot{\Phi}}\dot{\Phi} + C^{\pi}_{\dot{\Psi}}\dot{\Psi} 
    + \left(C^{\pi}_{\Phi} + C^{\pi}_{\Phi2}{\rm K}^2\right)\Phi + C^{\pi}_{\Psi2}{\rm K}^2\Psi = 0 \,, 
    \label{eqn:EoM-pi}
 \end{align}
\end{subequations}
where the $C^{X}_{Y}$ coefficients can be easily read off from Eqs.~(109)--(113) of \cite{Gleyzes2014} and are only a function of time. Note that not all the coefficients are dimensionless and we keep explicit the ${\rm K}^2$ term, to single out the hierarchy for the application of the QSA. The superscripts trl and tr stand for traceless and trace part, 
respectively. These equations represent the $00$, $0i$, traceless and trace part of the $ij$ components of the field equations, respectively, and the last expression is the equation of motion for the perturbed scalar field.

Here we describe in detail the procedure followed to derive the expressions for $\mu$ and $\eta$. We start from Eqs.~(\ref{eqn:EFE}) and combine Eq.~(\ref{eqn:EFE-00}) with Eq.~(\ref{eqn:EFE-0i}) to have on the right hand side the gauge-invariant matter density perturbation $\Delta_{\rm m}$; we solve for $\sigma_{\rm m}$ in Eq.~(\ref{eqn:EFE-ij-trl}) and substitute it in Eq.~(\ref{eqn:EFE-ij-tr}). In the remaining 3 equations [combined Eqs.~(\ref{eqn:EFE-00}) and (\ref{eqn:EFE-0i}), Eqs.~(\ref{eqn:EFE-ij-trl}) and (\ref{eqn:EFE-ij-tr}), Eq.~(\ref{eqn:EoM-pi})], we neglect the time derivatives of the potentials $\Phi$ and $\Psi$ and of the perturbed scalar field $\pi$. We also only consider terms proportional to ${\rm K}^2$, as this is appropriate on sub-horizon modes (${\rm K}\gg 1$). We recall that Eq.~(\ref{eqn:EoM-pi}) represents the equation of motion of the perturbed scalar field, therefore, we also include in our discussion the term $C^{\pi}_{\pi}$, as this represents a mass term which can be, in principle, of the same order of magnitude as ${\rm K}^2$ (this is the case, for example, for $f(R)$ models \cite{DeFelice2010}).

We note that as $\sigma_{\rm m}$ appears with the pre-factor ${\rm K}^2$ in Eq.~(\ref{eqn:EFE-ij-tr}), the only terms contributing after considering the QSA and the sub-horizon limit are those coming from Eq.~(\ref{eqn:EFE-ij-trl}). In other words, the relevant equations which contribute to $\mu$ and $\eta$ are Eqs.~(\ref{eqn:EFE-00})--(\ref{eqn:EFE-ij-trl}).

After applying the QSA to the field equations (\ref{eqn:EFE}), we are left with
\begin{subequations}\label{eqn:QSA_EFE}
 \begin{align}
  C^{00}_{\Psi}{\rm K}^2\Psi + C^{00}_{\pi2}{\rm K}^2\pi = &\, -\frac{\rho_{\rm m}\Delta_{\rm m}}{M^2}\,,\\
  C^{ij,{\rm trl}}_{\Phi}\Phi + C^{ij,{\rm trl}}_{\Psi}\Psi + C^{ij,{\rm trl}}_{\pi}\pi = &\, 0\,,\\
  C^{\pi}_{\Phi2}{\rm K}^2\Phi + C^{\pi}_{\Psi2}{\rm K}^2\Psi + 
  \left(C_{\pi}^{\pi}+C^{\pi}_{\pi2}{\rm K}^2\right)\pi = &\, 0 \,,\label{eqn:QSA_EFE_pi}
 \end{align}
\end{subequations}
and the coefficients read
\begin{subequations}
 \begin{align}
  C^{00}_{\Psi} = &\, 2H^2\,, \quad 
  C^{00}_{\pi2} = -2H^3\alpha_{\rm B}\,, \quad 
  C^{ij,{\rm trl}}_{\Phi} = 1\,, \quad 
  C^{ij,{\rm trl}}_{\Psi} = -(1+\alpha_{\rm T})\,, \\
  C^{ij,{\rm trl}}_{\pi} = &\, (\alpha_{\rm M}-\alpha_{\rm T})H\,, \quad 
  C^{\pi}_{\Phi2} = -2\alpha_{\rm B}H^3\,, \quad 
  C^{\pi}_{\Psi2} = 2(\alpha_{\rm M}-\alpha_{\rm T})H^3\,, \\
  C_{\pi}^{\pi} = &\, 6\left\{\left(\dot{H}+\frac{\rho_{\rm m}+P_{\rm m}}{2M^2}\right)\dot{H} +\dot{H} \alpha_{\rm B} 
                    \left[H^2(3+\alpha_{\rm M})+\dot{H}\right]+
                          H\frac{\partial(\dot{H} \alpha_{\rm B})}{\partial t}\right\}\,,\\
  C^{\pi}_{\pi2} = &\, -2H^2\left\{\dot{H}+\frac{\rho_{\rm m}+ P_{\rm m}}{2M^2} + 
                        H^2\left[\alpha_{\rm B}(1+\alpha_{\rm M})+\alpha_{\rm T}-\alpha_{\rm M}\right] + 
                        \frac{\partial(H\alpha_{\rm B})}{\partial t}\right\}\,.
 \end{align}
\end{subequations}

The system of equations above can be conveniently written in matrix form. Switching to the gauge-invariant notation of \cite{Pace2019a}, it reads
\begin{equation}
 \begin{pmatrix}
   0  & C^{00}_{\Psi}{\rm K}^2 & C^{00}_{\pi2}{\rm K}^2 \\
  C^{ij,{\rm trl}}_{\Phi} & C^{ij,{\rm trl}}_{\Psi} & C^{ij,{\rm trl}}_{\pi} \\
  C^{\pi}_{\Phi2}{\rm K}^2 & C^{\pi}_{\Psi2}{\rm K}^2 & C_{\pi}^{\pi}+C^{\pi}_{\pi2}{\rm K}^2
 \end{pmatrix}
 \begin{pmatrix}
  Y \\
  Z \\
  \pi
 \end{pmatrix}
 = -\frac{\rho_{\rm m}\Delta_{\rm m}}{M^2}
 \begin{pmatrix}
  1 \\
  0 \\
  0
 \end{pmatrix}
 \,.
\end{equation}
It is now simple to derive the expressions for the modified gravity parameters. By denoting with $\mathcal{M}$ the matrix of the coefficients, the solution to the previous equation is
\begin{equation}
 \begin{pmatrix}
  Y \\
  Z \\
  \pi
 \end{pmatrix}
 = -
 \begin{pmatrix}
  \left[\mathcal{M}^{-1}\right]_{11} \\
  \left[\mathcal{M}^{-1}\right]_{12} \\
  \left[\mathcal{M}^{-1}\right]_{13}
 \end{pmatrix}
 \frac{\rho_{\rm m}\Delta_{\rm m}}{M^2}\,,
\end{equation}
where
\begin{align}
 \left[\mathcal{M}^{-1}\right]_{11} = & 
      \frac{-C^{ij,{\rm trl}}_{\Psi}C_{\pi}^{\pi} + 
            \left(C^{ij,{\rm trl}}_{\pi} C^{\pi}_{\Psi2}-C^{ij,{\rm trl}}_{\Psi}C^{\pi}_{\pi2}\right){\rm K}^2}
           {C^{00}_{\Psi}C^{ij,{\rm trl}}_{\Phi}C_{\pi}^{\pi}{\rm K}^2 + 
            \left[C^{00}_{\Psi}
                  \left(C^{ij,{\rm trl}}_{\Phi}C^{\pi}_{\pi2} - C^{ij,{\rm trl}}_{\pi}C^{\pi}_{\Phi2}\right) + 
                  C^{00}_{\pi2}\left(C^{ij,{\rm trl}}_{\Psi}C^{\pi}_{\Phi2} - 
                                     C^{ij,{\rm trl}}_{\Phi}C^{\pi}_{\Psi2}\right)\right]{\rm K}^4}\,,\\
 \left[\mathcal{M}^{-1}\right]_{12} = & 
      \frac{C^{ij,{\rm trl}}_{\Phi}C_{\pi}^{\pi}+(C^{ij,{\rm trl}}_{\Phi}C^{\pi}_{\pi2}-C^{ij,{\rm trl}}_{\pi}
            C^{\pi}_{\Phi2}){\rm K}^2}
           {C^{00}_{\Psi}C^{ij,{\rm trl}}_{\Phi}C_{\pi}^{\pi}{\rm K}^2 + 
            \left[C^{00}_{\Psi}
                  \left(C^{ij,{\rm trl}}_{\Phi}C^{\pi}_{\pi2} - C^{ij,{\rm trl}}_{\pi}C^{\pi}_{\Phi2}\right) + 
                  C^{00}_{\pi2}\left(C^{ij,{\rm trl}}_{\Psi}C^{\pi}_{\Phi2} - 
                                     C^{ij,{\rm trl}}_{\Phi}C^{\pi}_{\Psi2}\right)\right]{\rm K}^4}\,,\\
 \left[\mathcal{M}^{-1}\right]_{13} = & 
      \frac{C^{ij,{\rm trl}}_{\Psi}C^{\pi}_{\Phi2}-C^{ij,{\rm trl}}_{\Phi}C^{\pi}_{\Psi2}}
           {C^{00}_{\Psi}C^{ij,{\rm trl}}_{\Phi}C_{\pi}^{\pi}{\rm K}^2 + 
            \left[C^{00}_{\Psi}
                  \left(C^{ij,{\rm trl}}_{\Phi}C^{\pi}_{\pi2} - C^{ij,{\rm trl}}_{\pi}C^{\pi}_{\Phi2}\right) + 
                  C^{00}_{\pi2}\left(C^{ij,{\rm trl}}_{\Psi}C^{\pi}_{\Phi2} - 
                                     C^{ij,{\rm trl}}_{\Phi}C^{\pi}_{\Psi2}\right)\right]{\rm K}^4}\,.
\end{align}

We can immediately see that $\mu_Z=2\left[\mathcal{M}^{-1}\right]_{12}H^2{\rm K}^2/\bar{M}^2$ and $\mu=2\left[\mathcal{M}^{-1}\right]_{11}H^2{\rm K}^2/\bar{M}^2$, where $\bar{M}=M/M_{\rm pl}$. Knowing $\mu_Z$ and $\mu$, the slip is $\eta=\mu_Z/\mu$ and reads
\begin{equation}
 \eta = \frac{C^{ij,{\rm trl}}_{\Phi}C_{\pi}^{\pi}+(C^{ij,{\rm trl}}_{\Phi}C^{\pi}_{\pi2}-C^{ij,{\rm trl}}_{\pi}
              C^{\pi}_{\Phi2}){\rm K}^2}
             {-C^{ij,{\rm trl}}_{\Psi}C_{\pi}^{\pi} + 
              \left(C^{ij,{\rm trl}}_{\pi} C^{\pi}_{\Psi2}-C^{ij,{\rm trl}}_{\Psi}C^{\pi}_{\pi2}\right){\rm K}^2}\,.
\end{equation}

\section{Derivation and coefficients from the metric potentials (MP)}
\label{sect:coefficientsMP}
It is instructive to consider the derivation of the equations for the metric potentials for a system made up of baryons, cold dark matter, photons and neutrinos, which we collectively call matter ({\rm m}), and the cosmological constant $\Lambda$. This will be a warm-up for the substantially more involved derivation of the full equations taking into account the scalar field contribution.

The field equations can be rewritten generically as (following the notation of \cite{Gleyzes2014})
\begin{subequations}\label{eqn:EFE_standard}
 \begin{align}
  \frac{k^2}{a^2}\Psi + 3H\left(\dot{\Psi} + H\Phi\right) = &\, -\frac{1}{2M_{\rm pl}^2}\delta\rho_{\rm m}\,, \label{eqn:00}\\
  \dot{\Psi} + H\Phi = &\, -\frac{1}{2M_{\rm pl}^2}q_{\rm m}\,, \label{eqn:0i}\\
  \Phi - \Psi = &\, -\frac{1}{M_{\rm pl}^2}\sigma_{\rm m}\,, \label{eqn:ij-trl}\\\
  \ddot{\Psi} + H\dot{\Phi} + 2\dot{H}\Phi + 3H\left(\dot{\Psi} + H\Phi\right) = &\, 
     \frac{1}{2M_{\rm pl}^2}\left(\delta P_{\rm m} - \frac{2}{3}\frac{k^2}{a^2}\sigma_{\rm m}\right)\,, 
     \label{eqn:ij-tr}
 \end{align}
\end{subequations}
where the constraint equation is simply given by the traceless part of the $ij$-components, Eq.~(\ref{eqn:ij-trl}). 
The equation describing the evolution of $\Psi$ is obtained by expressing $\Phi$ in terms of $\Psi$ and $\sigma_{\rm m}$ using Eq.~(\ref{eqn:ij-trl}) and $\dot{\Phi}$ by taking the time derivative of Eq.~(\ref{eqn:ij-trl}). This leads to
\begin{align}
 \ddot{\Psi} + 4H\dot{\Psi} + \left(3H^2+2\dot{H}\right)\Psi = &\, \frac{1}{2M_{\rm pl}^2}\left[
       \delta P_{\rm m} + 2\left(3H^2+2\dot{H}\right)\sigma_{\rm m} - 
       \frac{2}{3}H^2{\rm K}^2\sigma_{\rm m} + 2H\dot{\sigma}_{\rm m}\right]\,, \label{eqn:dyn_pot}\\
 \Psi - \Phi = &\, \frac{\sigma_{\rm m}}{M_{\rm pl}^2}\,,\label{eqn:constraint}
\end{align}
which agree with expressions (132) and (133) of \cite{Gleyzes2014} in the absence of the scalar field perturbations.

Assuming $\sigma_{\rm m}\approx 0$ (as it is the case for cold dark matter), 
$\delta P_{\rm m}=c_{\rm s}^2\delta\rho_{\rm m}$, and further considering Eq.~(\ref{eqn:00}) to express $\delta\rho_{\rm m}$ in terms of $\Psi$, in real space Eq.~(\ref{eqn:dyn_pot}) simply reads
\begin{equation}
 \ddot{\Psi} + \left(4+3c_{\rm s}^2\right)H\dot{\Psi} + 
 \left[3H^2\left(1+c_{\rm s}^2\right)+2\dot{H}\right]\Psi - c_{\rm s}^2\nabla^2\Psi = 0\,,
\end{equation}
which is equivalent to Eq.~(5.22) of \cite{Mukhanov1992} and Eq.~(5.30) of \cite{Bardeen1980} when $\Phi=\Psi$ and we consider flat spatial geometry.

Let us now consider again Eqs.~(\ref{eqn:EFE}). A similar procedure to the one outlined before can be applied to derive an equation for $\Psi$ in terms of the matter variables and a second one which expresses $\Phi$ in terms of $\Psi$, its first time derivative and the matter variables. To derive the equation for $\Psi$, one combines 
Eqs.~(\ref{eqn:EFE-00})--(\ref{eqn:EFE-ij-trl}) and solve for $\pi$, $\dot{\pi}$ and $\Phi$ in terms of $\Psi$, $\dot{\Psi}$ and the matter variables. To solve for $\dot{\Phi}$ one uses the time derivative of Eq.~(\ref{eqn:EFE-ij-trl}) and $\ddot{\pi}$ is expressed in terms of all the other quantities using the equation of motion of the scalar field (\ref{eqn:EoM-pi}). In this way, the scalar field fluctuation, $\pi$, and its time derivatives are expressed in terms of the potentials and the matter variables and the equation for $\Psi$ is derived. 
The constraint equation is then obtained by replacing $\pi$ and $\dot{\pi}$ in Eq.~(\ref{eqn:EFE-ij-tr}).

The two equations describing the evolution of the potentials read \citep{Bellini2014,Gleyzes2014}
\begin{subequations}\label{eqn:MP}
 \begin{align}
  & \ddot{\Psi} + C_{\dot{\Psi}}H\dot{\Psi} + C_{\Psi}H^2\Psi = -\frac{1}{2M^2}\left[
      C_{\delta\rho_{\rm m}}\delta\rho_{\rm m} + C_{q_{\rm m}}Hq_{\rm m} + C_{\sigma_{\rm m}}H^2\sigma_{\rm m} + 
      \frac{\alpha_{\rm K}}{\alpha}\delta P_{\rm m} - 2H\dot{\sigma}_{\rm m}\right]\,, \label{eqn:MP_Psi}\\
  & \alpha_{\rm B}^2{\rm K}^2\left[\Phi - \left(1+\alpha_{\rm T}+\frac{2\gamma_9}{\alpha\alpha_{\rm B}}\right)\Psi + 
      \frac{\sigma_{\rm m}}{M^2}\right] + 
    \beta_1\left[\Phi - \Psi(1+\alpha_{\rm T})\frac{\gamma_1}{\beta_1} + \frac{\sigma_{\rm m}}{M^2}\right] = 
    \nonumber\\
  & \qquad \frac{\gamma_9}{H^2M^2}\left[\frac{\alpha_{\rm B}}{\alpha}\left(\delta\rho_{\rm m}-3Hq_{\rm m}\right) + HM^2\dot{\Psi} + \frac{\alpha_{\rm K}}{2\alpha}Hq_{\rm m} - H^2\sigma_{\rm m}
  \right] \,. \label{eqn:MP_Phi}
 \end{align}
\end{subequations}
The $C_X$ coefficients depend on both time and space and can be easily read off from Eq.~(132) in \cite{Gleyzes2014}.

Applying a QSA as described in Section~\ref{sect:MP}, the two equations above simplify to
\begin{subequations}\label{eqn:QSA_MP}
 \begin{align}
  C_{\Psi}H^2Z = &\, -\frac{1}{2M^2} C_{\delta\rho_{\rm m}}\rho_{\rm m}\Delta_{\rm m}\,, 
  \label{eqn:MP_Psi_QSA}\\
  \alpha_{\rm B}^2{\rm K}^2\left[Y - \left(1+\alpha_{\rm T}+\frac{2\gamma_9}{\alpha\alpha_{\rm B}}\right)Z\right] = 
    &\,
  \frac{\gamma_9}{H^2M^2}\frac{\alpha_{\rm B}}{\alpha}\rho_{\rm m}\Delta_{\rm m} \,.\label{eqn:MP_Phi_QSA}
 \end{align}
\end{subequations}
The coefficients $C_X$ relevant for the QSA are
\begin{equation}
 C_{\Psi} =  \frac{\beta_1\beta_4+\beta_1\beta_5{\rm K}^2 + c_{\rm s}^2\alpha_{\rm B}^2{\rm K}^4}
                  {\beta_1+\alpha_{\rm B}^2{\rm K}^2}\,, \quad 
 C_{\delta\rho_{\rm m}} = \frac{\beta_1\beta_6+\beta_7\alpha_{\rm B}^2{\rm K}^2}
                               {\beta_1+\alpha_{\rm B}^2{\rm K}^2}\,,
\end{equation}
where the coefficients $\beta_i$ were previously introduced in \cite{Bellini2014,Gleyzes2014} and read
\begin{align*}
 \beta_1 & \equiv -\alpha_{\rm K}\frac{\rho_{\rm m}+P_{\rm m}}{4H^2M^2} - \frac{1}{2}\alpha
                    \left(\frac{\dot H}{H^2} + \alpha_{\rm T} - \alpha_{\rm M} \right) \,, \quad 
 \beta_2 \equiv 2(2+\alpha_{\rm M}) + 3 \Upsilon  \,, \\
 \beta_3 & \equiv 3+\alpha_{\rm M} + \frac{\alpha_{\rm B}^2}{H\alpha}\left(
                  \frac{\alpha_{\rm K}}{\alpha_{\rm B}^2}\right)^{\hbox{$\cdot$}}\,, \quad 
 \beta_4 \equiv (1+\alpha_{\rm T})\left[2\frac{\dot{H}}{H^2} + 3(1+\Upsilon) + \alpha_{\rm M} \right] + 
                \frac{\dot{\alpha}_{\rm T}}{H} \,, \\
 \beta_5 & \equiv c_{\rm s}^2 - \frac{2\alpha_{\rm B} (\beta_3-\beta_2)}{\alpha}
                  +\frac{\alpha_{\rm B}^2}{\beta_1}(1+\alpha_{\rm T})(\beta_3-\beta_2) + 
                   \frac{\alpha_{\rm B}^2\beta_4}{\beta_1}\,,\\
 \beta_6 & \equiv \beta_7 - 2 \frac{\alpha_{\rm B}(\beta_3 -\beta_2)}{\alpha} \,, \quad 
 \beta_7 \equiv c_{\rm s}^2 + 2 \frac{\alpha_{\rm B}^2(1+\alpha_{\rm T}) + 
                \alpha_{\rm B}(\alpha_{\rm T} - \alpha_{\rm M})}{\alpha} \,,
\end{align*}
with
\begin{equation}
 \begin{split}
  12\beta_1 H^3 M^2 \Upsilon \equiv &\, 2\alpha M^2 \left\{\left[\dot{H} + (\alpha_{\rm T} - \alpha_{\rm M}) H^2 
          \right]^{\hbox{$\cdot$}} + (3+\alpha_{\rm M})H\left[\dot{H} + (\alpha_{\rm T} - \alpha_{\rm M})
          H^2\right]\right\} \\
  & +\alpha_{\rm K}{\dot{P}_{\rm m}} - ({\rho_{\rm m} + P_{\rm m}})H(\alpha_{\rm K}-6\alpha_{\rm B})
     (\alpha_{\rm T} - \alpha_{\rm M}) + 6 ({\rho_{\rm m} + P_{\rm m}})\frac{\alpha_{\rm B}^4}{\alpha}
     \left(\frac{\alpha_{\rm K}}{\alpha_{\rm B}^2}\right)^{\hbox{$\cdot$}}\,,
 \end{split}
\end{equation}
and $\gamma_9=\alpha(\alpha_{\rm T}-\alpha_{\rm M})/2$. 
Note that for $\alpha_{\rm B}=0$, $\alpha_{\rm B}\beta_3=0$ and for models where only $\alpha_{\rm K}\neq 0$ (quintessence and $k$-essence) one has $\Upsilon=\mathrm{d}P_{\rm ds}/\mathrm{d}\rho_{\rm ds}=c_{\rm a,ds}^2$, i.e., the adiabatic sound speed for the dark sector component.

\section{Derivation and coefficients from the Equation of State approach (EoS)}
\label{sect:coefficientsEoS}
In this section we describe in detail the derivation of the attractor solution,  Eq.~(\ref{eqn:AIC}), linking dark sector ($\Delta_{\rm ds}$) and dark matter ($\Delta_{\rm m}$) perturbations, and provide explicit expressions for the coefficients necessary to evaluate it.

To derive the equations of state $w_{\rm ds}\Gamma_{\rm ds}$ and $w_{\rm ds}\Pi_{\rm ds}$, one starts from the field equations and expresses the scalar degree of freedom and its derivatives in terms of the fluid variables. 
More in detail, following \cite{Gleyzes2014}, we consider Eqs.~(\ref{eqn:EFE}) and their compact form in terms of matter and dark energy variables
\begin{subequations}\label{eqn:EFE_M_DS}
 \begin{align}
  \frac{k^2}{a^2}\Psi + 3H\left(\dot{\Psi} + H\Phi\right) = &\, 
  -\frac{1}{2M^2}\left(\delta\rho_{\rm m}+\delta\rho_{\rm de}^{\rm GLV}\right)\,, \label{eqn:00_M_DS}\\
  \dot{\Psi} + H\Phi = &\, -\frac{1}{2M^2}\left(q_{\rm m}+q_{\rm de}^{\rm GLV}\right)\,, \label{eqn:0i_M_DS}\\
  \Psi - \Phi = &\, \frac{1}{M^2}\left(\sigma_{\rm m}+\sigma_{\rm de}^{\rm GLV}\right)\,, \label{eqn:ij-trl_M_DS}\\\
  \ddot{\Psi} + H\dot{\Phi} + 2\dot{H}\Phi + 3H\left(\dot{\Psi} + H\Phi\right) = &\, 
  \frac{1}{2M^2}\left[\delta P_{\rm m} + \delta P_{\rm de}^{\rm GLV} - \frac{2}{3}\frac{k^2}{a^2}\left(\sigma_{\rm m} + \sigma_{\rm de}^{\rm GLV}\right)\right]\,. \label{eqn:ij-tr_M_DS}
 \end{align}
\end{subequations}

In Eqs.~(\ref{eqn:EFE_M_DS}), $\delta\rho_{\rm de}^{\rm GLV}$, $q_{\rm de}^{\rm GLV}$, $\sigma_{\rm de}^{\rm GLV}$ and $\delta P_{\rm de}^{\rm GLV}$ are the dark energy fluid variables which represent, collectively, the modifications to Einstein field equations induced by the modifications of gravity. Their expressions are given in Eqs.~(147)--(150) of \cite{Gleyzes2014}. Note that at the background level, the dark energy and the pressure component satisfy a non-standard continuity equation (see their Eqs.~(116) and (117)).

We then solve Eqs.~(\ref{eqn:EFE-00})--(\ref{eqn:EFE-ij-trl}) for $\Psi$, $\dot{\Psi}$ and $\dot{\pi}$ and plug these solutions in Eqs.~(\ref{eqn:00_M_DS}) and (\ref{eqn:0i_M_DS}) so that $\pi$ and $\Phi$ can be expressed in terms of the density ($\delta\rho$) and velocity perturbations ($q$) and matter anisotropic stress $\sigma_{\rm m}$. From the time derivative of Eq.~(\ref{eqn:EFE-ij-trl}) we derive an expression for $\dot{\Phi}$. Eqs.~(\ref{eqn:EFE-ij-tr}) and (\ref{eqn:EoM-pi}) are finally used to infer $\ddot{\Psi}$ and $\ddot{\pi}$. 
Combining all these expressions together allows to express $\delta P_{\rm de}^{\rm GLV}$ and $\sigma_{\rm de}^{\rm GLV}$ in terms of the other fluid variables.

At this point, we can express the equations of state in terms of gauge-invariant quantities using the following relations linking the variables used here and those in \cite{Gleyzes2014} ({\rm GLV}):
\begin{align*}
 \delta\rho_{\rm de}^{\rm GLV} & = \bar{M}^2\delta\rho_{\rm ds} + 
                                   \left(\bar{M}^2-1\right)\delta\rho_{\rm m}\,, \\
 \delta P_{\rm de}^{\rm GLV} & = \bar{M}^2\delta P_{\rm ds} + 
                                 \left(\bar{M}^2-1\right)\delta P_{\rm m}\,, \nonumber\\
 q_{\rm m}^{\rm GLV}+q_{\rm de}^{\rm GLV} & = -\bar{M}^2
                                               \frac{\rho_{\rm ds}\Theta_{\rm ds} + \rho_{\rm m}\Theta_{\rm m}}
                                                    {3H}\,, \\
 q_{\rm de}^{\rm GLV} & = -\frac{1}{3H}\left[\bar{M}^2
                           \rho_{\rm ds}\Theta_{\rm ds}+
                           \left(\bar{M}^2-1\right)\rho_{\rm m}\Theta_{\rm m}
                           \right]\,, \\
 \sigma_{\rm m}^{\rm GLV}+\sigma_{\rm de}^{\rm GLV} & = -\frac{a^2}{k^2}\bar{M}^2
                                                         (P_{\rm ds}\Pi_{\rm ds} + P_{\rm m}\Pi_{\rm m})\,, \\
 \sigma_{\rm de}^{\rm GLV} & = -\frac{a^2}{k^2}
                                \left[\bar{M}^2 P_{\rm ds}\Pi_{\rm ds}
                                      + \left(\bar{M}^2-1\right)P_{\rm m}\Pi_{\rm m}\right]\,,
\end{align*}
while at the background level we have
\begin{equation*}
 \rho_{\rm de}^{\rm GLV} = \rho_{\rm ds} + 3\left(M^2-M_{\rm pl}^2\right)H^2\,, \quad 
 P_{\rm de}^{\rm GLV} = P_{\rm ds} - \left(3H^2+2\dot{H}\right)\left(M^2-M_{\rm pl}^2\right)\,.
\end{equation*}
Note that entropy perturbations can be related to pressure perturbations via the relation 
$w\Gamma = \delta P/\rho - c_{\rm a}^2(\Delta - \Theta)$.

Dark sector variables satisfy the following continuity and Euler equations, respectively,
\begin{subequations}\label{eqn:EoM}
 \begin{align}
  \Delta_{\rm ds}^{\prime} - 3w_{\rm ds}\Delta_{\rm ds} - 2w_{\rm ds}\Pi_{\rm ds} + 
    g_{\rm K}\epsilon_{H}\Theta_{\rm ds} = &\, 3(1+w_{\rm ds})X\,, \label{eqn:deltadot}\\
  \Theta_{\rm ds}^{\prime} + 3\left(c_{\rm a,ds}^2-w_{\rm ds}+\frac{1}{3}\epsilon_{H}\right)\Theta_{\rm ds} - 
     3c_{\rm a,ds}^2\Delta_{\rm ds} - 2w_{\rm ds}\Pi_{\rm ds} - 3w_{\rm ds}\Gamma_{\rm ds} = &\, 3(1+w_{\rm ds})Y\,, 
     \label{eqn:Thetadot}
 \end{align}
\end{subequations}
where the prime $^{\prime}$ represents the derivative with respect to $\ln{a}$, $\epsilon_{H}=-H^{\prime}/H$, $g_{\rm K}=1+{\rm K}^2/(3\epsilon_{H})$ and 
$X=Z^{\prime} + Y = (\Omega_{\rm m}\Theta_{\rm m}+\Omega_{\rm ds}\Theta_{\rm ds})/2$ is a gauge-invariant quantity.

The perturbed equations of state $w_{\rm ds}\Gamma_{\rm ds}$ and 
$w_{\rm ds}\Pi_{\rm ds}$ are a linear combination of the matter (${\rm m}$) and dark sector (${\rm ds}$) perturbed fluid variables
\begin{subequations}\label{eqn:EoS}
 \begin{align}
  w_{\rm ds}\Gamma_{\rm ds} = &\, C_{\Gamma\Delta_{\rm ds}}\Delta_{\rm ds} + 
                                  C_{\Gamma\Theta_{\rm ds}}\Theta_{\rm ds} +
                                  \frac{\Omega_{\rm m}}{\Omega_{\rm ds}}C_{\Gamma\Delta_{\rm m}}\Delta_{\rm m} + 
                                  \frac{\Omega_{\rm m}}{\Omega_{\rm ds}}C_{\Gamma\Theta_{\rm m}}\Theta_{\rm m} + 
                                  \frac{\Omega_{\rm m}}{\Omega_{\rm ds}}C_{\Gamma\Gamma_{\rm m}}
                                 w_{\rm m}\Gamma_{\rm m}\,, \label{eqn:wGamma}\\
  w_{\rm ds}\Pi_{\rm ds} = &\, C_{\Pi\Delta_{\rm ds}}\Delta_{\rm ds} + 
                               C_{\Pi\Theta_{\rm ds}}\Theta_{\rm ds} + 
                               \frac{\Omega_{\rm m}}{\Omega_{\rm ds}}C_{\Pi\Delta_{\rm m}}\Delta_{\rm m} + 
                               \frac{\Omega_{\rm m}}{\Omega_{\rm ds}}C_{\Pi\Theta_{\rm m}}\Theta_{\rm m} + 
                               \frac{\Omega_{\rm m}}{\Omega_{\rm ds}}C_{\Pi\Pi_{\rm m}}w_{\rm m}\Pi_{\rm m}\,, \label{eqn:wPi}
 \end{align}
\end{subequations}
where $w_{\rm m}\Gamma_{\rm m}$ and $w_{\rm m}\Pi_{\rm m}$ are the matter entropy perturbations and anisotropic stress, respectively, and the coefficients $C_{XY}$ are a function of the scale factor $a$ and quadratic in the scale ${\rm K}$.

To derive the second order equation for $\Delta_{\rm ds}$, we take the derivative of Eq.~(\ref{eqn:deltadot}) with respect to $\ln{a}$ and replace the term $\Theta_{\rm ds}^{\prime}$ with the expression in Eq.~(\ref{eqn:Thetadot}). Considering scales ${\rm K}\gg 1$, where $\Theta\ll\Delta$ as shown in \cite{Pace2019a} and neglecting the terms $w_{\rm m}\Pi_{\rm m}$ and $w_{\rm m}\Gamma_{\rm m}$ because unimportant at late times ($w_{\rm m}\Gamma_{\rm m}\approx 0$ and $w_{\rm m}\Pi_{\rm m}\approx 0$), the two perturbed equations of state simplify to
\begin{equation}
 w_{\rm ds}\Gamma_{\rm ds} \approx C_{\Gamma\Delta_{\rm ds}}\Delta_{\rm ds} +
                                   \frac{\Omega_{\rm m}}{\Omega_{\rm ds}}
                                   C_{\Gamma\Delta_{\rm m}}\Delta_{\rm m}\,, \quad 
 w_{\rm ds}\Pi_{\rm ds} \approx C_{\Pi\Delta_{\rm ds}}\Delta_{\rm ds} + 
                                \frac{\Omega_{\rm m}}{\Omega_{\rm ds}}
                                 C_{\Pi\Delta_{\rm m}}\Delta_{\rm m}\,.
\end{equation}

Neglecting the $X$ and $Y$ terms, and time variation of the coefficients $C_{XY}$ in $w_{\rm ds}\Pi_{\rm ds}$ and $w_{\rm ds}\Gamma_{\rm ds}$, we find
\begin{equation}\label{eqn:2nd}
 \Delta_{\rm ds}^{\prime\prime} + \left(2+3c_{\rm a,ds}^2-6w_{\rm ds}+\frac{H^{\prime}}{H}-2C_{\Pi\Delta_{\rm ds}}
 \right)\Delta_{\rm ds}^{\prime} + 
 {\rm K}^2\left(c_{\rm a,ds}^2\Delta_{\rm ds}+\frac{2}{3}\Pi_{\rm ds}+\Gamma_{\rm ds}\right) = 0 \,.
\end{equation}
Plugging the simplified equations of state $w_{\rm ds}\Gamma_{\rm ds}$ and $w_{\rm ds}\Pi_{\rm ds}$ into Eq.~(\ref{eqn:2nd}) leads to
\begin{equation}\label{eqn:gf}
 \Delta_{\rm ds}^{\prime\prime} + \left(2+3c_{\rm a,ds}^2-6w_{\rm ds}+\frac{H^{\prime}}{H}-2C_{\Pi\Delta_{\rm ds}}
 \right)\Delta_{\rm ds}^{\prime} + \left(c_{\rm a,ds}^2+C_{\zeta\Delta_{\rm ds}}\right){\rm K}^2 \Delta_{\rm ds} 
  = -\frac{\Omega_{\rm m}}{\Omega_{\rm ds}}C_{\zeta\Delta_{\rm m}}{\rm K}^2\Delta_{\rm m}\,,
\end{equation}
where $C_{\zeta\Delta_{\rm ds}}=\tfrac{2}{3}C_{\Pi\Delta_{\rm ds}}+C_{\Gamma\Delta_{\rm ds}}$ and $C_{\zeta\Delta_{\rm m}}=\tfrac{2}{3}C_{\Pi\Delta_{\rm m}}+C_{\Gamma\Delta_{\rm m}}$.

Applying a QSA implies neglecting the time derivatives of $\Delta_{\rm ds}$. From a physical point of view, we are imposing that the time variation on cosmological time scales is small, as we did for the coefficients $C_{XY}$. We are then left with a relation between dark sector and matter density perturbations which manifest in the form of an attractor solution for Eq.~(\ref{eqn:gf})
\begin{equation}\label{eqn:AIC}
 \Omega_{\rm ds}\Delta_{\rm ds} = -\frac{C_{\zeta\Delta_{\rm m}}}{c_{\rm a,ds}^2+C_{\zeta\Delta_{\rm ds}}}\Omega_{\rm m}\Delta_{\rm m}\,.
\end{equation}

Note that it is possible to derive the attractor solution by neglecting the time derivatives of $\Delta_{\rm ds}$ and $\Theta_{\rm ds}$ in Eqs.~(\ref{eqn:EoM}), and substituting $\Theta_{\rm ds}$ as derived from the Euler equation into the continuity equation. The attractor solution is then given by considering only terms proportional to ${\rm K}^2$.

The coefficients of the simplified equations of state are
\begin{align}
 C_{\Gamma\Delta_{\rm ds}} & = \frac{\gamma_1 \gamma_2 + \tilde{\gamma}_3 {\rm K}^2}
                               {\gamma_1 + \alpha_{\rm B}^2 {\rm K}^2} - c_{\rm a,ds}^2	\,,\\
 C_{\Gamma\Delta_{\rm m}} & = \frac{\gamma_1 \gamma_2 + \tilde{\gamma}_3 {\rm K}^2}
                              {\gamma_1 + \alpha_{\rm B}^2 {\rm K}^2}\left(1-\frac{M_{\rm pl}^2}{M^2}\right)
                              +\gamma_7\frac{M_{\rm pl}^2}{M^2}
                              +\left(\frac{M_{\rm pl}^2}{M^2}\frac{\alpha_{\rm K}}{\alpha}-1\right)c_{\rm a,m}^2 \,,\\
 C_{\Pi\Delta_{\rm ds}} & = -\frac{1}{2}\frac{\gamma_1 \alpha_{\rm T} + \tilde{\gamma}_8 {\rm K}^2}
                             {\gamma_1 + \alpha_{\rm B}^2 {\rm K}^2}\,,\\
 C_{\Pi\Delta_{\rm m}} & = -\frac{1}{2}
                            \left[\frac{M_{\rm pl}^2}{M^2}\alpha_{\rm T}+
                            \frac{\gamma_1 \alpha_{\rm T} + \tilde{\gamma}_8 {\rm K}^2}
                            {\gamma_1 + \alpha_{\rm B}^2 {\rm K}^2}
                            \left(1-\frac{M_{\rm pl}^2}{M^2}\right)\right]\,,
\end{align}
where
\begin{align*}
  \gamma_1 & \equiv \alpha_{\rm K} \frac{\rho_{\rm ds}+P_{\rm ds}-2(M^2-M_{\rm pl}^2)\dot{H}}{4H^2M^2} 
                   -3\alpha_{\rm B}^2\frac{\dot{H}}{H^2}\,, \\
 \gamma_2 & \equiv c_s^2+\frac{\alpha_{\rm T}}{3} - 
                   2\frac{2\alpha_{\rm B}+\tilde{\Gamma}+(1+\alpha_{\rm B})(\alpha_{\rm M}-\alpha_{\rm T})}{\alpha}
                   \,,\\
 \gamma_3 & \equiv c_s^2 + \frac{\gamma_8}{3}\,, \quad \tilde{\gamma}_3 = \alpha_{\rm B}^2\gamma_3\,,\\
 \gamma_7 & \equiv \frac{\alpha_{\rm K} \alpha_{\rm M} - 6\alpha_{\rm B}^2}{3\alpha}-
                   \frac{(6\alpha_{\rm B}-\alpha_{\rm K})(\alpha_{\rm T}-\alpha_{\rm M})}{3\alpha}\,, \\
 \gamma_8 & \equiv \alpha_{\rm T} + \frac{\alpha_{\rm T}-\alpha_{\rm M}}{\alpha_{\rm B}}\,, \quad 
 \tilde{\gamma}_8 = \alpha_{\rm B}^2\gamma_8 = \alpha_{\rm B}^2\alpha_{\rm T} + 
                    \alpha_{\rm B}(\alpha_{\rm T}-\alpha_{\rm M})\,,
\end{align*}
and a dot stands for the derivative w.r.t. cosmic time $t$. We also have $\alpha=\alpha_{\rm K}+6\alpha_{\rm B}^2$ and
\begin{equation}\label{eqn:cs2}
 c_{\rm s}^2 = -\frac{2(1+\alpha_{\rm B})\left[\dot{H} + H^2\alpha_{\rm B}(1+\alpha_{\rm T}) - (\alpha_{\rm M}-\alpha_{\rm T})H^2\right] + 2H\dot{\alpha}_{\rm B} + \left(\rho_{\rm m} + P_{\rm m}\right)/M^2}{\alpha H^2}\,,
\end{equation}
is the sound speed of perturbations. We finally have $\gamma_1\tilde{\Gamma}=\gamma_1\alpha_{\rm B}\Gamma$, where
\begin{equation*}
 \Gamma = \frac{\alpha_{\rm B}^2}{H^3\gamma_1}\frac{\partial}{\partial t}\left(\frac{H^2\gamma_1}{\alpha_{\rm B}^2}\right)\,.
\end{equation*}

Note that in the derivation of the expressions for $\mu$ and $\eta$, we neglected the term proportional to $c_{\rm a,m}^2$ as it is negligible at late times.

\bibliographystyle{JHEP}
\bibliography{QSA,bbl}

\label{lastpage}

\end{document}


\label{firstpage}

\maketitle
\flushbottom

\section{General definitions}
Some of the following definitions have already been presented in the main manuscript, but we report them here as well for completeness and to have all the necessary expressions self-contained in a single document. As already discussed, the resulting expressions for the effective gravitational constant $\mu$ and the slip parameter $\eta$ in previous works starting from the field equations are substantially equivalent, even if the coefficients are written in a different form. An exception is given by the mass of the perturbed scalar field as derived by \cite{DeFelice2011}, since the degree of freedom is $\delta\phi$ rather than $\pi=\delta\phi/\dot{\phi}$. We will see later on how using the two variables leads to a mixing of the coefficients and ultimately to differences in the final expressions for the terms proving the limit for $k\rightarrow 0$.

We assume the following metric
\begin{equation}\label{eqn:metric}
 \mathrm{d}s^2 = -(1+2\Phi)\mathrm{d}t^2 + a(t)^2(1-2\Psi)\delta_{ij}\mathrm{d}x^i\mathrm{d}x^j\,,
\end{equation}
where $\Phi$ and $\Psi$ are the two metric potentials and the total matter stress-energy tensor is
\begin{align}
 {T^{0}}_0 \equiv &\, -(\rho_{\rm m} + \delta\rho_{\rm m})\,,\\
 {T^{0}}_i \equiv &\, (\rho_{\rm m}+P_{\rm m})\partial_{i}\upsilon_{\rm m} \equiv \partial_iq_{\rm m}\,,\\
 {T^{i}}_j \equiv &\, (P_{\rm m} + \delta P_{\rm m})\delta^{i}_{j} + 
                      \left(\partial^i\partial_j - \frac{1}{3}\delta^{i}_{j}\partial^2\right)\sigma_{\rm m}\,,
\end{align}
where $\rho_{\rm m}$ and $P_{\rm m}$ are the (total) matter background density and pressure, respectively; $\delta\rho_{\rm m}$ and $\delta P_{\rm m}$ density and pressure perturbations; $\upsilon_{\rm m}$ and $q_{\rm m}$ the 3-velocity and 3-momentum potentials, respectively and $\sigma_{\rm m}$ the anisotropic stress.

Using the standard EFT formalism for the action, limiting ourselves to the background part,
\citep{Gubitosi2013,Bloomfield2013a,Gleyzes2013,Gleyzes2014,Tsujikawa2015}
\begin{equation}\label{eqn:background}
 S = \int \mathrm{d}^4x\sqrt{-g}\left[\frac{M_{\ast}^2}{2}f(t)R-\Lambda(t)-c(t)g^{00}\right]\,,
\end{equation}
with $R=6(2H^2+\dot{H})$ the Ricci scalar, the background can be written as
\begin{equation}
 c + \Lambda = 3M_{\rm pl}^2\left(fH^2+\dot{f}H\right) - \rho_{\rm m}\,, \quad 
 \Lambda - c = M_{\rm pl}^2\left(2f\dot{H}+3fH^2+2\dot{f}H+\ddot{f}\right) + P_{\rm m}\,.
\end{equation}
One can, therefore, express $c(t)$ and $\Lambda(t)$ in terms of the other variables
\begin{align}
 c = &\, -\frac{1}{2}\left(\rho_{\rm m}+P_{\rm m}\right) 
         - M_{\ast}^2f\left(\dot{H}+\frac{1}{2}\frac{\ddot{f}}{f}-\frac{1}{2}H\frac{\dot{f}}{f}\right)\,,
         \label{eqn:c}\\
 \Lambda = &\, \frac{1}{2}(P_{\rm m}-\rho_{\rm m})+
               M_{\ast}^2f\left(3H^2+\dot{H}+\frac{5}{2}H\frac{\dot{f}}{f}+\frac{1}{2}\frac{\ddot{f}}{f}\right)\,,
               \label{eqn:Lambda}
\end{align}
where we used the continuity equation for matter $\dot{\rho}_{\rm m}+3H(\rho_{\rm m}+P_{\rm m})=0$. For later calculations, the following relation is also useful
\begin{equation*}
 \dot{c} + \dot{\Lambda} + 6Hc = 3M_{\ast}^2\dot{f}\left(2H^2 + \dot{H}\right)\,.
\end{equation*}

As the definition of the braiding function $\alpha_{\rm B}$ is related to that of the Horndeski Lagrangian, we write it as
\begin{align}
 \mathcal{L}_2 & \equiv G_2(\phi,X)\,,\label{eqn:G2}\\
 \mathcal{L}_3 & \equiv G_3(\phi,X)\Box\phi\,,\label{eqn:G3}\\
 \mathcal{L}_4 & \equiv G_4(\phi,X)R - 2G_{4,X}(\phi,X)
                     \left[\left(\Box\phi\right)^2 - \left(\nabla_{\mu}\nabla_{\nu}\phi\right)^2\right]\,,
                     \label{eqn:G4}\\
 \mathcal{L}_5 & \equiv G_5(\phi,X)G_{\mu\nu}\nabla^{\mu\nu}\phi + \frac{1}{3}G_{5,X}(\phi,X)
                     \left[
                       \left(\Box\phi\right)^3-
                       3\Box\phi\left(\nabla_{\mu}\nabla_{\nu}\phi\right)^2+
                       2\left(\nabla_{\mu}\nabla_{\nu}\phi\right)^3                    
                       \right]\,,\label{eqn:G5}
\end{align}
where $\phi$ is the scalar field and $X=g^{\mu\nu}\nabla_{\mu}\phi\nabla_{\nu}\phi$ the canonical kinetic term.

The variables building up the perturbed quantities are time-dependent operators in the quadratic EFT action
\begin{equation}\label{eqn:2ndaction}
 S^{(2)} = \int \! \mathrm{d}^4x \sqrt{-g} \left[\frac{M_2^4(t)}{2}(\delta g^{00})^2 - 
           \frac{m_3^3(t)}{2} \delta K \delta g^{00} -
           m_4^2(t)\left(\delta K^2 - \delta K^{\mu}_{\nu} \delta K^{\nu}_{\mu}\right) + 
           \frac{\tilde{m}_4^2(t)}{2}\delta R\delta g^{00}\right] \,.
\end{equation}

In terms of the coefficients used in Eqs.~(\ref{eqn:background}) and (\ref{eqn:2ndaction}), the $\alpha_{\rm X}$ functions read
\begin{align*}
 & M^2 = \, M_{\ast}^2f+2m_4^2\,, && 
   \alpha_{\rm K} = \, \frac{2c+4M_2^4}{M^2H^2}\,, &&
   \alpha_{\rm B} = \, \frac{M_{\ast}^2\dot{f}-m_3^3}{2M^2H}\,,\\
 & \alpha_{\rm M} = \, \frac{M_{\ast}^2\dot{f}+2\left(m_4^2\right)^{\hbox{$\cdot$}}}{M^2H}\,, &&
   \alpha_{\rm T} = \, -\frac{2m_4^2}{M^2}\,, &&
   \alpha_{\rm H} = \, \frac{2\left(\tilde{m}_4^2-m_4^2\right)}{M^2}\,.
\end{align*}
Note that in \cite{Bellini2014,Pogosian2016,Zumalacarregui2017}, the definition of $\alpha_{\rm B}$ differs by a factor $-2$, so that $\alpha_{\rm B}^{\rm there}=-2\alpha_{\rm B}^{\rm here}$.

Given this basic introduction of the notation adopted in this work, in Table~\ref{tab:comparisonEFT} we present a comparison of the variables used in other works both at the background and linear perturbation, with those adopted here.

\renewcommand{\arraystretch}{1.5}
\begin{table}[!t]
 \centering
 \begin{tabular}{|l c c c c c c c|}
  \hline
  & $c$ & $\Lambda$ & $M_{\ast}^2f$ & $M_2^4$ & $m_3^3$ & $m_4^2$ & $\tilde{m}_4^2$ \\
  \hline
  Refs.~\cite{Gleyzes2014,Gleyzes2013,Piazza2013} & $c$ & $\Lambda$ & $M_{\ast}^2f$ 
                                                  & $M_2^4$ & $m_3^3$ & $m_4^2$ &   
                                                  $\tilde{m}_4^2$ \\
                                      
  Refs.~\cite{Gubitosi2013,Tsujikawa2015} & $c$ & $\Lambda$ & $M_{\ast}^2f$ 
                                          & $M_2^4$ 
                                          & $\bar{m}_1^3$ 
                                          & $\frac{\bar{M}_2^2}{2}$ 
                                          & $\mu_1^2$ \\
                                      
  Refs.~\citep{Bloomfield2013a,Bloomfield2013b,Pogosian2016} & $c$ & $c-\Lambda$ & $m_0^2\Omega$ 
                                                & $M_2^4$ 
                                                & $\bar{M}_1^3$ 
                                                & $\frac{\bar{M}_2^2}{2}$
                                                & $\hat{M}^2$ \\
                                               Refs.~\citep{Lombriser2015,Lombriser2015a} & $M_{\rm pl}^2\frac{\Gamma}{2}$ 
                                             & $M_{\rm pl}^2\left(\Lambda+\frac{\Gamma}{2}\right)$ 
                                             & $M_{\rm pl}^2\Omega$ 
                                             & $M_{\rm pl}^2M_2^4$
                                             & $M_{\rm pl}^2\bar{M}_1^3$
                                             & $M_{\rm pl}^2\frac{\bar{M}_2^2}{2}$
                                             & $M_{\rm pl}^2\hat{M}^2$\\
                                             
  Refs.~\citep{Kennedy2017,Kennedy2018} & $M_{\rm pl}^2\frac{\Gamma}{2}$ 
                                             & $M_{\rm pl}^2\left(\Lambda+\frac{\Gamma}{2}\right)$ 
                                             & $M_{\rm pl}^2\Omega$ 
                                             & $M_2^4$
                                             & $\bar{M}_1^3$
                                             & $\frac{\bar{M}_2^2}{2}$
                                             & $\hat{M}^2$\\
  
  Refs.~\citep{Piazza2014,Perenon2015,Perenon2017} & $M^2\mathcal{C}$ & $M^2\lambda$ & $M^2$ 
                                                   & $M^2\mu_2^2$ 
                                                   & $M^2\mu_3$ 
                                                   & $\frac{M^2\epsilon_4}{2}$ 
                                                   & $\frac{M^2\epsilon_4}{2}$ \\
                    
  Ref.~\citep{Raveri2014} & $c$ & $c-\Lambda$ & $m_0^2(1+\Omega)$ 
                          & $M_2^4$ 
                          & $\bar{M}_1^3$ 
                          & $\frac{\bar{M}_2^2}{2}$
                          & $\hat{M}^2$ \\
                          
  Ref.~\citep{Bellini2018} & $c$ & $c-\Lambda$ & $M_{\rm Pl}^2(1+\Omega)$ 
                           & $M_2^4$ 
                           & $\bar{M}_1^3$ 
                           & $\frac{\bar{M}_2^2}{2}$
                           & $\hat{M}^2$\\
  \hline
 \end{tabular}
 \caption{Dictionary between the different EFT approaches for the background and the linear perturbation quantities.}
 \label{tab:comparisonEFT}
\end{table}

In the next sections we will write the explicit expressions used by other authors in terms of the $\alpha$ functions.

\section{Gubitosi, Piazza \& Vernizzi, 2013; Ref.~\texorpdfstring{\citep{Gubitosi2013}}{Gubitosi2013}}
\label{sect:Gubitosi2013}
Using the relation between the EFT parameters and the $\alpha$ functions and expressing $c$ in terms of the sound speed of perturbations when necessary, we can work out the expressions appearing in \citep{Gubitosi2013}. 
The sound speed and the effective gravitational constant of Eqs.~(69) and (70) are
\begin{align}
 c_{\rm s}^2 = &\, \frac{c+\frac{3}{4}M_{\ast}^2\dot{f}^2/f-\frac{1}{2}\bar{m}_1^3\dot{f}/f-
                         \frac{1}{4}\bar{m}_1^6/(M_{\ast}^2f)+\frac{1}{2}\left(\dot{\bar{m}}_1^3+H\bar{m}_1^3\right)}
                        {c+2M_2^4+\frac{3}{4}M_{\ast}^2\dot{f}^2/f-\frac{3}{2}\bar{m}_1^3\dot{f}/f
                         +\frac{3}{4}\bar{m}_1^6/(M_{\ast}^2f)} \,,\\
             = &\, \frac{-\frac{1}{2}M^2\left\{2(1+\alpha_{\rm B})
                          \left[\dot{H}+H^2(\alpha_{\rm B}-\alpha_{\rm M})\right]+2H\dot{\alpha}_{\rm B}+
                          \frac{\rho_{\rm m}+P_{\rm m}}{M^2}\right\}}
                        {\frac{1}{2}H^2M^2\alpha}\,,\nonumber\\
 8\pi G_{\rm N} = &\, \frac{1}{M_{\ast}^2f}
                      \frac{c+M_{\ast}^2\dot{f}^2/f+\frac{1}{2}\left(\dot{\bar{m}}_1^3+H\bar{m}_1^3\right)}
                           {c+\frac{3}{4}M_{\ast}^2\dot{f}^2/f-\frac{1}{2}\bar{m}_1^3\dot{f}/f-
                           \frac{1}{4}\bar{m}_1^6/(M_{\ast}^2f)+
                           \frac{1}{2}\left(\dot{\bar{m}}_1^3+H\bar{m}_1^3\right)}\,,
                      \\
                = &\, \frac{\frac{1}{2}H^2M^2\left[\alpha c_{\rm s}^2+2(\alpha_{\rm B}-\alpha_{\rm M})^2\right]}
                           {\frac{1}{2}H^2M^2\alpha c_{\rm s}^2}
                      \frac{1}{M^2}\,,\nonumber
\end{align}
identical to the expressions used in the main manuscript for $\alpha_{\rm T}=0$.

With the notation introduced in the main manuscript to denote the coefficients of the field equations, noting that \cite{Gubitosi2013} uses our same metric convention and $\nabla^2\rightarrow-k^2$, we find
\begin{subequations}
 \begin{align}
  C^{00}_{\dot{\Psi}} = &\, -3\left(M_{\ast}^2\dot{f}-\bar{m}_1^3+2HM_{\ast}^2f\right) 
                      = -6M^2H(1+\alpha_{\rm B})\,,\\
  C^{00}_{\Phi} = &\, -2\Lambda - \rho_{\rm m} + 4M_2^4 + 6H\bar{m}_1^3
                = -M^2H^2(6-\alpha_{\rm K}+12\alpha_{\rm B})\,,\\
  C^{00}_{\Psi} = &\, -2M_{\ast}^2f
                = -2M^2\,,\\
  C^{00}_{\dot{\pi}} = &\, -\left(2c+4M_2^4\right) + 3H\left(M_{\ast}^2\dot{f}-\bar{m}_1^3\right) 
                     = -(\alpha_{\rm K}-6\alpha_{\rm B})M^2H^2\,, \\
  C^{00}_{\pi} = &\, 3HM_{\ast}^2\left(\ddot{f}+H\dot{f}\right) - \left(\dot{c}+\dot{\Lambda}\right) + 
                     3\dot{H}\bar{m}_1^3
               = -6HM^2\left[(1+\alpha_{\rm B})\dot{H}+\frac{\rho_{\rm m}+P_{\rm m}}{2M^2}\right]\,, \\
  C^{00}_{\pi2} = &\, \left(M_{\ast}^2\dot{f}-\bar{m}_1^3\right) 
                = 2M^2H\alpha_{\rm B}\,,
 \end{align}
\end{subequations}
for the $00$-component [their Eqs.~(145) and (146)],\footnote{We corrected a typo in Eqs.~(145) and (146) of \cite{Gubitosi2013}, where $-3\dot{H}\pi$ should rather be $3\dot{H}\pi$.}
\begin{subequations}
 \begin{align}
  C^{0i}_{\dot{\Psi}} = &\, 2 M_{\ast}^2f = 2M^2\,, \\
  C^{0i}_{\Phi} = &\, 2HM_{\ast}^2f+\left(M_{\ast}^2\dot{f}-\bar{m}_1^3\right) 
                = 2(1+\alpha_{\rm B})HM^2\,, \\
  C^{0i}_{\dot{\pi}} = &\, -\left(M_{\ast}^2\dot{f}-\bar{m}_1^3\right) 
                     = -2HM^2\alpha_{\rm B}\,, \\
  C^{0i}_{\pi} = &\, -\left(\rho_{\rm D}+P_{\rm D}\right) 
               = -2c - M_{\ast}^2\left(\ddot{f}-H\dot{f}\right)
               = M^2\left(2\dot{H}+\frac{\rho_{\rm m}+P_{\rm m}}{M^2}\right)\,, 
 \end{align}
\end{subequations}
for the $0i$-components [their Eq.~(150)],\footnote{The $0i$-components should read 
$M_{\ast}^2\partial_i\left[2f\left(\dot{\Psi}+H\Phi\right)+\dot{f}\left(\Phi-\dot{\pi}\right)\right]
-\left(\rho_{\rm D}+P_{\rm D}\right)\partial_i\pi -\bar{m}_1^3\partial_i\left(\Phi-\dot{\pi}\right) = 
M_{\ast}^2\partial_i\left[2f\left(\dot{\Psi}+H\Phi\right)-\left(\ddot{f}-H\dot{f}\right)\pi+
\dot{f}\left(\Phi-\dot{\pi}\right)\right]-2c\partial_i\pi -\bar{m}_1^3\partial_i\left(\Phi-\dot{\pi}\right) = 
\delta T_{i0}$.}
\begin{align}
 C_{\Psi}^{ij,{\rm trl}} = -C_{\Phi}^{ij,{\rm trl}} = M_{\ast}^2f = M^2\,, \quad 
 C_{\pi} = -M_{\ast}^2\dot{f} = -\alpha_{\rm M}M^2H\,,
\end{align}
for the $ij$-traceless components [their Eq.~(149)], and
\begin{subequations}
 \begin{align}
  C^{ij,{\rm tr}}_{\ddot{\Psi}} = &\, 2M_{\ast}^2f
                                = 2M^2\,, \\
  C^{ij,{\rm tr}}_{\dot{\Psi}} = &\, 2M_{\ast}^2f\left(3H+\dot{f}/f\right)
                               = 2(3+\alpha_{\rm M})HM^2\,, \\
  C^{ij,{\rm tr}}_{\Psi2} = &\, \frac{2}{3}M_{\ast}^2f 
                          = \frac{2}{3}M^2 \,, \\
  C^{ij,{\rm tr}}_{\dot{\Phi}} = &\, 2HM_{\ast}^2f+M_{\ast}^2\dot{f}-\bar{m}_1^3
                               = 2HM^2(1+\alpha_{\rm B}) \,, \\
  C^{ij,{\rm tr}}_{\Phi} = &\, M_{\ast}^2\left[\ddot{f}+5H\dot{f}+2\left(3H^2+2\dot{H}\right)\right] + 
                               \rho_{\rm D}+P_{\rm D} - 3H\bar{m}_1^3\,, \\
                         = &\, 2M^2\left[\dot{H} - \frac{\rho_{\rm m}+P_{\rm m}}{2M^2} + 
                                         (H\alpha_{\rm B})^{\hbox{$\cdot$}} + 
                                         (3+\alpha_{\rm M})(1+\alpha_{\rm B})H^2\right]\,, \nonumber\\
  C^{ij,{\rm tr}}_{\Phi2} = &\, -\frac{2}{3}M_{\ast}^2f 
                          = -\frac{2}{3}M^2\,, \\
  C^{ij,{\rm tr}}_{\ddot{\pi}} = &\, -M_{\ast}^2\dot{f}+\bar{m}_1^3
                               = -2HM^2\alpha_{\rm B}\,, \\
  C^{ij,{\rm tr}}_{\dot{\pi}} = &\, -M_{\ast}^2\left(\ddot{f}+3H\dot{f}\right)-(\rho_{\rm D}+P_{\rm D})+
                                    3H\bar{m}_1^3 \,, \\
                              = &\, 2M^2\left[\dot{H}+\frac{\rho_{\rm m}+P_{\rm m}}{2M^2}-
                                              \left(H\alpha_{\rm B}\right)^{\hbox{$\cdot$}}-
                                              (3+\alpha_{\rm M})\alpha_{\rm B}H^2\right]\,, \nonumber\\
  C^{ij,{\rm tr}}_{\pi} = &\, -3H^2M_{\ast}^2f-\dot{P}_{\rm D}
                        = 2M^2\left[(3+\alpha_{\rm M})H\dot{H} + \frac{\dot{P}_{\rm m}}{2M^2} + \ddot{H}\right]\,,\\
  C^{ij,{\rm tr}}_{\pi2} = &\, -\frac{2}{3}M_{\ast}^2\dot{f} 
                         = -\frac{2}{3}M^2H\alpha_{\rm M}\,,
 \end{align}
\end{subequations}
for the $ij$-trace components [their Eqs.~(147) and (148)].

Once again, we can easily see that they agree with the expressions in Eqs.~(109)--(112) in \cite{Gleyzes2014} for the special case $\alpha_{\rm T}=\alpha_{\rm H}=0$. To convert dark energy variables into matter variables we used the relations $\rho_{\rm D}=3M^2H^2-\rho_{\rm m}$ and $P_{\rm D}=-M^2\left(2\dot{H}+3H^2\right)-P_{\rm m}$.

\section{Piazza \& Vernizzi, 2013; Ref.~\texorpdfstring{\citep{Piazza2013}}{Piazza2013}}\label{Piazza2013}
In a similar way as in the previous section, we can convert the relevant expressions of \citep{Piazza2013} into the notation adopted here. The sound speed [their Eq.~(79)] is
\begin{align}
 c_{\rm s}^2 = &\, \frac{c+\frac{3}{4}M_{\ast}^2\dot{f}^2/f-\frac{1}{2}m_3^3\dot{f}/f-
                         \frac{1}{4}m_3^6/(M_{\ast}^2f)+\frac{1}{2}\left(\dot{m}_3^3+Hm_3^3\right)}
                        {c+2M_2^4+\frac{3}{4}M_{\ast}^2\dot{f}^2/f-\frac{3}{2}m_3^3\dot{f}/f
                         +\frac{3}{4}m_3^6/(M_{\ast}^2f)} \,,\\
             = &\, \frac{-\frac{1}{2}M^2\left\{2(1+\alpha_{\rm B})
                          \left[\dot{H}+H^2(\alpha_{\rm B}-\alpha_{\rm M})\right]+2H\dot{\alpha}_{\rm B}+
                          \frac{\rho_{\rm m}+P_{\rm m}}{M^2}\right\}}
                        {\frac{1}{2}H^2M^2\alpha}\,,\nonumber
\end{align}
again for the more restrictive case of $\alpha_{\rm T}=0$. The expressions for $\alpha$ and $\beta$ of their Eqs.~(77) and (78), respectively, correspond to the denominator and the numerator of the sound speed. Together with the expressions introduced in their Eqs.~(74) and (75) (labelled with $C_{\dot{\zeta}}$ and $C_{\partial^2\zeta}$), they read\footnote{The numerator should read $c+2M_2^4$, rather than $c+4M_2^4$.}
\begin{align}
 A \equiv &\, H + \frac{M_{\ast}^2\dot{f}-m_3^3}{2\left(M_{\ast}^2f+2m_4^2\right)} 
   = H(1+\alpha_{\rm B})\,,\\
 C_{\dot{\zeta}} \equiv &\, \left[\frac{3}{2}(A-H)^2 + \frac{c+2M_2^4}{M_{\ast}^2f+2m_4^2}\right]\frac{1}{A^2} 
                 = \frac{\alpha_{\rm K}+3\alpha_{\rm B}^2}{2(1+\alpha_{\rm B})^2}\,,\\
 C_{\partial^2\zeta} \equiv &\, \frac{M_{\ast}^2f+2\tilde{m}_4^2}{M_{\ast}^2f+2m_4^2}\frac{1}{A} 
                     = \frac{1+\alpha_{\rm H}}{1+\alpha_{\rm B}}\frac{1}{H}\,, \\
 \alpha \equiv &\, \frac{1}{A^2}
                   \left[c+2M_2^4+\frac{3}{4}\frac{(M_{\ast}^2\dot{f}-m_3^3)^2}{M_{\ast}^2f+2m_4^2}\right]
        = \frac{1}{2}\frac{\alpha}{(1+\alpha_{\rm B})^2}M^2\,,\\
 \beta \equiv &\, -M_{\ast}^2f
                  +\frac{1}{2a}\frac{{\rm d}}{{\rm d}t}\left[
                  \frac{2\left(M_{\ast}^2f+2\tilde{m}_4^2\right)a}{A}\right]\,, \\
       = &\, -M^2\left\{1+\alpha_{\rm T}-
                    \frac{1}{H}\frac{{\rm d}}{{\rm d}t}\left(\frac{1+\alpha_{\rm H}}{1+\alpha_{\rm B}}\right)-
                    \frac{1+\alpha_{\rm H}}{1+\alpha_{\rm B}}
                    \left(1+\alpha_{\rm M}-\frac{\dot{H}}{H^2}\right)\right\}\,.\nonumber
\end{align}
The definition of $c_{\rm s}^2=\beta/\alpha$ leads to the same result of Eq.~(85) of \cite{Gleyzes2014} in absence of matter.

\section{Bellini \& Sawicki, 2014; Ref.~\texorpdfstring{\citep{Bellini2014}}{Bellini2014}}\label{sect:Bellini2014}
The expressions in Eqs.~(3.13) and (3.17)--(3.21) of \cite{Bellini2014} are equivalent to Eqs.~(85) and (109)--(113) 
of \cite{Gleyzes2014}, respectively, upon the following identifications and replacements
\begin{align*}
 & \Psi\rightarrow\Phi\,, && \Phi\rightarrow\Psi\,, && M_{\ast}^2\rightarrow M^2\,,\\
 & \tilde{\rho}_{\rm m}+\tilde{P}_{\rm m}\rightarrow\frac{\rho_{\rm m}+P_{\rm m}}{M^2}\,, && 
 \pi_{\rm m}\rightarrow-\sigma_{\rm m}\,, && \upsilon_X\rightarrow -\pi\,,
\end{align*}
and $\alpha_{\rm B}\rightarrow-2\alpha_{\rm B}$. For the functions appearing in the definition of the coefficients describing the evolution of the two metric potentials, one needs only to divide the $\beta_i$ and $\Upsilon$ by the appropriate powers of $H$ to make them dimensionless and redefine $\beta_1$ as $\beta_1/(4H^2)$.

Their expressions (4.9) correspond to
\begin{equation*}
 Z_{\rm QS} = \mu_{Z,\infty}\,, \qquad \bar{\eta}_{\rm QS} = \frac{2}{1+\eta_{\infty}}\,.
\end{equation*}

\section{Bloomfield et~al., 2013; Ref.~\texorpdfstring{\citep{Bloomfield2013a}}{Bloomfield2013a}}
\label{sect:Bloomfield2013a}
The relation between the $\alpha$s and the EFT parameters used in \citep{Bloomfield2013a} is
\begin{align*}
 & M^2 = m_0^2\Omega + \bar{M}_2^2\,, && 
   \alpha_{\rm K} = \frac{2c+4M_2^4}{H^2M^2}\,, &&  
   \alpha_{\rm B} = \frac{m_0^2\dot{\Omega} + \bar{M}_1^3}{2M^2H}\,, \\
 & \alpha_{\rm M} = \frac{m_0^2\dot{\Omega} + (\bar{M}_2^2)^{\hbox{$\cdot$}}}{M^2H}\,, &&  
   \alpha_{\rm T} = -\frac{\bar{M}_2^2}{M^2}\,, && 
   \alpha_{\rm H} = \frac{2\hat{M}^2-\bar{M}_2^2}{M^2}\,,
\end{align*}
and the coefficients in their Eqs.~(4.13a)--(4.13c) become
\begin{subequations}
 \begin{align}
  A_1 = &\, 2m_0^2\Omega + 4\hat{M}^2 
      = 2M^2(1+\alpha_{\rm H})\,,\\
  A_2 = &\, -m_0^2\dot{\Omega} - \bar{M}_1^3 + 2H\bar{M}_3^2
      = -2HM^2\alpha_{\rm B}\,,\\
  A_3 = &\, 0 \,,\\
  B_1 = &\, -1 - \frac{2\hat{M}^2}{m_0^2\Omega}
      = -\frac{1+\alpha_{\rm H}}{1+\alpha_{\rm T}}\,,\\
  B_2 = &\, 1\,,\\
  B_3 = &\, -\frac{\dot{\Omega}}{\Omega} + 
             \frac{\bar{M}_3^2}{m_0^2\Omega}\left(H+\frac{2\dot{\bar{M}}_3}{\bar{M}_3}\right)
      = \frac{\alpha_{\rm T}-\alpha_{\rm M}}{1+\alpha_{\rm T}}H\,,\\
  C_1 = &\, m_0^2\dot{\Omega} + 2H\hat{M}^2 + 4\hat{M}\dot{\hat{M}} \,,\\
      = &\, M^2\left\{H\left[\alpha_{\rm M}+(1+\alpha_{\rm M})\alpha_{\rm H}-\alpha_{\rm T}\right]+
                      \dot{\alpha}_{\rm H}\right\}\,,\nonumber\\
  C_2 = &\, -\frac{1}{2}m_0^2\dot{\Omega} - \frac{1}{2}\bar{M}_1^3 - \frac{3}{2}\bar{M}_2^2H - 
             \frac{1}{2}\bar{M}_3^2H + 2H\hat{M}^2
      = M^2H(\alpha_{\rm H}-\alpha_{\rm B})\,,\\
  C_3 = &\, c - \frac{1}{2}(H+\partial_t)\bar{M}_1^3 - 3\bar{M}_2^2\dot{H} + \bar{M}_2^2\frac{k^2}{2a^2} + 
            \bar{M}_3^2\left(\frac{k^2}{2a^2}-\dot{H}\right) + 2\left(H^2+\dot{H}+H\partial_t\right)\hat{M}^2\,,\\
      = &\, -M^2\left\{\dot{H}+\frac{\rho_{\rm m}+ P_{\rm m}}{2M^2}+
                       H^2\left[1+\alpha_{\rm B}(1+\alpha_{\rm M})+\alpha_{\rm T}-
                                (1+\alpha_{\rm H})(1+\alpha_{\rm M})\right] + 
                       [H(\alpha_{\rm B}-\alpha_{\rm H})]^{\hbox{$\cdot$}}\right\}\,,\nonumber\\
  M_{\pi}^2 = &\, \frac{1}{4}m_0^2\dot{\Omega}\dot{R}^{(0)} - 3c\dot{H} + 
                  \frac{3}{2}\left(3H\dot{H}+\dot{H}\partial_t+\ddot{H}\right)\bar{M}_1^3 + 
                  \frac{9}{2}\dot{H}^2\bar{M}_2^2 + \frac{3}{2}\dot{H}^2\bar{M}_3^2\,,\\
            = &\, 3M^2
                  \left\{\left(\dot{H}+\frac{\rho_{\rm m}+P_{\rm m}}{2M^2}\right)\dot{H} + \dot{H}\alpha_{\rm B} 
                  \left[ H^2(3+\alpha_{\rm M})+\dot{H}\right]+H(\dot{H} \alpha_{\rm B})^{\hbox{$\cdot$}}\right\}\,.
                  \nonumber
 \end{align}
\end{subequations}
We renamed the mass term for the scalar degree of freedom $\pi$ as $M_{\pi}^2$, to avoid confusion with the effective Planck mass $M^2$. In terms of the potentials, the following identification holds: $\phi^N\rightarrow\Psi$ and $\psi^N\rightarrow\Phi$ and to have only second order equations of motion for the scalar field we assumed $\bar{M}_3^2=-\bar{M}_2^2$. Assuming flat cosmologies, we set $k_0=0$. These expressions are in perfect agreement with 
those of \cite{Gleyzes2014}.

These general results, valid for models beyond Horndeski with $\alpha_{\rm H}\neq 0$, will be helpful for the next section where, assuming an Horndeski model, the authors derived the functional form of the modified gravity parameters $\mu$ and $\eta$ using the quasi-static approximation.

\section{Bloomfield, 2013; Ref.~\texorpdfstring{\citep{Bloomfield2013b}}{Bloomfield2013b}}
\label{sect:Bloomfield2013b}
Using the same relations between the EFT and the $\alpha$ functions presented in the previous section (\ref{sect:Bloomfield2013a}) also for \citep{Bloomfield2013b}, and considering that $\phi\rightarrow\Phi$ and $\psi\rightarrow\Psi$, as the work specifically considers Horndeski models where $\alpha_{\rm H}=0$, we find that the coefficients for the equation of motion for $\pi$ [their Eq.~(56)] are
\begin{subequations}
 \begin{align}
  C^{\pi}_{\ddot{\pi}} = &\, c+2M_2^4 
                 = \frac{1}{2}M^2H^2\alpha_{\rm K}\,,\\
  C^{\pi}_{\dot{\pi}} = &\, \dot{c} + 3Hc + 2M_2^4\left(3H+4\frac{\dot{M}_2}{M_2}\right) 
                = \frac{1}{2}M^2H\left\{\left[H^2(3+\alpha_{\rm M})+\dot{H}\right]\alpha_{\rm K} + 
                                        \left(H\alpha_{\rm K}\right)^{\hbox{$\cdot$}}\right\}\,,\\
  C^{\pi}_{\pi} = &\, \frac{1}{4}m_0^2\dot{\Omega}\dot{R}^{(0)} - 3c\dot{H} + 
                \frac{3}{2}(3H\dot{H}+\dot{H}\partial_t+\ddot{H})\bar{M}_1^3 + 3\dot{H}^2\bar{M}_2^2\\
          = &\, 3M^2
                \left\{\left(\dot{H}+\frac{\rho_{\rm m}+P_{\rm m}}{2M^2}\right)\dot{H} + \dot{H}\alpha_{\rm B} 
                \left[ H^2(3+\alpha_{\rm M})+\dot{H}\right]+H(\dot{H} \alpha_{\rm B})^{\hbox{$\cdot$}}\right\}
          \,, \nonumber\\
  C^{\pi}_{\pi2} = &\, c - \frac{1}{2}\left(H+\partial_t\right)\bar{M}_1^3+
                       \left(H^2-\dot{H}+H\partial_t\right)\bar{M}_2^2\,,\\
           = &\, -M^2\left\{\dot{H}+\frac{\rho_{\rm m}+ P_{\rm m}}{2M^2} + 
                            H^2\left[\alpha_{\rm B}(1+\alpha_{\rm M})+\alpha_{\rm T}-\alpha_{\rm M}\right] + 
                            (H\alpha_{\rm B})^{\hbox{$\cdot$}}\right\}\,,\nonumber \\
  C^{\pi}_{\ddot{\psi}} = &\, \frac{3}{2}\left(m_0^2\dot{\Omega}+\bar{M}_1^3\right) 
                  = 3HM^2\alpha_{\rm B}\,,\\
  C^{\pi}_{\dot{\psi}} = &\, -3c + 6Hm_0^2\dot{\Omega} + 
                       \frac{9}{2}\bar{M}_1^3\left(H+\frac{\dot{\bar{M}}_1}{\bar{M}_1}\right) + 
                       3\dot{H}\bar{M}_2^2\,, \\
                 =&\, 3M^2\left[\dot{H} + \frac{\rho_{\rm m}+ P_{\rm m}}{2M^2} + 
                                H^2\alpha_{\rm B}(3+\alpha_{\rm M})+(\alpha_{\rm B}H)^{\hbox{$\cdot$}}\right] \,, 
                 \nonumber\\
  C^{\pi}_{\dot{\phi}} = &\, -\left(c+2M_2^4\right) + \frac{3}{2}H\left(m_0^2\dot{\Omega}+\bar{M}_1^3\right) 
                 = \frac{1}{2}H^2M^2\left(6\alpha_{\rm B}-\alpha_{\rm K}\right)\,,\\
  C^{\pi}_{\phi} = &\, -\dot{c} - 6Hc +3m_0^2\dot{\Omega}\left(2H^2+\dot{H}\right) - 
                 2M_2^4\left(3H+4\frac{\dot{M}_2}{M_2}\right) + \nonumber\\
             &\, \frac{3}{2}\bar{M}_1^3\left(3H^2+2\dot{H}+3H\frac{\dot{\bar{M}}_1}{\bar{M}_1}\right) + 
                 3H\dot{H}\bar{M}_2^2\,,\\
           = &\, \frac{1}{2}M^2\left[6\left(\dot{H}+\frac{\rho_{\rm m}+P_{\rm m}}{2M^2}\right) + 
                                     H^2(6\alpha_{\rm B}-\alpha_{\rm K})(3+\alpha_{\rm M}) + 
                                     2(9\alpha_{\rm B} - \alpha_{\rm K})\dot{H} + 
                                     H(6\dot{\alpha}_{\rm B}-\dot{\alpha}_{\rm K})\right]H\,,\nonumber\\
  C^{\pi}_{\phi2} = &\, -\frac{1}{2}\left(m_0^2\dot{\Omega}+\bar{M}_1^3\right) 
            = -HM^2\alpha_{\rm B}\,, \\
  C^{\pi}_{\psi2} = &\, m_0^2\dot{\Omega} + \dot{\bar{M}}_2^2 + H\bar{M}_2^2 
            = HM^2(\alpha_{\rm M}-\alpha_{\rm T})\,,
 \end{align}
\end{subequations}
the coefficients for the Poisson equation [their Eq.~(54)] are
\begin{subequations}
 \begin{align}
  C^{00}_{\psi2} = &\, 2\left(m_0^2\Omega+\bar{M}_2^2\right)
            = 2M^2\,,\\
  C^{00}_{\dot{\psi}} = &\, 3\left(m_0^2\dot{\Omega} + \bar{M}_1^3\right) 
                 = 6H\alpha_{\rm B}\,,\\
  C^{00}_{\phi} = &\, 3H\left(m_0^2\dot{\Omega} + \bar{M}_1^3\right) - \left(2c+4M_2^4\right) 
           = (6\alpha_{\rm B}-\alpha_{\rm K})M^2H^2\,,\\
  C^{00}_{\dot{\pi}} = &\, 2c+4M_2^4
                = M^2H^2\alpha_{\rm K}\,,\\
  C^{00}_{\pi} = &\, 3\dot{H}\left(m_0^2\dot{\Omega} + \bar{M}_1^3\right) 
          = 6H\dot{H}M^2\alpha_{\rm B}\,,\\
  C^{00}_{\pi2} = &\, -\left(m_0^2\dot{\Omega} + \bar{M}_1^3\right) 
           = -2M^2H\alpha_{\rm B}\,,
 \end{align}
\end{subequations}
and the coefficients for the anisotropic shear stress equation [their Eq.~(55)] read
\begin{subequations}
 \begin{align}
  C^{ij}_{\psi} = &\, m_0^2\Omega 
           = M^2(1+\alpha_{\rm T})\,,\\
  C^{ij}_{\phi} = &\, -\left(m_0^2\Omega+\bar{M}_2^2\right) 
           = -M^2\,,\\
  C^{ij}_{\pi} = &\, -\left[m_0^2\dot{\Omega}+\bar{M}_2^2\left(H+\frac{2\dot{\bar{M}}_2}{\bar{M}_2}\right)\right] 
          = -(\alpha_{\rm M}-\alpha_{\rm T})M^2H\,.
 \end{align}
\end{subequations}

We can now work out the expressions for the modified gravity parameters $\mu$ and $\eta$. The relevant coefficients 
used in \cite{Bloomfield2013b} are defined in their Eqs.~(57)--(60): $A_{\psi}=C^{00}_{\psi2}$, $A_{\phi}=0$, 
$A_{\pi}=C^{00}_{\pi2}$, $B_{\psi}=-1$, $B_{\phi}=-C^{ij}_{\phi}/C^{ij}_{\psi}=1/(1+\alpha_{\rm T})$, 
$B_{\pi}=-C^{ij}_{\pi}/C^{ij}_{\psi}=H(\alpha_{\rm M}-\alpha_{\rm T})/(1+\alpha_{\rm T})$, $C_{\psi}=C^{\pi}_{\psi2}$, 
$C_{\phi}=C^{\pi}_{\phi2}$, $C_{\pi}=C^{\pi}_{\pi2}$, $C_{\pi2}=C^{\pi}_{\pi}$.

They are defined as [their Eq.~(66)]
\begin{equation}\label{eqn:MG}
 \mu = 2M_{\rm pl}^2\frac{f_1+f_2\frac{a^2}{k^2}}{f_3+f_4\frac{a^2}{k^2}}\,, \quad 
 \eta = \frac{f_5+f_6\frac{a^2}{k^2}}{f_1+f_2\frac{a^2}{k^2}}\,,
\end{equation}
where [their Eqs.~(64) and (65)]
\begin{subequations}
 \begin{align}
  f_1 = &\, B_{\pi}C_{\psi} - B_{\psi}C_{\pi} 
      = \frac{1}{2}\frac{\alpha c_{\rm s}^2\bar{M}^2\mu_{\infty}}{1+\alpha_{\rm T}}H^2M^2\,,\\
  f_2 = &\, -B_{\psi}C_{\pi2} 
      = 3M^2\left\{\left(\dot{H}+\frac{\rho_{\rm m}+P_{\rm m}}{2M^2}\right)\dot{H} + \dot{H}\alpha_{\rm B} 
                   \left[ H^2(3+\alpha_{\rm M})+\dot{H}\right]+H(\dot{H} \alpha_{\rm B})^{\hbox{$\cdot$}}\right\}\,,\\
  f_3 = &\, A_{\psi}(B_{\phi}C_{\pi}-B_{\pi}C_{\phi}) + A_{\pi}(B_{\psi}C_{\phi}-B_{\phi}C_{\psi}) 
      = \frac{\alpha c_{\rm s}^2}{1+\alpha_{\rm T}}H^2M^4 \,,\\
  f_4 = &\, A_{\psi}B_{\phi}C_{\pi2} 
      = \frac{6M^4}{1+\alpha_{\rm T}}
           \left\{\left(\dot{H}+\frac{\rho_{\rm m}+P_{\rm m}}{2M^2}\right)\dot{H} + \dot{H}\alpha_{\rm B} 
           \left[ H^2(3+\alpha_{\rm M})+\dot{H}\right]+H(\dot{H} \alpha_{\rm B})^{\hbox{$\cdot$}}\right\}\,,\\
  f_5 = &\, B_{\phi}C_{\pi}-B_{\pi}C_{\phi} 
      = \frac{1}{2}\frac{\alpha c_{\rm s}^2\bar{M}^2\mu_{Z,\infty}}{1+\alpha_{\rm T}}H^2M^2 \,,\\
  f_6 = &\, B_{\phi}C_{\pi2} 
      = \frac{3M^4}{1+\alpha_{\rm T}}
            \left\{\left(\dot{H}+\frac{\rho_{\rm m}+P_{\rm m}}{2M^2}\right)\dot{H} + \dot{H}\alpha_{\rm B} 
            \left[ H^2(3+\alpha_{\rm M})+\dot{H}\right]+H(\dot{H} \alpha_{\rm B})^{\hbox{$\cdot$}}\right\}\,.
 \end{align}
\end{subequations}

\section{Pogosian \& Silvestri, 2016; Ref.~\texorpdfstring{\citep{Pogosian2016}}{Pogosian2016}}
\cite{Pogosian2016} use the same notation for the EFT functions and the metric potentials of \cite{Bloomfield2013a}, and define the coefficients required to evaluate the effective gravitational constant and the slip in their Appendix~A. 
The expressions provided are given for generic models beyond Horndeski, whose equations of motion are higher than second order. We report them here for completeness, but, focusing on beyond-Horndeski where $\alpha_{\rm H}\neq 0$ without higher spatial derivatives, to make a direct comparison with \cite{Gleyzes2014}. We will thus assume $m_2^2=0$.

With the help of their conversion table (and with the replacement $M_{\ast}^2\to M^2$)\footnote{We corrected a typo in their (A18) where it should read $H^2M_{\ast}^2\alpha_{\rm K}$, rather than $HM_{\ast}^2\alpha_{\rm K}$.}
\begin{subequations}
 \begin{align}
  M^2 = &\, m_0^2\Omega + \bar{M}_2^2\,, & 
  M^2H^2\alpha_{\rm K} = &\, 2c + 4M_2^4\,, & 
  2M^2H\alpha_{\rm B} = &\, m_0^2\dot{\Omega} + \bar{M}_1^3\,,\\
  M^2H\alpha_{\rm M} = &\, m_0^2\dot{\Omega} + \dot{\bar{M}}_2^2\,, & 
  M^2\alpha_{\rm T} = &\, -\bar{M}_2^2\,, & 
  M^2\alpha_{\rm H} = &\, 2\hat{M}^2 - \bar{M}_2^2\,,
 \end{align}
\end{subequations}
and $\bar{M}_2^2+\bar{M}_3^2=0$, the coefficients [their Eqs.~(A4)] read as
\begin{subequations}
 \begin{align}
  A_1 = &\, 2m_0^2\Omega + 4\hat{M}^2 
      = 2M^2(1+\alpha_{\rm H}) \,,\\
  A_2 = &\, -m_0^2\dot{\Omega} - \bar{M}^3_1 + 2H\bar{M}^2_3 + 4H\hat{M}^2 
      = 2HM^2(\alpha_{\rm H}-\alpha_{\rm B}) \,,\\
  A_3 = &\, -8m_2^2 
      = 0\,,\\
  B_1 = &\, -1-\frac{2\hat{M}^2}{m_0^2\Omega} 
      = -\frac{1+\alpha_{\rm H}}{1+\alpha_{\rm T}} \,,\\
  B_2 = &\, 1 \,,\\
  B_3 = &\, -\frac{\dot{\Omega}}{\Omega} + 
             \frac{\bar{M}^2_3}{m_0^2\Omega}\left(H+\frac{2\dot{\bar{M}}_3}{\bar{M}_3}\right) 
      = -\frac{\alpha_{\rm M}-\alpha_{\rm T}}{1+\alpha_{\rm T}}H \,,\\
  C_1 = &\, m_0^2\dot{\Omega} + 2H\hat{M}^2 + 4\hat{M}\dot{\hat{M}} 
      = M^2\left\{H\left[\alpha_{\rm M} + \alpha_{\rm H}\left(1+\alpha_{\rm T}\right) - \alpha_{\rm T}\right] + \dot{\alpha}_{\rm H}\right\}\,, \\
  C_2 = &\, -\frac{m_0^2}{2}\dot{\Omega} - \frac{1}{2}\bar{M}_1^3 - \frac{3}{2}H\bar{M}_2^2 - \frac{1}{2}H\bar{M}_3^2 + 2H\hat{M}^2 
      = HM^2\left(\alpha_{\rm H}-\alpha_{\rm B}\right)\,, \\
  C_3 = &\, c - \frac{1}{2}(H+\partial_t)\bar{M}_1^3 + \left(\frac{k^2}{2a^2}-3\dot{H}\right)\bar{M}_2^2 + 
            \left(\frac{k^2}{2a^2} - \dot{H}\right)\bar{M}_3^2 + 2\left(H^2+\dot{H}+H\partial_t\right)\hat{M}^2 \,,\\
      = &\, -M^2\left\{\dot{H}+\frac{\rho_{\rm m}+ P_{\rm m}}{2M^2} + 
                       H^2\left[1 + \alpha_{\rm B}\left(1+\alpha_{\rm M}\right) + \alpha_{\rm T} - \left(1+\alpha_{\rm H}\right)\left(1+\alpha_{\rm M}\right)\right] + \left[H\left(\alpha_{\rm B} - \alpha_{\rm H}\right)\right]^{\hbox{$\cdot$}}\right\} \,, \nonumber \\
  C_{\pi} = &\, \frac{m_0^2}{4}\dot{\Omega}\dot{R}^{(0)} - 3c\dot{H} + 
                \frac{3}{2}\left(3H\dot{H} +\dot{H}\partial_t+\ddot{H}\right)\bar{M}_1^3 + 
                \frac{9}{2}\dot{H}^2\bar{M}_2^2 + \frac{3}{2}\dot{H}^2\bar{M}_3^2\,,\\
          = &\, 3M^2
                \left\{\left(\dot{H}+\frac{\rho_{\rm m}+P_{\rm m}}{2M^2}\right)\dot{H} + \dot{H}\alpha_{\rm B} 
                \left[ H^2(3+\alpha_{\rm M})+\dot{H}\right]+H(\dot{H} \alpha_{\rm B})^{\hbox{$\cdot$}}\right\}\,.
 \end{align}
\end{subequations}
These coefficients are in perfect agreement with \cite{Gleyzes2014}.

With respect to the coefficients used in \cite{Bloomfield2013b}, we have the following identifications (with $\alpha_{\rm H}=0$)
\begin{align*}
 & A_1 = A_{\psi}\,, && A_2 = A_{\pi}\,, && A_3 = A_{\phi}\,, && B_1 = -B_{\phi} \,, && B_2 = -B_{\psi}\,, \\
 & B_3 = -B_{\pi}\,, && C_1 = C_{\psi}\,, && C_2 = C_{\phi}\,, && C_3 = C_{\pi}\,, && C_{\pi} = C_{\pi 2}\,,
\end{align*}
and defining $\mu$ and $\eta$ as in Eq.~(\ref{eqn:MG}), we find [their Eqs.~(A8)]
\begin{subequations}
 \begin{align}
  f_1 = &\, B_2C_3 - C_1B_3 = B_{\pi}C_{\psi} - B_{\psi}C_{\pi} \,,\\
  f_2 = &\, B_2C_{\pi} = -B_{\psi}C_{\pi2} \,,\\
  f_3 = &\, A_1(B_3C_2-B_1C_3) + A_2(B_1C_1-B_2C_2) + A_3(B_2C_3-B_3C_1) \,,\\
      = &\, A_{\psi}(B_{\phi}C_{\pi}-B_{\pi}C_{\phi}) + A_{\pi}(B_{\psi}C_{\phi}-B_{\phi}C_{\psi}) \nonumber \,,\\
  f_4 = &\, (A_3B_2-A_1B_1)C_\pi = A_{\psi}B_{\phi}C_{\pi2}\,,\\
  f_5 = &\, B_3C_2-B_1C_3 = B_{\phi}C_{\pi}-B_{\pi}C_{\phi}\,,\\
  f_6 = &\, -B_1 C_{\pi} = B_{\phi}C_{\pi2}\,,
 \end{align}
\end{subequations}
which clearly shows that the final outcome is the same for \cite{Pogosian2016} and \cite{Bloomfield2013b} and, as already stated, in agreement with our results for $\mu$ and $\eta$ when derived from the field equations. Besides the dimensionfull form of the previous expressions, they are in agreement with our Eq.~(3.1) of the main test.

\section{Piazza et~al., 2014; Ref.~\texorpdfstring{\citep{Piazza2014}}{Piazza2014} and 
Perenon et~al., 2015, 2017; Ref.~\texorpdfstring{\citep{Perenon2015,Perenon2017}}{Perenon20152017}}
\label{sect:PiazzaPerenon}
In this section we will translate the expression for $\mu$ and $\eta$ of \cite{Piazza2014,Perenon2015,Perenon2017}, as they adopt the same notation in terms of the metric potentials (as in Eq.~(\ref{eqn:metric})) and for the EFT dynamical variables. The only difference is in the short-hand notation to indicate the time derivative of 
$M_{\ast}^2f$ ($M^2$ in their notation). \cite{Piazza2014} and \cite{Perenon2015} define [their Eqs.~(5) and (6), respectively] $\mu\equiv\frac{\mathrm{d}\ln{M^2}}{\mathrm{d}t}$, while \cite{Perenon2017} uses $\mu_1\equiv\frac{\mathrm{d}\ln{M^2}}{\mathrm{d}t}$ [their Eq.~(2)]. We, therefore, have $\mu=\mu_1=\dot{f}/f$. In the following, for simplicity, we will use as variable $\mu_1$, to avoid confusion with the effective gravitational constant $\mu$ and replace $M^2$ with the equivalent quantity, in our notation, $M_{\ast}^2f$.

The relation between the EFT variables and the $\alpha$s is \citep{Perenon2015,Perenon2017}
\begin{align}
 & M^2 = M_{\ast}^2f(1+\epsilon_4)\,, && 
   \alpha_{\rm M} = \frac{1}{H}\left[\frac{\dot{\epsilon}_4}{1+\epsilon_4}+\mu_1\right]\,, && 
   \mu_1 = \frac{\dot{f}}{f}\,, \nonumber\\
 & \alpha_{\rm K} = \frac{2\mathcal{C}+4\mu_2^2}{H^2(1+\epsilon_4)}\,, && 
   \alpha_{\rm B} = \frac{\mu_1-\mu_3}{2H(1+\epsilon_4)}\,, && 
   \alpha_{\rm T} = -\frac{\epsilon_4}{1+\epsilon_4}\,,
\end{align}
and background variables are defined as
\begin{equation}
 \mathcal{C} = \frac{1}{2}\left(H\mu_1 - \dot{\mu}_1 - \mu_1^2\right) + 
               \frac{\rho_{\rm D}+P_{\rm D}}{2M_{\ast}^2f}\,, \quad
 \lambda = \frac{1}{2}(5H\mu_1 + \dot{\mu}_1 + \mu_1^2) + \frac{\rho_{\rm D}+P_{\rm D}}{2M_{\ast}^2f}\,,
\end{equation}
which differ from the quantities defined in other works.

We start our conversion with the coefficients $A$ and $B$ used to define the sound speed of perturbations $c_{\rm s}^2=B/A$. They are defined in Eqs.~(19) of \cite{Piazza2014}, (22) and (23) of \cite{Perenon2015}, and (23) and (24) of \cite{Perenon2017}, together with the auxiliary variables $\mathring{\mu}_3$ and $\mathring{\epsilon}_4$
\begin{align}
 \mathring{\mu}_3 \equiv &\, \dot{\mu}_3 + \mu_1\mu_3 + H\mu_3 \,, \qquad 
 \mathring{\epsilon}_4 \equiv \dot{\epsilon}_4 + \mu_1\epsilon_4 + H\epsilon_4 \,,\\
 A = &\, \left(\mathcal{C} + 2\mu_2^2\right)(1+\epsilon_4) + \frac{3}{4}(\mu_1-\mu_3)^2 
   = \frac{1}{2}\frac{\alpha}{(1+\alpha_{\rm T})^2}H^2\,,\\
 B = &\, \left(\mathcal{C} + \frac{\mathring{\mu}_3}{2} - \dot{H}\epsilon_4 +  H\mathring{\epsilon}_4\right)
              (1+\epsilon_4) - 
              (\mu_1-\mu_3)\left[\frac{\mu_1-\mu_3}{4(1+\epsilon_4)} - \mu_1 - \mathring{\epsilon}_4\right] 
   = \frac{1}{2}\frac{\alpha c_{\rm s}^2}{(1+\alpha_{\rm T})^2}H^2\,.
\end{align}

We can now analyse the expressions for $\mu$ and $\eta$. While \cite{Perenon2015,Perenon2017} only consider the limit on small scales [their Eqs.~(14) and (17), (15) and (20), respectively], \cite{Piazza2014}, instead, performs a full analysis including scale dependence. 
Here, we will, therefore, only consider the expressions provided by \cite{Piazza2014}, as the other two works agree with \cite{Piazza2014} on small scales.

The two quantities read [their Eqs.~(11), (14) and (16)]\footnote{We corrected a typo in the definition of the slip parameter $\gamma_{\rm sl}=\eta$.}
\begin{equation}\label{eqn:Geff}
 \mu = \frac{a_0 + Y_{\rm IR}}{b_0 + Y_{\rm IR}}\frac{1+\alpha_{\rm T}}{\bar{M}^2} \,, \quad 
 \eta = \frac{c_0 + (1+\epsilon_4)Y_{\rm IR}}{a_0 + Y_{\rm IR}} \,,
\end{equation}
where\footnote{We corrected the definition of $Y_{\rm IR}$, as the authors say it is equivalent to that of \cite{Gleyzes2013}.}
\begin{subequations}
\begin{align}
 a_0 = &\, 2\mathcal{C} + \mathring{\mu}_3 - 2\dot{H}\epsilon_4 + 2H\mathring{\epsilon}_4 + 
           2(\mu_1 + \mathring{\epsilon}_4)^2 \,,\\
     = &\, 
     \frac{\alpha c_{\rm s}^2(1+\alpha_{\rm T})+2[\alpha_{\rm B}(1+\alpha_{\rm T})+\alpha_{\rm T}-\alpha_{\rm M}]^2}
          {(1+\alpha_{\rm T})^2}H^2 
     = \frac{\alpha c_{\rm s}^2\bar{M}^2\mu_{\infty}}{(1+\alpha_{\rm T})^2}H^2 \,,\nonumber\\
 b_0 = &\, 2\mathcal{C} + \mathring{\mu}_3 - 2\dot{H}\epsilon_4 + 2H\mathring{\epsilon}_4 + 
           2\dfrac{(\mu + \mathring{\epsilon}_4)(\mu_1 - \mu_3)}{1+\epsilon_4} - 
           \dfrac{(\mu_1 - \mu_3)^2}{2(1+\epsilon_4)^2} \,,\\
     = &\, \frac{\alpha c_{\rm s}^2}{1+\alpha_{\rm T}}H^2 \,,\nonumber\\
 c_0 = &\, (1+\epsilon_4)(2\mathcal{C} + \mathring{\mu}_3 - 2\dot{H}\epsilon_4 + 2H\mathring{\epsilon}_4) - 
           (\mu_1 + \mathring{\epsilon}_4)(\mu_1 + \mu_3 + 2\mathring{\epsilon}_4) + 
           2(\mu_1 + \mathring{\epsilon}_4)^2 \,,\\
     = &\, \frac{\alpha c_{\rm s}^2 + 2\alpha_{\rm B}[\alpha_{\rm B}(1+\alpha_{\rm T})+\alpha_{\rm T}-\alpha_{\rm M}]}
                {(1+\alpha_{\rm T})^2}H^2
     = \frac{\alpha c_{\rm s}^2\bar{M}^2\mu_{Z,\infty}}{(1+\alpha_{\rm T})^2}H^2 \,,\nonumber\\
 Y_{\rm IR} \equiv &\, 3\left(\frac{a}{k}\right)^2 \left[-2\dot{H}\mathcal{C} - \dot{H}\mathring{\mu}_3 + 
                       2\dot{H}^2\epsilon_4 + \ddot{H}(\mu_1 - \mu_3) - 2H\dot{H}\mu_3 + 4H\dot{H}\mu_1\right]\,,\\
            = &\, 6\left\{\left(\dot{H}+\frac{\rho_{\rm m}+P_{\rm m}}{2M^2}\right)\dot{H} + \dot{H}\alpha_{\rm B} 
                          \left[H^2(3+\alpha_{\rm M})+\dot{H}\right]+H(\dot{H}\alpha_{\rm B})^{\hbox{$\cdot$}}
                   \right\}/(1+\alpha_{\rm T})\,.\nonumber
\end{align}
\end{subequations}
The quantity often appearing in the previous expressions becomes
\begin{equation*}
 2\mathcal{C} + \mathring{\mu}_3 - 2\dot{H}\epsilon_4 + 2H\mathring{\epsilon}_4 = 
 \frac{H^2}{1+\alpha_{\rm T}}
 \left\{\alpha c_{\rm s}^2 + 2\alpha_{\rm B}\left[\alpha_{\rm B}(1+\alpha_{\rm T})+2(\alpha_{\rm T}-\alpha_{\rm M})
        \right]\right\}\,.
\end{equation*}

Finally, it is interesting to convert the expressions for the ratios of the several definitions of the effective gravitational constant defined in \cite{Perenon2015}. The authors identify: $i)$ $\mu_{\rm gw}=1/\bar{M^2}$ as the coupling of gravitational waves to matter; $ii)$ $\mu$ as the gravitational coupling between two objects in the linear regime under the quasi-static approximation (therefore relevant on cosmological scales) without the effect of any screening mechanism; and $iii)$ $\mu_{\rm sc}=(1+\alpha_{\rm T})/\bar{M^2}$ as the gravitational coupling between two objects under the quasi-static approximation and with the effects of screening mechanisms (therefore relevant for Solar System scales). One can write $\frac{G_{\rm eff}}{G_{\rm N}} = \frac{G_{\rm eff}}{G_{\rm sc}}\frac{G_{\rm sc}}{G_{\rm N}} \Longrightarrow \mu = \frac{\mu}{\mu_{\rm sc}}\mu_{\rm sc}$, where
\begin{align}
 \frac{\mu}{\mu_{\rm sc}} = &\, 1 + \frac{1+\epsilon_4}{B}
                            \left[\frac{\mu_1-\mu_3}{1+\epsilon_4}-\left(\mu_1+\mathring{\epsilon}_4\right)\right]^2
                          = 1 + \frac{2\left[2\alpha_{\rm B}(1+\alpha_{\rm T}+\alpha_{\rm T}-\alpha_{\rm M})\right]^2}
                                     {\alpha c_{\rm s}^2}\,.
\end{align}

\section{Amendola et~al., 2019; Ref.~\texorpdfstring{\citep{Amendola2019}}{Amendola2019}}\label{sect:Amendola}
The quasi-static approximation has also been investigated by \cite{Amendola2019}, providing time- and scale-dependent expressions based on \cite{DeFelice2011}. The expressions read (see also \cite{Amendola2013a}):
\begin{equation}
 \mu_Z \equiv h_1h_2\left(\frac{1+h_4{\rm K}^2}{1+h_3{\rm K}^2}\right)\,, \quad 
 \mu \equiv h_1\left(\frac{1+h_5{\rm K}^2}{1+h_3{\rm K}^2}\right)\,, \quad 
 \eta \equiv h_2\left(\frac{1+h_4{\rm K}^2}{1+h_5{\rm K}^2}\right)\,,
\end{equation}
where, by defining
\begin{subequations}
 \begin{align}
  \alpha_1 = &\, -2[\alpha_{\rm B}(1+\alpha_{\rm T})+\alpha_{\rm T}-\alpha_{\rm M}]\,,\\
  \alpha_2 = &\, -2\left[(1+\alpha_{\rm B})\xi+\frac{\dot{\alpha}_{\rm B}}{H}+
                         \frac{\rho_{\rm m}+P_{\rm m}}{2H^2M^2}\right]\,,\\
  \mu^2 = &\, -3\left\{\xi\alpha_2 - 2\alpha_{\rm B}\left[2\xi^2+\dot{\xi}/H+(3+\alpha_{\rm M})\xi\right]\right\}\,.
 \end{align}
\end{subequations}
with $\xi=\dot{H}/H^2$,
\begin{subequations}
 \begin{align}
  h_1 = &\, \frac{1+\alpha_{\rm T}}{\bar{M}^2}\,, \quad 
  h_2 = \frac{1}{1+\alpha_{\rm T}}\,,\\
  h_3 = &\, \frac{1}{\mu^2}[(1+\alpha_{\rm B})\alpha_1+\alpha_2] = \alpha c_{\rm s}^2/\mu^2\,,\\
  h_4 = &\, \frac{1}{\mu^2}(\alpha_1+\alpha_2) = \alpha c_{\rm s}^2\bar{M}^2\mu_{Z,\infty}/\mu^2\,,\\
  h_5 = &\, \frac{1}{\mu^2}\left(\frac{1+\alpha_{\rm M}}{1+\alpha_{\rm T}}\alpha_1+\alpha_2\right) = 
            \frac{1}{\mu^2}\frac{\alpha c_{\rm s}^2\bar{M}^2\mu_{Y,\infty}}{1+\alpha_{\rm T}}\,.
 \end{align}
\end{subequations}

The limits for ${\rm K}\rightarrow\infty$ and ${\rm K}\rightarrow 0$, are, respectively,
\begin{align*}
 \mu_{Z,\infty} \equiv \frac{h_1h_2h_4}{h_3}\,, \quad 
 \mu_{\infty} \equiv \frac{h_1h_5}{h_3}\,, \quad 
 \eta_{\infty} \equiv \frac{h_2h_4}{h_5}\,; \quad 
 \mu_{Z,0} \equiv h_1h_2 \,, \quad 
 \mu_{0} \equiv h_1 \,, \quad 
 \eta_{0} \equiv h_2\,.
\end{align*}
These expressions are identical to those found starting from EFE, as \\
$\mu^2=\tilde{f}_4=6\left\{\left(\dot{H}+\frac{\rho_{\rm m}+P_{\rm m}}{2M^2}\right)\dot{H} +\dot{H} 
\alpha_{\rm B}\left[ H^2(3+\alpha_{\rm M})+\dot{H}\right]+H(\dot{H} \alpha_{\rm B})^{\hbox{$\cdot$}}\right\}/H^4$.

\section{Gleyzes et~al., 2013; Ref.~\texorpdfstring{\citep{Gleyzes2013}}{Gleyzes2013}}\label{Gleyzes2013}
This approach is based on the EFT formalism and uses a similar set of variables of 
\cite{Bloomfield2013a,Bloomfield2013b,Pogosian2016}. The coefficients and expressions are provided for beyond Horndeski models, therefore, several coefficients will be set to zero to match the Horndeski expressions.

For Horndeski models
\begin{equation*}
 \bar{m}_4^2 = \bar{m}_5 = \bar{\lambda} = 0\,, \quad \bar{M}_3^2 = -\bar{M}_2^2\,, \quad 
 \bar{M}_2^2 = 2m_4^2\,, \quad \tilde{m}_4^2 = m_4^2\,,
\end{equation*}
and one defines $M_4^2 \equiv 2m_4^2+3\bar{m}_4^2$. Therefore, in the following expressions, the only relevant functions will be $f(t)$, $\Lambda(t)$, $c(t)$, $M_2^4$, $m_3^3$ and $m_4^2$. We will, although, also consider $\tilde{m}_4^2 \neq m_4^2$ to include terms dependent on $\alpha_{\rm H}$ in the field equations, to be able to make a more direct comparison with \cite{Gleyzes2014}.

In the following, we will present the coefficients of the field equations, which will later be used for the functions in the quasi-static approximation. The functions $A_x$ are the coefficients in the $00$-component of the field equations:
\begin{subequations}
 \begin{align}
  A_{\Phi} & = 2c + 4M_2^4 - 6H\left[fH \MM^2 + \MM^2 \dot{f} - m_3^3 + H \Mfs\right] 
             = -H^2M^2(6-\alpha_{\rm K}+12\alpha_{\rm B})\,,\\
  A_{\dot{\Psi}} & = -3\left[2H\left(\MM^2f + \Mfs\right) + \MM^2\dot{f} - m_3^3\right] 
                   = -6HM^2(1+\alpha_{\rm B})\,, \\
  A_{\pi} & = 3H^2\MM^2\dot{f} + 3m_3^3\dot{H} - \dot{c} - \dot{\Lambda} - 6\Mfs H\dot{H} +3H\MM^2\ddot{f}\,, \\ 
          & = 6Hc - 3\left(H^2+\dot{H}\right)\MM^2\dot{f} + 3H\MM^2\ddot{f} + 3m_3^3\dot{H} - 6\Mfs H\dot{H}\,, 
              \nonumber \\
          & = -6HM^2\left[(1+\alpha_{\rm B})\dot{H} + \frac{\rho_{\rm m}+P_{\rm m}}{2M^2}\right] \,,\nonumber\\
  A_{\dot{\pi}} & = -2c - 4M_2^4 - 3H\left(m_3^3-\MM^2\dot{f}\right) 
                  = H^2M^2(6\alpha_{\rm B}-\alpha_{\rm K})\,, \\
  A^{(2)}_{\Psi} & = -2f\MM^2 + 6H\bmfi - 4\tmfs 
                   = -2M^2(1+\alpha_{\rm H})\,, \\ 
  A^{(2)}_{\pi} & = \MM^2 \dot{f} - m_3^3 + 2H\Mfs - 4H\tmfs + 6H^2\bmfi 
                  = 2M^2H(\alpha_{\rm B}-\alpha_{\rm H})\,;
 \end{align}
\end{subequations}
the functions $B_x$ are the coefficients in the $0i$-component of the field equations:
\begin{subequations}
 \begin{align}
  B_{\Phi} & = -m_3^3 + 2H\left(f\MM^2 + \Mfs\right)+\MM^2\dot{f} 
             = 2HM^2(1+\alpha_{\rm B})\,, \\
  B_{\dot{\Psi}} & = 2\left(\MM^2f + \Mfs\right) 
                   = 2M^2\,, \\ 
  B_{\pi} & = -2c + 2\Mfs\dot{H} + \MM^2\left(H\dot{f} - \ddot{f}\right) 
            = 2M^2\left(\dot{H} + \frac{\rho_{\rm m}+P_{\rm m}}{2M^2}\right)\,, \\
  B_{\dot{\pi}} & =  m_3^3-\MM^2\dot{f} 
                  = -2M^2H\alpha_{\rm B}\,, \\
  B^{(2)}_{\Psi} & = -2\bmfi 
                   = 0\,, \\
  B^{(2)}_{\pi} & = -2(\bmfs + H\bmfi) 
                  = 0 \,;
 \end{align}
\end{subequations}
the functions $C_x$ are the coefficients of the $ij$-trace component:
\begin{subequations}
 \begin{align}
  C_{\Phi} & = 3\left[2c + 2\left(3H^2+\dot{H}\right)\Mfs + 2\MM^2f\left(3H^2+2\dot{H}\right) - 
               \left(m_3^3\right)^{\hbox{$\cdot$}} + \right.\nonumber\\
           & \quad\quad  \left. H\left(-3m_3^3 + 4\MM^2\dot{f} + 
               2\left(\Mfs\right)^{\hbox{$\cdot$}}\right) + 2\MM^2\ddot{f}\right]\,, \\
           & = 6M^2\left[\dot{H} - \frac{\rho_{\rm m}+P_{\rm m}}{2M^2} + (H\alpha_{\rm B})^{\hbox{$\cdot$}} + 
                         (3+\alpha_{\rm M})(1+\alpha_{\rm B})H^2\right]\,, \\
  C_{\dot{\Phi}} & = -3m_3^3 + 6H\left(\MM^2f + \Mfs\right) + 3\MM^2\dot{f} 
                   = 6HM^2(1+\alpha_{\rm B})\,,\\
  C_{\dot{\Psi}} & = 6 \left[3H\MM^2f + 3H\Mfs + \MM^2\dot{f} + \left(\Mfs\right)^{\hbox{$\cdot$}}\right] 
                   = 6HM^2(3+\alpha_{\rm M})\,,\\
  C_{\ddot{\Psi}} & = 6\left(\MM^2f + \Mfs\right) 
                    = 6M^2\,,\\
  C_{\pi} & = -3\left[\dot{c} - \dot{\Lambda} - 2\dot{H}\left(\Mfs\right)^{\hbox{$\cdot$}} -
               2\left(\ddot{H} + 3H\dot{H}\right)\Mfs + \MM^2\left(2\dot{H}\dot{f} + \dddot{f} + 3H^2\dot{f} + 
               2H\ddot{f}\right)\right]\,, \\
          & = -3\left[2\dot{c} + 6Hc - 2\dot{H}(\Mfs)^{\hbox{$\cdot$}} - 2(\ddot{H} + 3H\dot{H})\Mfs + 
                      \MM^2\left(\dddot{f} - (3H^2+\dot{H})\dot{f} + 2H\ddot{f}\right)\right]\,,\nonumber \\
          & = 6M^2\left[\frac{\dot{P}_{\rm m}}{2M^2} + \ddot{H} + (3+\alpha_{\rm M})H\dot{H}\right]\,, \nonumber\\
  C_{\dot{\pi}} & = 3\left[-2c + 3Hm_3^3 - 2H\MM^2\dot{f} + 2\Mfs\dot{H} + \left(m_3^3\right)^{\hbox{$\cdot$}} - 
                    2\MM^2\ddot{f}\right] \\
                & = 2M^2\left[\dot{H} + \frac{\rho_{\rm m}+P_{\rm m}}{2M^2} - 
                              \left(H\alpha_{\rm B}\right)^{\hbox{$\cdot$}} - 
                              (3+\alpha_{\rm M})\alpha_{\rm B}H^2\right]\,,\nonumber\\
  C_{\ddot{\pi}} & = 3\left(m_3^3 - \MM^2\dot{f}\right) 
                   = -2M^2H\alpha_{\rm B} \,, \\
  C^{(2)}_{\Phi} & = -\left(2\MM^2f - 6H\bmfi + 4\tmfs\right) 
                   =  -2M^2(1+\alpha_{\rm H})\,,  \\
  C^{(2)}_{\Psi} & = 2\MM^2f - 6H\bmfi - 6\dot{\bar{m}}_5 
                   = 2M^2(1+\alpha_{\rm T}) \,, \\
  C^{(2)}_{\pi} & = -2\left[\MM^2\dot{f} + \left(\Mfs\right)^{\hbox{$\cdot$}} + H\Mfs + 3H^2\bmfi + 
                    3H\dot{\bar{m}}_5\right] 
                  = -2HM^2(\alpha_{\rm M}-\alpha_{\rm T}) \,,\\
  C^{(2)}_{\dot{\pi}} & = -2\left(\Mfs - 2\tmfs + 3H\bmfi\right) 
                        = 2M^2\alpha_{\rm H} \,,\\
  C^{(4)}_{\Psi} & = 16\bar{\lambda} 
                   = 0 \,, \\
  C^{(4)}_{\pi} & = -2\bmfi + 16H\bar{\lambda} 
                  = 0\,;
 \end{align}
\end{subequations}
the functions $D_x$ are the coefficients of the $ij$-traceless component of the field equations:
\begin{subequations}
 \begin{align}
  D^{(2)}_{\Phi} & = \MM^2f + 2\tmfs - 3H\bmfi 
                   = M^2(1+\alpha_{\rm H})\,, \\
  D^{(2)}_{\Psi} & = -\MM^2f 
                   = -M^2(1+\alpha_{\rm T})\,, \\
  D^{(2)}_{\dot{\Psi}} & = -3\bmfi 
                         = 0 \,, \\
  D^{(2)}_{\pi} & = \MM^2\dot{f} + 2\mfs H + 2\left(m_4^2\right)^{\hbox{$\cdot$}} - 3\dot{H}\bmfi 
                  = M^2H(\alpha_{\rm M}-\alpha_{\rm T}) \,, \\
  D^{(2)}_{\dot{\pi}} & = 2\mfs - 2\tmfs 
                        = -M^2\alpha_{\rm H}\,, \\
  D^{(4)}_{\Psi} & = -8\bar{\lambda} 
                   = 0 \,, \\
  D^{(4)}_{\pi} & = \bmfi - 8H\bar{\lambda} 
                  = 0\,;
 \end{align}
\end{subequations}
and the functions $E_x$ are the coefficients of the equation of motion for the perturbed scalar field $\pi$
\begin{subequations}
 \begin{align}
  E_{\Phi} & = 6Hc + \dot{c} + H^2\left(9m_3^3 - 6\MM^2\dot{f}\right) + 
               3\dot{H}\left(2m_3^3 - \MM^2\dot{f}\right) - \dot{\Lambda} + \nonumber\\
           & \quad~ 3H\left[4M_2^4 - 2\Mfs\dot{H} + (m_3^3)^{\hbox{$\cdot$}}\right] + 
                    4\left(M_2^4\right)^{\hbox{$\cdot$}} \\
           & = 12Hc + 2\dot{c} + 3m_3^3(3H^2 + 2\dot{H}) - 6\MM^2\dot{f}(2H^2 + \dot{H}) + 
               3H\left[4M_2^4 - 2\Mfs\dot{H} + ( m_3^3)^{\hbox{$\cdot$}}\right] + 4(M_2^4)^{\hbox{$\cdot$}}\,,
               \nonumber\\
           & = -HM^2\left[6\left(\dot{H} + \frac{\rho_{\rm m}+P_{\rm m}}{2M^2}\right) + 
                          H^2(6\alpha_{\rm B}-\alpha_{\rm K})(3+\alpha_{\rm M}) + 
                          2(9\alpha_{\rm B}-\alpha_{\rm K})\dot{H} + 
                          H(6\dot{\alpha}_{\rm B}-\dot{\alpha}_{\rm K})\right]\,, \nonumber \\
  E_{\dot{\Phi}} & = 2c + 4M_2^4 + 3H(m_3^3-\MM^2\dot{f}) 
                   = M^2H^2(\alpha_{\rm K} - 6\alpha_{\rm B})\,, \\
  E_{\Psi} & = 3\left[6cH + \dot{c} + \dot{\Lambda} - 3\MM^2\dot{f}(2H^2 + \dot{H})\right] 
             = 0 \,, \\
  E_{\dot{\Psi}} & = 3\left[2c + 3Hm_3^3 - 4H\MM^2\dot{f} -2\Mfs\dot{H} + (m_3^3)^{\hbox{$\cdot$}}\right] \,,\\
                 & = -6M^2\left[\dot{H} + \frac{\rho_{\rm m}+P_{\rm m}}{2M^2} + H^2\alpha_{\rm B}(3+\alpha_{\rm M}) + 
                                (\alpha_{\rm B}H)^{\hbox{$\cdot$}}\right]\,,\nonumber\\
  E_{\ddot{\Psi}} & = 3(m_3^3-\MM^2\dot{f}) 
                    = -6M^2H\alpha_{\rm B}\,,\\
  E_{\pi} & = -\left[6\Mfs\dot{H}^2 - 3(m_3^3)^{\hbox{$\cdot$}}\dot{H} + 6H\dot{c} - 9H\dot{H}m_3^3 + \ddot{c} - 
                     3\MM^2\dot{H}\ddot{f} - 6H^2\MM^2\ddot{f} - 3m_3^3\ddot{H} + \ddot{\Lambda}\right]\,, \\ 
          & = -\left[6\Mfs\dot{H}^2 - 3(m_3^3)^{\hbox{$\cdot$}}\dot{H} - 6\dot{H}c + 
                     3\MM^2(\ddot{H} + 4H\dot{H})\dot{f} - 9H\dot{H}m_3^3 - 3m_3^3\ddot{H}\right]\,,\nonumber \\
          & = -6M^2\left\{\left(\dot{H} + \frac{\rho_{\rm m}+P_{\rm m}}{2M^2}\right)\dot{H} + \dot{H}\alpha_{\rm B}\left[H^2(3+\alpha_{\rm M}) + \dot{H}\right] + 
            H(\dot{H}\alpha_{\rm B})^{\hbox{$\cdot$}}\right\}\,,\nonumber\\
  E_{\dot{\pi}} & = -2\left[3H\left(c+2 M_2^4\right) + \dot{c} + 2(M_2^4)^{\hbox{$\cdot$}}\right] \,,\\
                & = -M^2H\left\{\left[H^2(3+\alpha_{\rm M})+\dot{H}\right]\alpha_{\rm K} + 
                                \left(H\alpha_{\rm K}\right)^{\hbox{$\cdot$}}\right\}\,,\nonumber\\
  E_{\ddot{\pi}} & = -2\left(c+2 M_2^4\right) 
                   = -M^2H^2\alpha_{\rm K}\,, \\
  E^{(2)}_{\Phi} & = -\left[m_3^3 + H\left(-2\Mfs - 6H\bmfi + 4\tmfs\right) - \MM^2\dot{f}\right] 
                   = -2M^2H(\alpha_{\rm H} - \alpha_{\rm B})\,,  \\
  E^{(2)}_{\Psi} & = -2\left[2H\tmfs + \MM^2\dot{f} - 3\bmfi\dot{H} + 2(\tilde{m}_4^2)^{\hbox{$\cdot$}}\right] 
                   = -2M^2\left\{H[\alpha_{\rm M} - \alpha_{\rm T} + \alpha_{\rm H}(1+\alpha_{\rm M})] + 
                                 \dot{\alpha}_{\rm H}\right\}\,,\\
  E^{(2)}_{\dot{\Psi}} & = 2\Mfs + 6H\bmfi - 4\tmfs 
                         = -M^2\alpha_{\rm H}\,,\\
  E^{(2)}_{\pi} & = -\left[2c - 4\Mfs\dot{H} + 4\tmfs\dot{H} + (m_3^3)^{\hbox{$\cdot$}} + 4H^2\tmfs + Hm_3^3 - 12\bmfi H\dot{H} + 4H(\tilde{m}_4^2)^{\hbox{$\cdot$}}\right]\,, \\
                & = 2M^2\left\{\dot{H} + \frac{\rho_{\rm m}+P_{\rm m}}{2M^2} + H^2\left[
                               1 + \alpha_{\rm B}(1+\alpha_{\rm M}) + \alpha_{\rm T} - 
                               (1+\alpha_{\rm H})(1+\alpha_{\rm M})\right] + 
                               [H(\alpha_{\rm B}-\alpha_{\rm H})]^{\hbox{$\cdot$}}\right\}\,, \nonumber\\ 
  E^{(4)}_{\Psi} & = -2\bmfi + 16H\bar{\lambda} 
                   = 0\,,\\
  E^{(4)}_{\pi} & = -2(\bmfs + 2H\bmfi) + 16H^2\bar{\lambda} 
                  = 0 \,.
 \end{align}
\end{subequations}
Finally, the functions $F_x$ are the coefficients of the generalised Poisson equation \\
$F_{\Phi}\Phi + F_{\dot{\Psi}}\dot{\Psi} + F_{\pi}\pi + F_{\dot{\pi}}\dot{\pi} + 
H^2{\rm K}^2(F^{(2)}_{\Psi}\Psi + F^{(2)}_{\pi}\pi) = \rho_{\rm m}\Delta_{\rm m}$,
\begin{subequations}
 \begin{align}
  F_{\Phi} & = -3\MM^2H\dot{f} + 2c + 4M_2^4 + 3Hm_3^3 
             = M^2H^2(\alpha_{\rm K}-6\alpha_{\rm B})\,, \\
  F_{\dot{\Psi}} & = -3\MM^2\dot{f} + 3m_3^3 
                   = -6HM^2\alpha_{\rm B}\,,\\
  F_{\pi} & = 6\MM^2H^2\dot{f} - 6Hc - \dot{c} - \dot{\Lambda} + 3m_3^3\dot{H} 
         \, = -3\dot{H}\left(\MM^2\dot{f} - m_3^3\right) 
            = -6H\dot{H}M^2\alpha_{\rm B}\,, \\
  F_{\dot{\pi}} & = -(2c + 4M_2^4) 
                  = -H^2M^2\alpha_{\rm K}\,, \\
  F^{(2)}_{\Psi} & = -(2\MM^2f + 4\tmfs) 
                   = -2M^2(1+\alpha_{\rm H})\,, \\
  F^{(2)}_{\pi} & = \MM^2\dot{f} - m_3^3 + 4H\mfs - 4H\tmfs 
                  = 2HM^2(\alpha_{\rm B}-\alpha_{\rm H})\,.
 \end{align}
\end{subequations}

We now consider the effective gravitational constant and the slip. The effective gravitational constant $\mu$ is
\begin{equation}
 4\pi G_{\rm eff}(k) = \frac{a_{-2}(k/a)^{-2} + a_0 + a_2(k/a)^2 + a_4(k/a)^4}
                            {b_{-2}(k/a)^{-2} + b_0 + b_2(k/a)^2}\,,
\end{equation}
where, for $\alpha_{\rm H}=0$,
\begin{subequations}
 \begin{align}
  a_{-2} & =  D^{(2)}_{\Psi}E_{\pi}
           = 6(1+\alpha_{\rm T})M^4\left\{\left(\dot{H} + \frac{\rho_{\rm m}+P_{\rm m}}{2M^2}\right)\dot{H} + 
                         \dot{H}\alpha_{\rm B}\left[H^2(3+\alpha_{\rm M}) + \dot{H}\right] + 
                         H(\dot{H}\alpha_{\rm B})^{\hbox{$\cdot$}}\right\}\,, \\
  a_0 & = D^{(2)}_{\Psi}E^{(2)}_{\pi} - D^{(2)}_{\pi}E^{(2)}_{\Psi} + D^{(4)}_{\Psi}E_{\pi} 
        = H^2M^4\alpha c_{\rm s}^2\bar{M}^2\mu_{Y,\infty} \,, \\
  a_2 & = D^{(2)}_{\Psi}E^{(4)}_{\pi} - D^{(4)}_{\pi}E^{(2)}_{\Psi} - D^{(2)}_{\pi}E^{(4)}_{\Psi} + 
          D^{(4)}_{\Psi}E^{(2)}_{\pi} 
        = 0 \,, \\
  a_4 & = -D^{(4)}_{\pi}E^{(4)}_{\Psi} + D^{(4)}_{\Psi}E^{(4)}_{\pi} 
        = 0 \;, \\
  b_{-2} & = D^{(2)}_{\Phi}E_{\pi}F^{(2)}_{\Psi} 
           = 12M^6\left\{\left(\dot{H} + \frac{\rho_{\rm m}+P_{\rm m}}{2M^2}\right)\dot{H} +
                         \dot{H}\alpha_{\rm B}\left[H^2(3+\alpha_{\rm M}) + \dot{H}\right] +
                         H(\dot{H}\alpha_{\rm B})^{\hbox{$\cdot$}}\right\}\,, \\
  b_0 & = D^{(2)}_{\Psi}E^{(2)}_{\Phi}F^{(2)}_{\pi} - D^{(2)}_{\Phi}E^{(2)}_{\Psi}F^{(2)}_{\pi} - 
          D^{(2)}_{\pi}E^{(2)}_{\Phi}F^{(2)}_{\Psi} + D^{(2)}_{\Phi}E^{(2)}_{\pi}F^{(2)}_{\Psi} 
        = 2H^2M^6\alpha c_{\rm s}^2\,, \\
  b_2 & = -D^{(2)}_{\Phi}E^{(4)}_{\Psi}F^{(2)}_{\pi} - D^{(4)}_{\pi}E^{(2)}_{\Phi}F^{(2)}_{\Psi} + 
           D^{(2)}_{\Phi}E^{(4)}_{\pi}F^{(2)}_{\Psi} + D^{(4)}_{\Psi}E^{(2)}_{\Phi}F^{(2)}_{\pi} 
        = 0 \,.
 \end{align}
\end{subequations}

The slip $\eta \equiv \Psi/\Phi$ is given by
\begin{equation}
 \eta = \frac{ c_{-2} (k/a)^{-2} + c_0  + c_2 (k/a)^2 }{ a_{-2} (k/a)^{-2} + a_0  + a_2 (k/a)^2 + a_4 (k/a)^4} \, ,
\end{equation}
with
\begin{subequations}
 \begin{align}
  c_{-2} & = -D^{(2)}_{\Phi}E_{\pi}
           = 6M^4\left\{\left(\dot{H} + \frac{\rho_{\rm m}+P_{\rm m}}{2M^2}\right)\dot{H} + 
                          \dot{H}\alpha_{\rm B}\left[H^2(3+\alpha_{\rm M}) + \dot{H}\right] + 
                          H(\dot{H}\alpha_{\rm B})^{\hbox{$\cdot$}}\right\}\,, \\
  c_0 & = D^{(2)}_{\pi}E^{(2)}_{\Phi} - D^{(2)}_{\Phi}E^{(2)}_{\pi}
        = H^2M^4\alpha c_{\rm s}^2\bar{M}^2\mu_{Z,\infty} \,, \\
  c_2 & = D^{(4)}_{\pi}E^{(2)}_{\Phi} - D^{(2)}_{\Phi}E^{(4)}_{\pi} 
        = 0 \,.
 \end{align}
\end{subequations}

The expressions for $G_{\rm eff}$ and $\eta$ generalise those given in \cite{DeFelice2011} in absence of higher derivative operators, in which case $a_2 = a_4 = b_2 = c_2 = 0$. When also $a_{-2}=b_{-2}=c_{-2}=0$ we recover the results of \cite{Gubitosi2013}. 
These expressions coincide with the expressions derived from the field equations, after some simple simplifications.

\section{De Felice et~al., 2011; Ref.~\texorpdfstring{\citep{DeFelice2011}}{DeFelice2011}}\label{sect:DeFelice2011}
In this section we convert the expressions for the perturbations in generic Horndeski models written in terms of the $G_i$ functions to expressions written in terms of the $\alpha_{\rm X}$ functions. For simplicity, we will not write again the full expressions in terms of the $G_i$ functions, but we refer to \cite{DeFelice2011} for them.

The first thing to take into account is that in \cite{DeFelice2011} a different definition is used for the metric potentials ($\Psi\rightarrow\Phi$, $\Phi\rightarrow-\Psi$, $\chi\rightarrow-\chi$). Secondly, the authors use the perturbed scalar field as degree of freedom. As discussed in the main manuscript, the following relations between the perturbed scalar field $\delta\phi$ and $\pi$ hold:
\begin{align*}
 \pi & = \frac{\delta\phi}{\dot{\phi}}\,, \quad & 
 \dot{\pi} & = \frac{\dot{\delta\phi}}{\dot{\phi}} - 
               \frac{\ddot{\phi}}{\dot{\phi}}\frac{\delta\phi}{\dot{\phi}}\,, \quad & 
 \ddot{\pi} & = \frac{\ddot{\delta\phi}}{\dot{\phi}} - 
                2\frac{\ddot{\phi}}{\dot{\phi}}\frac{\dot{\delta\phi}}{\dot{\phi}} + 
                \left[2\left(\frac{\ddot{\phi}}{\dot{\phi}}\right)^2-\frac{\dddot{\phi}}{\dot{\phi}}\right]
                \frac{\delta\phi}{\dot{\phi}}\,,\\
 \delta\phi & = \dot{\phi}\pi\,, \quad & 
 \dot{\delta\phi} & = \ddot{\phi}\pi + \dot{\phi}\dot{\pi}\,, \quad & 
 \ddot{\delta\phi} & = \dot{\phi}\ddot{\pi} + 2\ddot{\phi}\dot{\pi} + \dddot{\phi}\pi\,.
\end{align*}

The background quantities are defined as
\begin{equation*}
 \frac{1}{a^3}\frac{\mathrm{d}}{\mathrm{d}t}(a^3J) = P_{\phi}\,, \quad 
 \mathcal{E} \equiv \sum_{i=2}^5 \mathcal{E}_i = \rho_{\rm ds} - 3H^2M_{\rm pl}^2 = -\rho_{\rm m}\,, \quad 
 \mathcal{P} \equiv \sum_{i=2}^5 \mathcal{P}_i = P_{\rm ds} + \left(3H^2+2\dot{H}\right)M_{\rm pl}^2 = -P_{\rm m}\,,
\end{equation*}
and their explicit expressions can be found in \cite{DeFelice2011,Pace2019a}.

One defines
\begin{equation}
 \mathcal{F}_{\rm T} \equiv M^2(1+\alpha_{\rm T})\,, \quad 
 \mathcal{G}_{\rm T} \equiv M^2 = \frac{1}{2}\frac{\partial\mathcal{P}}{\partial\dot{H}}\,,
\end{equation}
such that $c_{\rm T}^2 \equiv 1+\alpha_{\rm T} = \mathcal{F}_{\rm T}/\mathcal{G}_{\rm T}$. Moreover,
\begin{align}
 \Theta & \equiv -\frac{1}{6}\frac{\partial\mathcal{E}}{\partial H} 
          = M^2H(1+\alpha_{\rm B})\,,\\
 \Sigma & \equiv X\frac{\partial\mathcal{E}}{\partial X} + \frac{1}{2}H\frac{\partial\mathcal{E}}{\partial H}
          = \frac{1}{2}M^2H^2(\alpha_{\rm K}-12\alpha_{\rm B}-6)\,,
\end{align}
and
\begin{align}
 \mathcal{F}_{\rm S} & \equiv \frac{1}{a}\frac{\mathrm{d}}{\mathrm{d}t}
                              \left(a\frac{\mathcal{G}_{\rm T}^2}{\Theta}\right) - \mathcal{F}_{\rm T} 
                       = -M^2\frac{(1+\alpha_{\rm B})[\dot{H}-(\alpha_{\rm M}-\alpha_{\rm T})H^2+
                                   \alpha_{\rm B}(1+\alpha_{\rm T})H^2]+H\dot{\alpha}_{\rm B}}
                                  {H^2(1+\alpha_{\rm B})^2}\,,\\
 \mathcal{G}_{\rm S} & \equiv \Sigma\frac{\mathcal{G}_{\rm T}^2}{\Theta^2}+3\mathcal{G}_{\rm T}
                       = \frac{1}{2}M^2\frac{\alpha}{(1+\alpha_{\rm B})^2}\,,
\end{align}
which allow to write the sound speed for perturbations as
\begin{equation}
 c_{\rm s}^2 = \frac{\mathcal{F}_{\rm S}}{\mathcal{G}_{\rm S}} - \frac{\rho_{\rm m}+P_{\rm m}}{H^2M^2\alpha}\,.
\end{equation}

Taking into account the differences for the variables appearing in the perturbed metric, the coefficients of the $00$-component of the perturbed field equations are
\begin{subequations}
 \begin{align}
  A_{1} & = 6\Theta 
          = 6HM^2(1+\alpha_{\rm B})\,, \\
  A_{2} & = -2(\Sigma+3H\Theta)/\dot{\phi} 
          = -M^2H^2(\alpha_{\rm K}-6\alpha_{\rm B})/\dot{\phi} \,, \\
  A_{3} & = 2\mathcal{G}_{\rm T} 
          = 2M^2 \,, \\
  A_{4} & = 2\Sigma + \rho_{\rm m} + P_{\rm m} 
          = -M^2H^2(6-\alpha_{\rm K}+12\alpha_{\rm B}) + \rho_{\rm m} + P_{\rm m}\,, \\ 
  A_{5} & = -2\Theta 
          = -2HM^2(1+\alpha_{\rm B})\,, \\
  A_{6} & = 2(\Theta-H\mathcal{G}_{\rm T})/\dot{\phi} 
          = 2HM^2\alpha_{\rm B}/\dot{\phi}\,;
 \end{align}
\end{subequations}
the coefficients of the trace of the $ij$-component of the perturbed field equations (more precisely three times the trace minus $2k^2/a^2$ the traceless part, using Eqs.~(A.1) from the appendix in the main manuscript)
\begin{subequations}
 \begin{align}
  B_{1} & = 6\mathcal{G}_{\rm T} 
          = 6M^2\,, \\
  B_{2} & = 6(\Theta-H\mathcal{G}_{\rm T})/\dot{\phi} 
          = 6HM^2\alpha_{\rm B}/\dot{\phi}\,, \\
  B_{3} & = 6(\dot{\mathcal{G}}_{\rm T}+3H\mathcal{G}_{\rm T}) 
          = 6HM^2(3+\alpha_{\rm M})\,, \\
  B_{4} & = 3\left[\left(4H\ddot{\phi}-4\dot{H}\dot{\phi}-6H^{2}\dot{\phi}\right)\mathcal{G}_{\rm T} - 
                   2H\dot{\phi}\dot{\mathcal{G}}_{\rm T} - 
                   \left(4\ddot{\phi}-6H\dot{\phi}\right)\Theta + 2\dot{\phi}\dot{\Theta} - 
                   (\rho_{\rm m}+P_{\rm m})\dot{\phi}\right]/\dot{\phi}^{2}\,, \\
        & = -6M^2\left\{2H\alpha_{\rm B}\ddot{\phi} + \left[
                        \dot{H} + \frac{\rho_{\rm m} + P_{\rm m}}{2 M^2} - (\alpha_{\rm B} H)^{\hbox{$\cdot$}} - 
                        (3+\alpha_{\rm M})\alpha_{\rm B}H^2\right]\dot{\phi}\right\}/\dot{\phi}^{2} \,, \nonumber\\
  B_{5} & = -6\Theta 
          = -6HM^2(1+\alpha_{\rm B})\,, \\
  B_{6} & = 2\mathcal{F}_{\rm T} 
          = 2M^2(1+\alpha_{\rm T})\,, \\
  B_{7} & = 2\left[\dot{\mathcal{G}}_{\rm T}+H\left(\mathcal{G}_{\rm T}-\mathcal{F}_{\rm T}\right)\right]/\dot{\phi} 
          = 2HM^2(\alpha_{\rm M}-\alpha_{\rm T})/\dot{\phi}\,, \\
  B_{8} & = 2\mathcal{G}_{\rm T} 
          = 2M^2\,, \\
  B_{9} & = -6(\dot{\Theta}+3H\Theta) 
          = -6M^2\left[\dot{H} + (\alpha_{\rm B} H)^{\hbox{$\cdot$}} + 
                       (3+\alpha_{\rm M})(1+\alpha_{\rm B})H^2\right] \,, \\
  B_{10} & = -2\mathcal{G}_{\rm T} 
           = -2M^2\,, \\ 
  B_{11} & = -2(\dot{\mathcal{G}}_{\rm T}+H\mathcal{G}_{\rm T}) 
           = -2HM^2(1+\alpha_{\rm M})\,;
 \end{align}
\end{subequations}
the coefficients of the $0i$-component of the perturbed Einstein field equations are
\begin{subequations}
 \begin{align}
  C_{1} & = 2\mathcal{G}_{\rm T} 
          = 2M^2 \,, \\ 
  C_{2} & = 2(\Theta-H\mathcal{G}_{\rm T})/\dot{\phi} 
          = 2HM^2\alpha_{\rm B}/\dot{\phi}\,, \\
  C_{3} & = -2\Theta 
          = -2HM^2(1+\alpha_{\rm B})\,, \\ 
  C_{4} & = \left[2(H\ddot{\phi}-\dot{H}\dot{\phi})\mathcal{G}_{\rm T} - 2\ddot{\phi}\Theta - 
                  (\rho_{m}+P_{\rm m})\dot{\phi}\right]/\dot{\phi}^{2} \,,\\
        & = -2M^2\left[H\alpha_{\rm B}\ddot{\phi} + 
                       \left(\dot{H} + \frac{\rho_{\rm m}+P_{\rm m}}{2M^2}\right)\dot{\phi}\right]/\dot{\phi}^2 \,;
            \nonumber
 \end{align}
\end{subequations}
the coefficients of the equation of motion of the perturbed scalar field are
\begin{subequations}
\begin{align}
 D_{1} & = 6(\Theta-H\mathcal{G}_{\rm T})/\dot{\phi} 
         = 6HM^2\alpha_{\rm B}/\dot{\phi}\,,\\
 D_{2} & = 2\left(3H^{2}\mathcal{G}_{\rm T}-6H\Theta-\Sigma\right)/\dot{\phi}^{2}
         = -H^2M^2\alpha_{\rm K}/\dot{\phi}^{2}\,, \\
 D_{3} & = -3\left[2H\left(\dot{\mathcal{G}}_{\rm T}+3H\mathcal{G}_{\rm T}\right) - 
                   2\left(\dot{\Theta}+3H\Theta\right) - \left(\rho_{\rm m}+P_{\rm m}\right)\right]/\dot{\phi}\,, \\
       & = 6M^2\left[\dot{H} + \frac{\rho_{\rm m}+P_{\rm m}}{2M^2} + H^2\alpha_{\rm B}(3+\alpha_{\rm M}) + 
                     (\alpha_{\rm B}H)^{\hbox{$\cdot$}}\right]/\dot{\phi} \,, \nonumber\\
 D_{4} & = 2\left\{3H\left[\left(3H^{2}+2\dot{H}\right)\dot{\phi}-2H\ddot{\phi}\right]\mathcal{G}_{\rm T} + 
                           3H^{2}\dot{\phi}\dot{\mathcal{G}}_{\rm T} + 
                           6\left[2H\ddot{\phi} - \left(3H^{2}+\dot{H}\right)\dot{\phi}\right]\Theta - 
                           6H\dot{\phi}\dot{\Theta} + \right.\nonumber\\
       & \qquad  \left.\left(2\ddot{\phi}-3H\dot{\phi}\right)\Sigma - 
                   \dot{\phi}\dot{\Sigma}\right\}/\dot{\phi}^{3}\,, \\
       & = HM^2\left\{2H\alpha_{\rm K}\ddot{\phi} - 
                      \left[H^2(3+\alpha_{\rm M})+\dot{H}\right]\alpha_{\rm K}\dot{\phi} - 
                      \dot{\phi}(H\alpha_{\rm K})^{\hbox{$\cdot$}}\right\}/\dot{\phi}^{3} \,,\nonumber\\
 D_{5} & = 2(\Sigma+3H\Theta)/\dot{\phi} 
         = H^2M^2(\alpha_{\rm K}-6\alpha_{\rm B})/\dot{\phi}\,, \\
 D_{6} & = -2(\Theta-H\mathcal{G}_{\rm T})/\dot{\phi} 
         = -2HM^2\alpha_{\rm B}/\dot{\phi}\,,\\
 D_{7} & = 2\left[\dot{\mathcal{G}}_{\rm T}+H\left(\mathcal{G}_{\rm T}-\mathcal{F}_{\rm T}\right)\right]/\dot{\phi} 
         = 2HM^2(\alpha_{\rm M}-\alpha_{\rm T})/\dot{\phi}\,, \\
 D_{8} & = 3\left[6\left(\dot{H}\dot{\phi}-H\ddot{\phi}\right)\Theta - 2\ddot{\phi}\Sigma + 
                  3H(\rho_{\rm m}+P_{\rm m})\dot{\phi} - \mu\dot{\phi}^{2}\right]/\dot{\phi}^{2}\,, \\
       & = 3HM^2\left\{6\dot{\phi}\left[\dot{H}(1+\alpha_{\rm B})+\frac{\rho_{\rm m}+P_{\rm m}}{2M^2}\right] - 
                      H\ddot{\phi}(\alpha_{\rm K}-6\alpha_{\rm B})\right\}/\dot{\phi}^{2} - 3\mu\,, \nonumber\\
       & = 0\,,\nonumber\\
 D_{9} & = \left[2H^{2}\mathcal{F}_{\rm T} - 4H\left(\dot{\mathcal{G}}_{\rm T} + H\mathcal{G}_{\rm T}\right) + 
 2\left(\dot{\Theta}+H\Theta\right) + (\rho_{\rm m}+P_{\rm m})\right]/\dot{\phi}^{2}\,, \\
       & = 2M^2\left\{\dot{H} + \frac{\rho_{\rm m}+P_{\rm m}}{2M^2} + 
                      H^2\left[\alpha_{\rm B}(1+\alpha_{\rm M}) + \alpha_{\rm T} - \alpha_{\rm M}\right] + 
                      (H\alpha_{\rm B})^{\hbox{$\cdot$}}\right\}/\dot{\phi}^{2}\,, \nonumber\\
 D_{10} & = 2(\Theta-H\mathcal{G}_{\rm T})/\dot{\phi} 
          = 2HM^2\alpha_{\rm B}/\dot{\phi}\,, \\
 D_{11} & = \left\{6\left[\left(3H^{2}+\dot{H}\right)\dot{\phi} - H\ddot{\phi}\right]\Theta + 6H\dot{\phi}\dot{\Theta} + 2(3H\dot{\phi}-\ddot{\phi})\Sigma + 
   2\dot{\phi}\dot{\Sigma} - \mu\dot{\phi}^{2}\right\}/\dot{\phi}^{2}\,, \\
        & = M^2\left\{\dot{\phi}H\left[(\alpha_{\rm K}-6\alpha_{\rm B})(3+\alpha_{\rm M})H^2     + 2\dot{H}(\alpha_{\rm K}-6\alpha_{\rm B}) + 
              H(\dot{\alpha}_{\rm K}-6\dot{\alpha}_{\rm B})\right] - 
              \ddot{\phi}H^2(\alpha_{\rm K}-6\alpha_{\rm B})\right\}/\dot{\phi}^{2} - \mu\,, \nonumber\\
        & = -M^2\left[6\left(\dot{H}+\frac{\rho_{\rm m}+P_{\rm m}}{2M^2}\right) + 
                      H^2(6\alpha_{\rm B}-\alpha_{\rm K})(3+\alpha_{\rm M}) + 
                      2(9\alpha_{\rm B}-\alpha_{\rm K})\dot{H} + 
                      H(6\dot{\alpha}_{\rm B}-\dot{\alpha}_{\rm K})\right]H/\dot{\phi}\,, \nonumber\\
 D_{12} & = \left[2H\left(\dot{\mathcal{G}}_{\rm T}+H\mathcal{G}_{\rm T}\right) - 2\left(\dot{\Theta}+H\Theta\right) - 
                  (\rho_{\rm m}+P_{\rm m})\right]/\dot{\phi}\,, \\
        & = -2M^2\left\{\dot{H} + \frac{\rho_{\rm m}+P_{\rm m}}{2M^2} + 
              H^2\alpha_{\rm B}(1+\alpha_{\rm M}) + (H\alpha_{\rm B})^{\hbox{$\cdot$}}
                 \right\}/\dot{\phi}\,.\nonumber
\end{align}
\end{subequations}

We finally have (we renamed the mass of the scalar field as $M_{\delta\phi}^2$ to avoid confusion with the effective Planck mass $M^2$)
\begin{subequations}
\begin{align}
 \mu & = {\cal E}_{,\phi} 
      = -\left[2\ddot{\phi}\Sigma+6\Theta\left(H\ddot{\phi}-\dot{H}\dot{\phi}\right)+
                3H\dot{\phi}\left(\mathcal{E}+\mathcal{P}\right)\right]/\dot{\phi}^2\\
     & = -M^2H\left\{H(\alpha_{\rm K}-6\alpha_{\rm B})\ddot{\phi}-
                     6\left[(1+\alpha_{\rm B})\dot{H}+\frac{\rho_{\rm m}+P_{\rm m}}{2M^2}\right]\dot{\phi}
              \right\}/\dot{\phi}^2\,,\nonumber\\
 \nu & = {\cal P}_{,\phi} 
       = -2\left[
           \ddot{H}\mathcal{G}_{\rm T}+\dot{H}\left(\dot{\mathcal{G}}_{\rm T}+3H\mathcal{G}_{\rm T}\right)
           \right]/\dot{\phi} - 2\left(\Theta-H\mathcal{G}_{\rm T}\right)\dddot{\phi}/\dot{\phi}^2 - \\
      &\qquad\qquad
      \left[\left(4H\ddot{\phi}-4\dot{H}\dot{\phi}-6H^{2}\dot{\phi}\right)\mathcal{G}_{\rm T} - 2H\dot{\phi}\dot{\mathcal{G}}_{\rm T} - 
           \left(4\ddot{\phi}-6H\dot{\phi}\right)\Theta + 
                    2\dot{\phi}\dot{\Theta}\right]\ddot{\phi}/\dot{\phi}^{3} - \nonumber\\
      &\qquad\qquad \left(\mathcal{E}+\mathcal{P}\right)\ddot{\phi}/\dot{\phi}^{2} + 
                 \dot{\mathcal{P}}/\dot{\phi}\,,\nonumber\\
     & = -2HM^2\alpha_{\rm B}\left(\dot{\phi}\dddot{\phi}-2\ddot{\phi}^2\right)/\dot{\phi}^3 + 2M^2\left[\dot{H} + \frac{\rho_{\rm m} + P_{\rm m}}{2 M^2} - (\alpha_{\rm B} H)^{\hbox{$\cdot$}} - 
       (3+\alpha_{\rm M})\alpha_{\rm B}H^2\right]\ddot{\phi}/\dot{\phi}^{2} \nonumber\\
     &\quad~    -2M^2\left[(3+\alpha_{\rm M})H \dot H+\frac{\dot{P}_{\rm m}}{2 M^2} + \ddot{H}\right]/\dot{\phi}\,, \nonumber\\
 M_{\delta\phi}^{2} & = \left[\dot{\mu}+3H(\mu+\nu)\right]/\dot{\phi} \\
                    & = 6M^2
                        \left\{\left(\dot{H} + \frac{\rho_{\rm m} + P_{\rm m}}{2 M^2}\right)\dot{H} + 
                               \alpha_{\rm B}\dot{H}\left[(3+\alpha_{\rm M})H^2+\dot{H}\right]+
                               H(\alpha_{\rm B}\dot{H})^{\hbox{$\cdot$}}\right\}/\dot{\phi}^2 \nonumber\\
                &\quad -\alpha_{\rm K}H^2M^2\dddot{\phi}/\dot{\phi}^3 + 
                        M^2H\left\{2\alpha_{\rm K}H\ddot{\phi} - 
                        \alpha_{\rm K}\left[(3+\alpha_{\rm M})H^2+\dot{H}\right]\dot{\phi}-
                        \dot{\phi}(\alpha_{\rm K}H)^{\hbox{$\cdot$}}
                        \right\}\ddot{\phi}/\dot{\phi}^4\,.\nonumber
\end{align}
\end{subequations}

Applying the QSA approximation, the field equations reduce to
\begin{align}
 & B_8H^2{\rm K}^2Z - A_6H^2{\rm K}^2\delta\phi = -\bar{\rho}_{\rm m}\Delta_{\rm m}\,,\\
 & B_6Z - B_7\delta\phi - B_8Y = 0\,,\\
 & B_7H^2{\rm K}^2Z - \left(D_9H^2{\rm K}^2-M_{\delta\phi}^2\right)\delta\phi - A_6H^2{\rm K}^2Y = 0\,,
\end{align}
and, as explained in the main manuscript, we have $\mu_Z = 2[\mathcal{M}^{-1}]_{12}H^2M_{\rm pl}^2{\rm K}^2$ and $\mu = 2[\mathcal{M}^{-1}]_{11}H^2M_{\rm pl}^2{\rm K}^2$, where $\mathcal{M}$ is the matrix of the coefficients.

The inverse matrix coefficients are
\begin{align}
 \left[\mathcal{M}^{-1}\right]_{11} = & \frac{B_6M_{\delta\phi}^2+\left(B_7^2-B_6D_9\right)H^2{\rm K}^2}
                                             {B_8^2M_{\delta\phi}^2H^2{\rm K}^2+
                                              (2A_6B_7B_8 - A_6^2B_6 - B_8^2D_9)H^4{\rm K}^4}\,,\\
 \left[\mathcal{M}^{-1}\right]_{12} = & \frac{B_8M_{\delta\phi}^2+\left(A_6B_7-B_8D_9\right)H^2{\rm K}^2}
                                             {B_8^2M_{\delta\phi}^2H^2{\rm K}^2+
                                              (2A_6B_7B_8 - A_6^2B_6 - B_8^2D_9)H^4{\rm K}^4}\,,\\
 \left[\mathcal{M}^{-1}\right]_{13} = & \frac{A_6B_6-B_7B_8}
                                             {B_8^2M_{\delta\phi}^2+
                                              (2A_6B_7B_8 - A_6^2B_6 - B_8^2D_9)H^2{\rm K}^2}\,,
\end{align}
and the previous set of coefficients read
\begin{align}
 & \left(B_7^2-B_6D_9\right)H^2 = 2H^4M^4 \alpha c_{\rm s}^2\bar{M}^2\mu_{Y,\infty}/\dot{\phi}^2\,,\\
 & B_6M_{\delta\phi}^2 = 2M^2(1+\alpha_{\rm T})M_{\delta\phi}^2\,,\\
 & B_8^2M_{\delta\phi}^2H^2 = 4H^2M^4M_{\delta\phi}^2\,,\\
 & (2A_6B_7B_8 - A_6^2B_6 - B_8^2D_9)H^4 = 4H^6M^6\alpha c_{\rm s}^2/\dot{\phi}^2\,,\\
 & \left(A_6B_7-B_8D_9\right)H^2  = 2H^4M^4 \alpha c_{\rm s}^2\bar{M}^2\mu_{Z,\infty}/\dot{\phi}^2\,,\\
 & B_8M_{\delta\phi}^2 = 2M^2M_{\delta\phi}^2\,,\\
 & A_6B_6-B_7B_8 = 4HM^4[\alpha_{\rm B}(1+\alpha_{\rm T})+\alpha_{\rm T}-\alpha_{\rm M}]/\dot{\phi} \,.
\end{align}

Following the notation of \cite{Pogosian2016}, we define
\begin{align*}
 & f_1 = 2H^4\left(B_7^2-B_6D_9\right)\,, \quad 
   f_2 = 2H^2B_6M_{\delta\phi}^2\,, \quad 
   f_3 = (2A_6B_7B_8 - A_6^2B_6 - B_8^2D_9)H^4\,, \\
 & f_4 = B_8^2M_{\delta\phi}^2H^2\,, \quad 
   f_5 = 2H^4\left(A_6B_7-B_8D_9\right)\,, \quad 
   f_6 = 2H^2B_8M_{\delta\phi}^2\,, \quad 
   f_7 = 3H^2(A_6B_6-B_7B_8)\,,
\end{align*}
and using the following dimensionless coefficients
\begin{align*}
 & \tilde{f}_1 = \alpha c_{\rm s}^2\bar{M}^2\mu_{Y,\infty}\,, && 
   \tilde{f}_5 = \alpha c_{\rm s}^2\bar{M}^2\mu_{Z,\infty}\,, && 
   \tilde{f}_3 = \alpha c_{\rm s}^2\,,\\
 & \tilde{f}_2 = (1+\alpha_{\rm T})\frac{\dot{\phi}^2M_{\delta\phi}^2}{H^4M^2}\,, && 
   \tilde{f}_4 = \frac{\dot{\phi}^2M_{\delta\phi}^2}{H^4M^2}\,, && 
   \tilde{f}_6 = \frac{\dot{\phi}^2M_{\delta\phi}^2}{H^4M^2}\,,\\
 & \tilde{f}_7 = 3[\alpha_{\rm B}(1+\alpha_{\rm T})+\alpha_{\rm T}-\alpha_{\rm M}]\,,
\end{align*}
the modified gravity functions and the relation between the perturbed scalar field and matter density perturbations read
\begin{align*}
 \mu_Z \equiv \frac{\tilde{f}_6+\tilde{f}_5{\rm K}^2}{\tilde{f}_4+\tilde{f}_3{\rm K}^2}\frac{1}{\bar{M}^2}\,, \quad 
 \mu \equiv \frac{\tilde{f}_2+\tilde{f}_1{\rm K}^2}{\tilde{f}_4+\tilde{f}_3{\rm K}^2}\frac{1}{\bar{M}^2}\,, \quad
 \eta \equiv \frac{\tilde{f}_6+\tilde{f}_5{\rm K}^2}{\tilde{f}_2+\tilde{f}_1{\rm K}^2}\,, \quad 
 H\frac{\delta\phi}{\dot{\phi}} \equiv H\pi = 
                                \frac{\tilde{f}_7}{\tilde{f}_4+\tilde{f}_3{\rm K}^2}\frac{1}{\bar{M}^2}
                                \Omega_{\rm m}\Delta_{\rm m}\,.
\end{align*}

Also these expressions are in agreement with what was found in previous works and in ours when using the field equations. We verified that under the assumption that $\alpha_{\rm T}=0$, these expressions coincide with those of \cite{Arjona2019b}. Note though, that even if the limits for ${\rm K}\rightarrow 0$ coincide with what generally found in the literature, the mass of the scalar field differs from ours and previous works, due to the use of $\delta\phi$ rather than $\pi$. We verified that this choice does not affect $f(R)$ models.

\section{Arjona, Cardona \& Nesseris, 2019; Ref.~\texorpdfstring{\citep{Arjona2019b}}{Arjona2019b}}
\label{sect:Arjona2019}
In \cite{Arjona2019b}, the authors apply an effective fluid approach, using exactly the same notation of \cite{DeFelice2011} which we also will follow in this section, under the basic idea that dark sector quantities ($\delta P_{\rm de}$, $\delta_{\rm de}$ and $\Theta_{\rm de}$) can be written in terms of the potentials, of the scalar field perturbations and their derivatives. To the resulting equations one applies the quasi-static approximation.

With respect to \cite{Arjona2019b}, we will consider a generic Horndeski model, taking, therefore, into account also the $G_5$ Lagrangian and we will express the resulting expressions in terms of the coefficients $A_i$, $B_i$, $C_i$ and $D_i$ presented in the previous section. We will limit ourselves to the case where the perturbations of the scalar field are given by $\delta\phi$, rather than $\pi$.

The idea is that the field equations can be written in the standard form where the right-hand-side is given by matter and dark sector perturbations. Therefore, subtracting from the equations expressed in terms of the $A_i$, $B_i$, $C_i$ and $D_i$ the general relativistic result, one obtains the expressions for the dark sector perturbations in terms of the potential, the scalar field perturbations and their derivatives. From the traceless part of the $ij$-component of the field equations, one expresses the scalar field perturbations in terms of the two potentials and can evaluate its first and second time derivative. Plugging them in the expressions for the dark sector perturbations, one can apply the QSA and neglect time derivatives and express the perturbations in terms of the potentials. Ultimately, the potentials can be related to matter perturbations as also explained in the main manuscript and the final result is that dark sector perturbations are expressed in terms of matter perturbations.

For completeness, we report here the relevant equations needed to infer the final expressions. From the traceless part of the $ij$-component
\begin{equation}\label{eqn:deltaphi}
 B_6\Phi + B_8\Psi + B_7\delta\phi = 0\,,
\end{equation}
one expresses $\delta\phi$ in terms of the two potentials $\Phi$ and $\Psi$ and evaluates its first and second time derivatives. Applying the QSA to the field equations and using the relation between the scalar field perturbations and the potentials, we find
\begin{align}
 \Psi = &\, \frac{\left(B_7^2-B_6D_9\right)H^2{\rm K}^2+B_6M_{\delta}^2}
                 {\left(A_6^2B_6-2A_6B_7B_8+B_8^2D_9\right)H^4{\rm K}^4-B_8M_{\delta}^2H^2{\rm K}^2}\delta\rho_{\rm m}\,,\\
 \Phi = &\, -\frac{\left(A_6B_7-B_8D_9\right)H^2{\rm K}^2+B_8 M_{\delta}^2}
                 {\left(A_6^2B_6-2A_6B_7B_8+B_8^2 D_9\right)H^4{\rm K}^4 - 
                 B_8M_{\delta}^2H^2{\rm K}^2}\delta\rho_{\rm m}\,,\\
 \delta\phi = &\, \frac{A_6B_6-B_7B_8}
                       {\left(A_6^2B_6-2A_6B_7 B_8+B_8 D_9\right)H^2{\rm K}^2 - B_8M_{\delta}^2}\delta\rho_{\rm m}\,.
\end{align}

The expressions for the dark sector density, velocity and pressure perturbations, the effective anisotropic stress, the sound speed and the effective sound speed are\footnote{We corrected a typo in the expression of $c_{\rm s,de}^2$, as in \cite{Arjona2019b} the coefficient $\mathcal{F}_4$ is missing.}
\begin{align}
 \frac{\delta P_{\rm de}}{\rho_{\rm de}} & \simeq 
      \frac{1}{3\mathcal{F}_4}
      \frac{\mathcal{F}_3 + \mathcal{F}_2H^2{\rm K}^2 + \mathcal{F}_1H^4{\rm K}^4}
           {\mathcal{F}_6H^2{\rm K}^2 + \mathcal{F}_5H^4{\rm K}^4}
      \frac{\rho_{\rm m}}{\rho_{\rm de}}\delta_{\rm m} \,,\\
 & \simeq \frac{1}{3\mathcal{F}_4}
          \frac{\mathcal{F}_3 + \mathcal{F}_2H^2{\rm K}^2 + \mathcal{F}_1H^4{\rm K}^4}
               {\mathcal{F}_9 + \mathcal{F}_8H^2{\rm K}^2 + \mathcal{F}_7H^4{\rm K}^4}
          \,\delta_{\rm de}\,, \nonumber\\
 \delta_{\rm de} & \simeq 
        \frac{\mathcal{F}_9 + \mathcal{F}_8H^2{\rm K}^2 + \mathcal{F}_7H^4{\rm K}^4}
              {\mathcal{F}_6H^2{\rm K}^2 + \mathcal{F}_5H^4{\rm K}^4}
        \frac{\rho_{\rm m}}{\rho_{\rm de}}\delta_{\rm m}\,,\\
 V_{\rm de} & \simeq 
   a\frac{\mathcal{F}_{11} + \mathcal{F}_{10}H^2{\rm K}^2}
         {\mathcal{F}_6 + \mathcal{F}_5H^2{\rm K}^2}
    \frac{\rho_{\rm m}}{\rho_{\rm de}}\delta_{\rm m} \,,\\
 \pi_{\rm de} & = M_{\rm pl}^2\frac{\frac{k^2}{a^2}(\Phi-\Psi)}{\rho_{\rm de}} 
  \simeq -\frac{1}{2\mathcal{F}_4}
          \frac{\mathcal{F}_{12}+\mathcal{F}_1H^2{\rm K}^2}
               {\mathcal{F}_6 + \mathcal{F}_5H^2{\rm K}^2}
               \frac{\rho_{\rm m}}{\rho_{\rm de}}\delta_{\rm m} \,,\\
 & \phantom{ = M_{\rm pl}^2\frac{\frac{k^2}{a^2}(\Phi-\Psi)}{\rho_{\rm de}}}\,\,
 \simeq -\frac{1}{2\mathcal{F}_4}
         \frac{\mathcal{F}_{12}H^2{\rm K}^2+\mathcal{F}_1H^4{\rm K}^4}
              {\mathcal{F}_9 + \mathcal{F}_8H^2{\rm K}^2 + \mathcal{F}_7H^4{\rm K}^4}
         \delta_{\rm de}\,, \nonumber\\
 c_{\rm s,de}^2 & \equiv \frac{\delta P_{\rm de}}{\delta\rho_{\rm de}} 
              = \frac{1}{3\mathcal{F}_4}
                \frac{\mathcal{F}_3 + \mathcal{F}_2H^2{\rm K}^2 + \mathcal{F}_1H^4{\rm K}^4}
                     {\mathcal{F}_9 + \mathcal{F}_8H^2{\rm K}^2 + \mathcal{F}_7H^4{\rm K}^4} \,,\\
 c_{\rm s,eff}^2 & \equiv c_{\rm s,de}^2 - \frac{2}{3}\frac{\pi_{\rm de}}{\delta_{\rm de}}
                    = \frac{1}{3\mathcal{F}_4}
                      \frac{\mathcal{F}_3 + \left(\mathcal{F}_2+\mathcal{F}_{12}\right)H^2{\rm K}^2 + 
                            2\mathcal{F}_1 H^4{\rm K}^4}
                           {\mathcal{F}_9 + \mathcal{F}_8H^2{\rm K}^2 + \mathcal{F}_7H^4{\rm K}^4}\,,
\end{align}

Note that we have the following equivalence for the velocity perturbations:
\begin{equation*}
 \underbrace{a(\rho_{\rm m}+P_{\rm m})V_{\rm m}/k^2}_{\text{\citep{Arjona2019b}}} = 
 \underbrace{-(\rho_{\rm m}+P_{\rm m})v_{\rm m}}_{\text{\cite{DeFelice2011}}} = 
 \underbrace{(\rho_{\rm m}+P_{\rm m})v_{\rm m}=q_{\rm m}}_{\text{\citep{Gleyzes2014}}} = \underbrace{-\rho_{\rm m}\Theta_{\rm m}/(3H)}_{\text{\citep{Pace2019a}}}\,.
\end{equation*}

The $\mathcal{F}_i$ coefficients are then:
\begin{subequations}
 \begin{align}
 \mathcal{F}_{1} = &\, 2M_{\rm pl}^2 B_7^2 \left[B_7 \left(A_6 - B_7\right) - D_9\left(B_8 - B_6\right)\right]\,,\\
 \mathcal{F}_{2} = &\, 2M_{\rm pl}^2 B_7^2 \left[3\left(3H^2+2\dot{H}\right) \left(B_7^2 - B_6 D_9\right) + M_{\delta}^2\left(B_8 - B_6\right)\right] + 3B_7^2\nu\left(A_6 B_6 - B_7 B_8\right) + \nonumber\\
 &\, \left(B_7^2 - B_6 D_9\right)\left(B_7^2 B_9 - B_4 B_7 \dot{B}_8 + 2 B_2 \dot{B}_7 \dot{B}_8 - B_2 B_7 \ddot{B_8}\right) - \nonumber\\
 &\, \left(A_6 B_7 - B_8 D_9\right)\left(2B_2 \dot{B}_6 \dot{B}_7 - B_4 B_7 \dot{B}_6 - B_2 B_7 \ddot{B}_6\right) - \nonumber\\
 &\, \left(A_6 B_6 - B_7 B_8\right) \left[B_7\left(B_2 \ddot{B}_7 + B_4 \dot{B}_7\right) - 2B_2 \dot{B}_7^2\right]\,,\\
 \mathcal{F}_{3} = &\, M_{\delta}^2 \left\{\left(B_4 B_7 - 2 B_2 \dot{B}_7\right) \left(B_8 \dot{B}_6 - B_6 \dot{B}_8\right) + B_2 B_7\left(B_8 \ddot{B}_6 - B_6 \ddot{B}_8\right) + \right.\nonumber\\ 
 &\, \qquad \left. B_6 B_7^2\left[B_9+6M_{\rm pl}^2\left(3H^2+2\dot{H}\right)\right]\right\} \,,\\
 \mathcal{F}_{4} = &\, B_7\,,\\
 \mathcal{F}_{5} = &\, B_7 \left(A_6^2 B_6 - 2 A_6 B_7 B_8 + B_8^2 D_9\right)\,,\\
 \mathcal{F}_{6} = &\, -B_7 B_8 M_{\delta}^2\,,\\
 \mathcal{F}_{7} = &\, B_7 \left[A_6 \left(A_6 B_6 - B_7 B_8\right) + \left(B_8 D_9 - A_6 B_7\right)\left(A_3 - 2M_{\rm pl}^2\right)\right]\,,\\
 \mathcal{F}_{8} = &\, \left(B_7^2 - B_6 D_9\right)\left(A_4 B_7 - A_2 \dot{B}_8 + 6 M_{\rm pl}^2H^2 B_7\right) - A_2\left(B_8 D_9 - A_6 B_7\right)\dot{B}_6 - \nonumber\\
 &\, \left(A_6 B_6 - B_7 B_8\right) \left(A_2 \dot{B}_7 + B_7\,\mu\right) - B_7 B_8 M_{\delta}^2\left(A_3 - 2M_{\rm pl}^2\right)\,,\\
 \mathcal{F}_{9} = &\, M_{\delta}^2\left\{A_2 B_8 \dot{B}_6 + B_6\left[B_7 \left(A_4 + 6M_{\rm pl}^2H^2\right) - A_2 \dot{B}_8\right]\right\}\,,\\
 \mathcal{F}_{10} = &\, B_7^3 \left(C_3 + 2M_{\rm pl}^2H\right) - B_7^2\left(B_8 C_4 + C_2 \dot{B}_8\right) - C_2\left[B_8 D_9 \dot{B}_6 + B_6\left(A_6 \dot{B}_7 - D_9 \dot{B}_8\right)\right] + \nonumber\\
 &\, B_7\left[A_6 \left(B_6 C_4 + C_2 \dot{B}_6\right) + B_8 C_2 \dot{B}_7 - B_6 D_9 \left(C_3 + 2M_{\rm pl}^2H\right)\right]\,,\\
 \mathcal{F}_{11} = &\, M_{\delta}^2 \left\{B_8 C_2 \dot{B}_6 + B_6\left[B_7 \left(C_3 + 2 M_{\rm pl}^2 H\right) - C_2\dot{B}_8\right]\right\}\,,\\
 \mathcal{F}_{12} = &\, 2M_{\rm pl}^2 B_7^2 M_{\delta}^2\left(B_8 - B_6\right)\,,
 \end{align}
\end{subequations}
where $A_4\to A_4-(\rho_{\rm m}+P_{\rm m})$ and we added $F_{12}$ for compactness of notation.

In the limit for $\alpha_{\rm T}=0$, our expressions might differ from those in \cite{Arjona2019b}, as their $G_{4\phi}$ terms come from the anisotropic stress equation and the coefficient $B_7$ is only replaced with its expression in this case. Note also that the original coefficients in \cite{Arjona2019b} are expressed in terms of $G_2$ and $G_3$ and then rewritten to use the coefficients in the field equations.\footnote{We corrected typos in the expressions for $\mathcal{F}_1$ and $\mathcal{F}_3$.}

\bibliographystyle{JHEP}
\bibliography{Supplementary_Materials.bbl}

\label{lastpage}